\documentclass[10pt]{article}
\usepackage{amsmath}
\usepackage{amssymb}
\usepackage{epsfig}
\renewcommand{\thesection}{\arabic{section}}

\newtheorem{theorem}{Theorem}[section]

\newtheorem{proposition}[theorem]{Proposition}
\newtheorem{lemma}[theorem]{Lemma}

\newcommand{\qed}{\hbox{\hskip 1pt \vrule width 7pt height 7pt
            depth 1.5pt \hskip 1pt}}

\newcommand{\R}{{\BR}}
\newcommand{\D}{{\cal D}}
\newcommand{\BC}{{\mathbb C}}

\newcommand{\BR}{{\mathbb R}}
\newcommand{\bfone}{{\bf 1}}
\newcommand{\id}{{\bfone}}
\newcommand{\dom}{{\rm Dom}}
\newcommand{\cA}{{\cal{A}}}
\newcommand{\cD}{{\cal{D}}}
\newcommand{\cH}{{\cal{H}}}
\newcommand{\cHo}{{\cal{H}}_0}

\newcommand{\fW}{\frak W}

\newcommand{\rIm}{{\rm{Im\, }}}
\newcommand{\rRan}{{\rm{Ran\, }}}
\newcommand{\rRe}{{\rm{Re\, }}}
\newcommand{\dist}{{\rm{dist\, }}}
\newcommand{\anal}{{\rm{anal}}}
\newcommand{\Ker}{{\rm{Ker\, }}}

\newcommand{\rTr}{{\rm{Tr\, }}}
\newcommand{\vep}{{{\varepsilon}}}
\newcommand{\btheta}{{\overline{\theta}}}

\newcommand{\oP}{{\overline{P}}}
\newcommand{\oR}{{\overline{R}}}

\newcommand{\thetabar}{{\overline{\theta}}}

\newcommand{\Ll}{{L^{(\ell)}}}

\newcommand{\innerprod}[2]{\left\langle {#1}, {#2}\right\rangle}
\newcommand{\bra}[1]{\left< #1 \right|}
\newcommand{\ket}[1]{\left| #1 \right>}
\newcommand{\set}[2]{\left\{ #1\vphantom{#2} \right. \left| \vphantom{#1} #2 \right\}}
\newcommand{\norm}[1]{\left\| #1 \right\|}

\newcommand{\GL}{{\rm GL}}
\newcommand{\Danal}{D^{\rm anal}}

\renewcommand{\d}{{\rm d}}
\newcommand{\bbbone}{\mathchoice {\rm 1\mskip-4mu l} {\rm 1\mskip-4mu l}
{\rm 1\mskip-4.5mu l} {\rm 1\mskip-5mu l}}
\newcommand{\lless}{<\!\!<}
\newcommand{\scalprod}[2]{\left\langle {#1}, {#2}\right\rangle}
\newcommand{\fer}[1]{(\ref{#1})}

\newcommand{\res}{{\cal R}}
\newcommand{\cF}{{\cal{F}}}
\newcommand{\cB}{{\cal{B}}}
\newcommand{\sOmega}{\Omega^{\#}}
\newcommand{\sI}{I^{\#}}
\newcommand{\stheta}{\theta}
\newcommand{\sstheta}{\theta^{\#}}
\newcommand{\szeta}{\zeta^{\#}}
\newcommand{\sK}{K^{\#}}
\newcommand{\betamax}{\beta}
\newcommand{\g}{g_{_{\!\delta\beta}}}

\renewcommand{\theequation}{\thesection.\arabic{equation}}
\let\subs\subsection
\renewcommand\subsection{\setcounter{equation}{0}
\gdef\theequation{\thesubsection.\arabic{equation}}\subs}
\let\sect\section
\renewcommand\section{\setcounter{equation}{0}
\gdef\theequation{\thesection.\arabic{equation}}\sect}

\begin{document}
{\allowdisplaybreaks[1]
\title{Theory of Non-Equilibrium Stationary
States\\ as a Theory of Resonances
}

\author{
M. Merkli\footnote{Department of Mathematics and Statistics, Memorial University of Newfoundland,  St. John's, NL,
Canada A1C 5S7;
merkli@math.mun.ca;  http://www.math.mun.ca/$\sim$merkli}
\thanks{Partly supported by an NSERC PDF, the Institute of Theoretical Physics of ETH Z\"urich, Switzerland, the Departments of Mathematics of McGill University and the University of Torotno, Canada.}
\and M. M\"uck \footnote{Fachbereich Mathematik, Johannes
Gutenberg-Universit\"at, D-55128 Mainz, Germany; mueck@mathematik.uni-mainz.de}
\thanks{Supported by DAAD under
grant HSP III.} \and I.M. Sigal \footnote{Department of Mathematics,
University of Toronto, Toronto, ON, M5S 2E4, Canada;
www.math.toronto.edu/sigal }
\thanks{Supported by NSERC under grant NA7901 }}
\maketitle

\begin{abstract}
We study a small quantum system (e.g. a simplified model for an atom
or molecule) interacting with two bosonic or fermionic reservoirs
(say, photon or phonon fields).
We show that the combined system has a family of stationary states,
parametrized by two numbers $T_1$, $T_2$ (``reservoir temperatures'').
If $T_1\neq T_2$, then these states are non-equilibrium, stationary
states  (NESS). In the latter case we show that they have
nonvanishing heat fluxes and positive entropy production.
Furthermore, we show that these states are dynamically
asymptotically stable. The latter means that the evolution with an
initial condition, normal with respect to any state where the
reservoirs are in equilibria at temperatures $T_1$ and $T_2$,
converges to the corresponding NESS. Our results are valid for the
temperatures satisfying the bound $\min(T_1, T_2)
> g^{2+\alpha}$, where $g$ is the coupling constant and $0<
\alpha<1$ is a power related to the infra-red behaviour of the
coupling functions.
\end{abstract}

\section{Introduction}\label{intro1}

\indent The present paper is a contribution to rigorous quantum
statistical mechanics. Key problems here are dynamical stability
of equilibrium states, and characterization (if not the
definition) and stability of non-equilibrium stationary states
(NESS).

While our understanding of the quantum equilibrium states, the
subject of equilibrium statistical mechanics (see \cite{BR, HHW,
Haagbook, Ru:StatMechRigResults}), and the recent progress in
proving their dynamical stability (\cite{JP:QFII, BFS:RtE,
DJP:PertTheoryWDynamics, DJ:SpecTheoryPauliFierz, Me:PosComm,
FM:rte, Me:Cond}) are satisfactory, results on non-equilibrium
stationary states are just beginning to emerge. The problem is that
we do not have a simple stationary characterization of NESS similar
to the principle of maximum of entropy or the KMS characterization
for equilibrium states. Thus it is remarkable that certain
characterizations of NESS and their stability were recently shown
for (idealized) particle systems coupled to Fermi reservoirs at high
temperatures, $\min(T_1,T_2)> C[\ln\frac 1g]^{-1}$, in
\cite{JP:NESS}, for XY-chains \cite{AP} and for coupled Fermi
reservoirs in \cite{DFG, FMU:DissipativeTransport, WABF, WAB}. Here,
$T_{1,2}$ are the temperatures of the reservoirs, and $g$ is the
coupling constant, which is assumed to be sufficiently small.

There are two rigorous approaches to non-equilibrium, quantum
statistical mechanics. One is based on scattering theory - wave (or
M{\o}ller) morphisms - and the other, on the theory of resonances via
complex deformations.
In this paper we follow the second approach which we believe applies
to a wider class of physical models.

In this paper we establish a spectral characterization of the
NESS and prove their dynamical stability for (idealized)
particle systems coupled to two Bose reservoirs (e.g. photons or
phonons) for reservoir temperatures satisfying $\min(T_1,T_2)> Cg^{2+\alpha}$, where $C$ is a constant and $0<\alpha<1$ ($\alpha=\frac{\mu-1/2}{\mu+1/2}$, where $\mu>1/2$ is given in Condition (B) below). Our approach applies to
an arbitrary finite number of bosonic or fermionic reservoirs; in
the latter case, it gives an extension of the results of
\cite{JP:NESS} to the temperature range mentioned above. Moreover,
we develop a perturbation theory for NESS and use it to
prove that the entropy production is strictly positive.

An appropriate iteration of our estimates in the spirit of the
spectral renormalization group of \cite{BCFS:SmoothFeshbach,
BFS:QED, BFS:RG, BFS:RtE} would give the above results for {\it all
temperatures}.
This extension will be presented elsewhere.

Similarly to \cite{JP:NESS}, we construct a NESS
from a zero (non-degenerate) resonance eigenvector of a certain
non-self-adjoint Liouville operator, $K$, acting on a positive temperature Hilbert space. The operator $K$ is the analogue of a $C$-Liouvillean in the terminology for $C^*$-dynamical systems \cite{JP:NESS}.

To show dynamical stability of a NESS we have to establish certain
long-time (ergodic)  properties of the evolution, $U(t)$, generated
by $K$. The operator $K$ does not belong to a class for which the
evolution is a priori known to exist (e.g. a class of normal or
accretive operators). To overcome this problem we establish a direct
connection between desired ergodic properties of $U(t)$ and certain
spectral properties of a complex deformation, $K_\theta$, $\theta
\in {\mathbb C}^2$, of $K$.  For technical reasons we can use
neither the complex deformations introduced in \cite{JP:QFII} nor
those introduced in \cite{BFS:RtE} but we combine both types, hence
$\theta$ is in ${\mathbb C}^2$ rather than in ${\mathbb C}$. (Such a
combination was already mentioned in \cite{BFS:RtE}). In order to
establish the desired spectral characteristics of the operator
family $K_\theta$, we use the method of the Feshbach map,
as developed in
\cite{BCFS:SmoothFeshbach,BFS:QED,BFS:RG}.

The present paper suffers from the main weaknesses shared by all the
works in the area, except, in some aspects, of \cite{BFS:RtE}:
\begin{itemize}
\item[(i)] The particle system has a finite-dimensional state space;
\item[(ii)] The restriction on the coupling functions is
severe;
\item[(iii)] Temperatures considered are  high.
\end{itemize}
To overcome the first limitation one would have to go beyond, or at
least significantly extend, the present approach.
The second limitation is due to use of translation analyticity
(which in our case is combined with the dilation analyticity), see
Remark 3 in Section 3. This analyticity is used in the present work
in a single place - in controlling the nonsingular part of the
resolvent of the operator $K_\theta$ near the zero resonance pole
by rendering this pole isolated and therefore the nonsingular part
of the resolvent analytic (see the estimate \eqref{eq8.1}). Without
it
the zero resonance of the operator $K_\theta$ is not separated from
the continuous spectrum and sits exactly at a threshold of the
latter.
Hence to control the nonsingular part of the resolvent near the zero
threshold becomes a delicate matter.

The paper \cite{BFS:RtE} has rather mild restrictions on the
coupling functions due to using the dilation analyticity. Since
\cite{BFS:RtE} deals with the dynamics near equilibrium, the
operator $K$ in this case is self-adjoint and an analogue of Eqn
\eqref{eq8.1} is obtained with help of an abstract spectral theory
of self-adjoint operators. Furthermore, \cite{BFS:RtE} handles
arbitrary temperatures
by employing the spectral renormalization group.  In the present
paper we take the first step in removing the high temperature
restriction. To this end we use the Feshbach map of \cite{BFS:RG}.
Already a single application of the Feshbach map considered in this
paper improves the temperature bounds yielding the results mentioned
above.
We also set the stage for the iteration of this map - the spectral
renormalization group method -
%
%
which would remove the restriction on the temperature altogether.
The iteration procedure will be carried out elsewhere.
(Note that the works \cite{DFG, FMU:DissipativeTransport} deal with
arbitrary temperatures, but the scattering approach they use seems
to be inapplicable to the models considered in this paper.)

A more detailed outline of our approach and of the organization of
the paper is given in Section 2.

\section{Model and Approach}\label{Sect2}

We consider a system consisting of a particle system, described by
a self-adjoint Hamiltonian $H_p$ on a Hilbert space $\cH_p$, and two bosonic
reservoirs, at inverse temperatures $\beta_1$ and $\beta_2$,
described by the Hamiltonians $H_{r1}$ and $H_{r2}$ acting on
Hilbert spaces $\cH_{r1}$ and $\cH_{r2}$, respectively. The full
Hamiltonian is
\begin{equation}\label{eq2.1}
H := H_0 + gv \ ,
\end{equation}
acting on the tensor product space $\cHo := \cH_p \otimes \cH_{r1}
\otimes \cH_{r2}$. Here
\begin{equation}
\label{hnot} H_0 := H_p \otimes \bfone \otimes \bfone + \bfone
\otimes H_{r1} \otimes \bfone + \bfone \otimes \bfone \otimes
H_{r2}
\end{equation}
is the unperturbed Hamiltonian, $v$ is an operator on $\cHo$
describing the interaction and $g\in{\mathbb R}$ is a coupling
constant.

For our key results we have to assume that ${\cal H}_p$ is finite-dimensional, but some of our results hold for $\dim{\cal H}_p=\infty$.

The operators $H_{rj}$ describe free scalar
(or vector, if wished) quantum fields on $\cH_{rj}$, the bosonic Fock spaces over the one-particle space $L^2({\mathbb R}^3,d^3k)$,
\begin{equation}
  H_{rj} = \int \omega(k) a^*_j(k) a_j(k) \, d^3k,
\label{fieldhamilt}
\end{equation}
where $a^*_j(k)$ and $a_j(k)$ are creation and annihilation
operators on $\cH_{rj}$ and $\omega(k) = |k|$ is the dispersion
relation for relativistic massless bosons. The interaction
operator is given by
\begin{equation}
\label{intop} v=\sum_{j=1}^2 v_j \mbox{\ \ \ with \ \ $v_j =
a_j(G_j) + a^*_j(G_j)$.}
\end{equation}
Its choice is motivated by standard models of particles
interacting with the quantized electromagnetic field or with
phonons.

In \fer{intop},  $G_j : k \mapsto G_j(k)$ is a map from $\R^3$ into ${\cal
B}({\cal H}_p)$,  the algebra of bounded operators on $\cH_p$, and
\begin{equation} \label{ahah}
 a_j(G_j) := \int  G_j(k)^* \otimes a_j(k) \, d^3k \qquad
 \text{and} \qquad a_j^*(G_j) := a_j(G_j)^*.
\end{equation}

If the coupling operators $G_j$  are such that
\begin{equation} \label{eqn_cond_sa}
g^2\int\limits_{\R^3} \left( 1 + |k|^{-1} \right) \norm{G_j(k)}^2
\, dk \qquad \text{is sufficiently small},
\end{equation}
then the operator $H$ is self-adjoint (see e.g. \cite{BFS:RtE}).

Now we set up a mathematical framework for non-equilibrium
statistical mechanics. Operators on the Hilbert space ${\cal H}_0$
will be called observables. (Only certain
self-adjoint operators on ${\cal H}_0$ are actually {\it physical} observables.)
As an algebra of observables describing the system we take the
$C^*$-algebra
\begin{equation} \label{calA}
{\cal A}={\cal B}({\cal H}_p)\otimes{\frak
W}(L^{2}_0)\otimes\frak{W}(L^{2}_0),
\end{equation}
where ${\frak W}(L^{2}_0)$ denotes the Weyl CCR algebra over
$L^{2}_0 := L^2({\mathbb R}^3, (1+|k|^{-1}) d^3k)$, i.e. the
$C^*-$algebra generated by the Weyl operators
$W_j(f):=e^{i\phi_j(f)}$,
$\phi_j(f):=\frac{1}{\sqrt{2}}\left(a^*_j(f)+a_j(f)\right)$, with $f
\in L^{2}_0$, see e.g. \cite{BR}.  States of the system are positive
linear (`expectation') functionals $\psi$ on the algebra ${\cal A}$,
normalized as $\psi(\bfone)= 1$.

The reason we chose ${\cal A}$ rather than ${\cal B}({\cal H}_0)$
is that the algebra ${\cal A}$ supports states in which each
reservoir is at a thermal equilibrium at its own temperature. More
precisely, consider the evolution for the $j$-th reservoir given
by
\begin{equation}
 \alpha_{rj}^t (A) := e^{iH_{rj}t} A e^{-iH_{rj}t}  .
\end{equation}
Then there are stationary states on the $j$-th reservoir algebra of
observables, $\frak{W}(L^{2}_0)$, which describe (single-phase) thermal
equilibria. These states are parametrized by the inverse temperature $\beta_j=1/T_j$ and their generating functional is given by
\begin{equation}
\label{mm31} \omega^{(\beta_j)}_{rj} \left(W_{j}(f)\right)=
\exp\left\{-\frac{1}{4}\int_{{\mathbb R}^3}\frac{e^{\beta_j
|k|}+1}{e^{\beta_j |k|}-1}|f(k)|^2 d^3k\right\}.
\end{equation}
The choice of the space $L^{2}_0$ above is dictated by
the need to have the r.h.s. of this functional finite. These
states are characterized by the KMS condition and are called the
$(\alpha^t_{rj} , \beta_j )$-KMS states.

{\it Remark.\ } It is convenient to define states $\psi$ on products
$a^{\#}(f_1) \ldots a^{\#}(f_n)$ of the creation and annihilation
operators, where $a^\#$ denotes either $a$ or $a^*$.  This is done
using derivatives $\partial_{s_k}$ of its values on the Weyl operators
$W(s_1f_1) \ldots W(s_nf_n)$ (see \cite{BR}, Section 5.2.3 and (\ref{eq2.15})).

Consider states (on ${\cal A}$) of the form
\begin{equation}\label{eq2.2}
\omega_0 := \omega_p \otimes\omega_{r1}^{(\beta_1)} \otimes\omega_{r2}^{(\beta_2)},
\end{equation}
where $\omega_p$ is a state of the particle system and
$\omega^{(\beta)}_{ri}$ is the $(\alpha^t_{ri} , \beta )$-KMS state
of the $i$-th reservoir. The set of states which are normal w.r.t.
$\omega_0$ is the same for any choice of $\omega_p$. A state normal
w.r.t. $\omega_0$ will be called a $\beta_1\beta_2$-normal state.

In the particular case $\omega_p(\cdot)={\rm Tr}(e^{-\beta_p H_p}\,\cdot)/{\rm Tr}(e^{-\beta_p H_p})$ we call $\omega_0$ a {\it reference state}.

The Hamiltonian $H$ generates the dynamics of observables $A\in
{\cal B}({\cal H}_0)$ according to the rule
\begin{equation}\label{eq2.6}
A \mapsto \alpha^t (A) := e^{iHt} A e^{-iHt} \ .
\end{equation}
Eqn (\ref{eq2.6}) defines a group of *-automorphisms of
$\cB(\cH_0)$. However, $\alpha^t$ does not map the
subalgebra ${\cal A}\subset{\cal B}({\cal H}_0)$ into itself, so \fer{eq2.6} does not define a dynamics on $\cal A$. To circumvent this problem we define the
interacting evolution of a class of {\it states on ${\cal A}$} by using the Araki-Dyson
expansion. Namely, we define the evolution of a state $\psi$ on $\cA$ which is normal w.r.t. $\omega_0$ by
\begin{equation}
\label{dyn1}
\psi^t(A):=\lim_{n \rightarrow \infty} \sum_{m=0}^\infty(ig)^m\int_0^tdt_1
\cdots\int_0^{t_{m-1}} dt_m \
\psi_n^{t,t_1,\ldots,t_m}(A),
\end{equation}
where the term with $m=0$ is $\psi(\alpha_0^{t}(A))$, and, for $m\geq 1$,
\begin{equation*}
\psi_n^{t,t_1,\ldots,t_m}(A) :=
\psi\left([\alpha_0^{t_m}(v_{n}),
\cdots [\alpha_0^{t_1}(v_{n}),\alpha_0^{t}(A)]\cdots]\right).
\end{equation*}
Here, $v_n=v^*_n \in
\cA$ is an approximating sequence for the operator $v$,
satisfying the relation
\begin{equation}
\label{eq2.13}
 \lim_{n\rightarrow\infty}\omega_{0}(A^{*}(v_{n}^* - v^*)(v_n - v)A)=  0,
\end{equation}
for all $A$ polynomials in $a_j^*(f)$, $j=1,2$, $f \in
L^{2}_0$. Such a sequence is constructed as follows. Let $\{e_m\}$  be an
orthonormal basis
in $L^{2}_0$.
We define the approximate creation operators
\begin{equation} \label{eq2.14}
 a_{j,n}^*(G_j) = \sum_{m=1}^{\mu}\langle{e}_{m},G_{j}\rangle
 b_{j,\lambda}^*(e_{m}),
 \end{equation}
where $n = (\lambda,\mu)$,
and, for any $f\in L^{2}({\mathbb R}^3)$ and $\lambda > 0$,
\begin{equation}
\label{eq2.15}
b^*_{j,\lambda}(f):=\frac{\lambda}{\sqrt 2 i}\left\{
W_j(f/\lambda) - \bfone - i W_j(if/\lambda) + i \bfone \right\}.
\end{equation}
Similarly we define the approximate annihilation operators
$a_{j,n}(G_j)$. These operators belong to $ \cA$. Via the above
construction we obtain the family of interactions $v_n$ which
belongs to $\cA$ and, as can be easily shown using \fer{mm31},
satisfies \fer{eq2.13}.

We show in Appendix \ref{appA} that under condition \fer{eq2.13}
the integrands on the r.h.s. of \fer{dyn1} are continuous
functions in $t_1,\ldots,t_m$, that the series is absolutely convergent and that the limit exists and is independent of the approximating sequence $v_n$.

Our goal is to understand stationary states of the interacting
system originating from $\beta_1 \beta_2$-normal states either by a
perturbation theory or through an ergodic limit of the full
evolution $\alpha^t$. These states are not equilibrium (KMS) states
states.  They will be called {\it non-equilibrium stationary states}
or NESS for short. Their main feature is that the energy (heat)
fluxes between the reservoirs and the particle system do not vanish.

Assuming certain smoothness and smallness conditions on the
coupling operators $gG_j(k)$ and assuming that the particle system
is effectively coupled to the reservoirs, we show that, starting
initially in any $\beta_1\beta_2$-normal state $\psi$, the system
converges, under the evolution $\alpha^t$, to a state $\eta$:
\begin{equation}
\psi^t\longrightarrow \eta \ \ \ \mbox{as $t\rightarrow\infty$}.
\label{new2.10}
\end{equation}
The convergence \fer{new2.10} is understood in the weak$*$ sense
on the sub-$C^*$-algebra of ``analytic observables''
\begin{equation}\label{eqn_4_9}
\cA_1 = \cB(\cH_p) \otimes \fW(\Danal) \otimes \fW(\Danal).
\end{equation}
Here, $\fW(\Danal)$ is the Weyl CCR algebra over the dense set
$\Danal\subset L^{2}_0$ which we define in Appendix~\ref{AppD}.
Roughly speaking, $\Danal$ consists of vectors from the space
$\bigcap_{b\ge 0} e^{-b|k|}L^{2}_0$ which have some analyticity
properties in $|k|$ and a certain behaviour at $k=0$. The density of
$\Danal\subset L^{2}_0$ implies that ${\cal A}_1$ is strongly dense
in $\cal A$. The construction of the state $\eta$ and the proof of
its stability, \eqref{new2.10}, will rely on the theory of
resonances for the evolution $\psi \rightarrow \psi^t$.

As mentioned in the introduction, so far, we do not have a simple
characterization of NESS. However, there is a key physical quantity
which differentiates between equilibrium and non-equilibrium
stationary states -- the collection of heat fluxes. In our case, the
heat flux, or more precisely the heat flow rate (i.e. the energy
flow rate due to thermal contact), $\phi_j$, $j = 1,2$, into the
$j$-th reservoir is given by
\begin{equation}\label{eq3.2}
\phi_j := \frac{\partial}{\partial t} \big|_{t=0} \alpha^t
(H_{rj}),
\end{equation}
and the heat flux, $\phi_0$,
into the particle system is defined as
\begin{equation}\label{eq3.6}
\phi_0 := \frac{\partial}{\partial t} \big|_{t=0} \ \alpha^t (H_p).
\end{equation}

The heat fluxes can be combined into a single quantity - the entropy
production. Motivated by the second law of thermodynamics ($dS=\sum
\beta_j dQ_j$) we introduce the observable of entropy production
(rate) as
\begin{equation}\label{eq3.1}
s := \sum\limits_{j=0}^2 \beta_j \phi_j \ ,
\end{equation}
where, for notational convenience, we write $\beta_0 := \beta_p$.
The entropy production, $EP (\omega)$, in a state $\omega$ is
defined as (see \cite{SL2, OHI, O, Ru:NaturalNES, Ru:EP,
Ru:DefineEP, Ru:TopicsStQM, JP:EP, JP:MathTheoryNEStQM, JP:NESS})
\begin{equation}\label{eq3.5}
EP (\omega ) = \omega (s ) \ .
\end{equation}
Since $s$ is not a bounded operator, we have to use an
approximation procedure similar to the one mentioned in the remark
after (\ref{mm31}) in order to define the r.h.s. of \fer{eq3.5} for sufficiently regular states.

The entropy production $EP(\eta)$ of the NESS is independent of the
particle state $\omega_p$ entering Definition (\ref{eq2.2}) of the
state of the decoupled system, since $\eta$ is independent of
$\omega_p$. Notice that $\eta (\phi_0) = \partial_t \big|_{t=0} \eta
\big(\alpha^t (H_p )\big)$, since $H_p$ is a proper observable (here
we assume $H_p\in {\cal B}({\cal H}_p)$) and $\eta$ is a continuous
and stationary state. Hence
\begin{equation}\label{eq3.7}
\eta (\phi_0) = 0 \ .
\end{equation}
Therefore, writing $\sum\limits_{j=1}^2 \eta (\phi_j) =
\sum\limits_{j=0}^2 \eta (\phi_j) = \eta
(\partial_t|_{t=0}\alpha^t(H_0)) = -\partial_t \big|_{t=0} \eta
(\alpha^t (v) ) =0$, we obtain
\begin{equation}\label{eq3.8}
\sum\limits_{j=1}^2 \eta (\phi_j ) = 0.
\end{equation}

Observe that the zero total flow relation (\ref{eq3.8}) and
Definition \fer{eq3.1} for the entropy production rate imply that
\begin{equation}\label{eq3.9}
EP(\eta ) = (\beta_1 - \beta_2) \eta (\phi_1) \ .
\end{equation}
Thus, the relation $EP (\eta) > 0$ is equivalent to
\begin{equation}\label{eq3.10}
\eta (\phi_1 ) > 0 {\hbox{ whenever }} T_2 > T_1  \ ,
\end{equation}
where $T_j = \beta^{-1}_j$ is the temperature of the $j$-th
reservoir. In other words, in the state $\eta$ the energy flows
from the hotter to the colder reservoir.

A general result due to \cite{JP:EP} shows that
$EP(\omega)\geq 0$ for any NESS
$\omega$. We show that for the NESS $\eta$,
\begin{equation*}
EP(\eta)>0\ \ \ \mbox{iff\ \ \ $\beta_1\neq\beta_2$},
\end{equation*}
see Theorem \ref{theorem3.2} and Section \ref{Sect10} for a precise statement of this result.
Moreover, we develop a perturbation theory for the NESS and
compute $EP(\eta)$ in leading order in the coupling constant $g$.

Let us outline the main steps of our proof of the convergence
\fer{new2.10} (c.f. \cite{JP:NESS}). We pass to the Araki-Woods
GNS representation of $({\cal A},\omega_0)$, with
$\omega_0$ of the form (\ref{eq2.2}) and $\omega_p (A) :=
{\rm Tr}(e^{-\beta_pH_p}A) / {\rm Tr} (e^{-\beta_pH_p})$;
\begin{equation*}
(\cA , \omega_0) \to (\cH , \pi , \Omega_0),
\end{equation*}
where $\cH$, $\pi$ and $\Omega_0$ are a Hilbert space, a
representation of the algebra $\cA$ by bounded operators on $\cH$,
and a cyclic element in $\cH$ (meaning that $ \overline{\pi(\cA) \Omega_0}
= \cH)$ s.t.
\begin{equation*}
\omega_0(A) = \innerprod{ \Omega_0 }{ \pi (A) \Omega_0}.
\end{equation*}

The GNS representation provides us with a Hilbert space framework which we use to convert the dynamical
problem described above into a spectral problem for a certain non-self-adjoint
operator $K$ on the Hilbert space $\cH$.
With the free evolution $ \alpha_0^{t}(A):=e^{itH_0}Ae^{-itH_0}$
one associates the unitary one-parameter group $U_0 (t) = e^{i t L_{0}}$ on
$\cH$ s.t.
\begin{equation}
\pi (\alpha^t_0 (A) ) = U_0(t) \pi (A) U_0 (t)^{-1}
\label{2.25a}
\end{equation}
and $U_0 (t) \Omega_0 = \Omega_0 $. Define the
operator $L^{(\ell)}:=L_0+g\pi(v)$ on the dense domain $\dom(L_0)\cap\dom(\pi(v))$. Here $\pi(v)$ can be defined either using
explicit formulae for $\pi$ in the Araki-Woods representation given below or
by using the approximation, $v_n\ \in \mathcal{A}$,
for the operator $v$ constructed above.
By the Glimm-Jaffe-Nelson commutator
theorem the operator $L^{(\ell)}$ is essentially self-adjoint; we denote its
self-adjoint closure again by the same symbol $L^{(\ell)}$. The operator
$L^{(\ell)}$ induces the one-parameter group $\sigma^t$ on
$\pi({\cal A})''$, the weak closure of  $\pi({\cal A})$,
\begin{equation}
\label{eq2.26}
\sigma^{t}(B):=e^{itL^{(\ell)}} B e^{-itL^{(\ell)}}
\end{equation}
for any $B \in \pi(\cA)''$. Let $\psi$ be a state on the algebra
$\cA$ normal w.r.t. $\omega_0$, i.e.
\begin{equation}\label{eq2.27}
\psi(A)= {\rm Tr} (\rho  \pi(A))
\end{equation}
for some positive trace class operator $\rho$ on $\cH$ of trace one. It is shown
in Appendix \ref{appA} that for $\psi$ as above the limit on the r.h.s. of
\fer{dyn1} exists and equals
\begin{equation}\label{eq2.28}
\psi^t(A)= {\rm Tr} (\rho \sigma^{t}(\pi(A))).
\end{equation}
In particular, the limit is independent of the choice of the approximating family $v_n$.

Due to \fer{eq2.28} the dynamics on normal states, defined in \fer{dyn1}, gives rise to the dynamics on the
Hilbert space $\cH$, determined by a one-parameter group $U(t)$, satisfying
\begin{equation}\label{eq4.1'}
{\rm Tr}(\rho U (t) \pi (A) U (t)^{-1}) = \psi^t(A),
\quad \forall A \in \cA.
\end{equation}
Due to the fact that the von Neumann algebra $\pi({\cal A})''$ has a large commutant (which isomorphic to $\pi({\cal A})''$, as is known from Tomita-Takesaki theory), relation \fer{eq4.1'} does not define $U(t)$ uniquely; however, if we
impose in addition to \fer{eq4.1'} the invariance condition
\begin{equation}
U(t)\Omega =\Omega, \label{**}
\end{equation}
where $\Omega$ is a fixed cyclic and separating vector, then
$U(t)$ is uniquely determined. (The vector $\Omega$ is called cyclic if $\pi({\cal A})\Omega$ is dense in $\cal H$ and separating if $\pi({\cal A})'\Omega$ is dense in $\cal H$, the prime denoting the commuant.)
If $\Omega$
were the vector representing an equilibrium state then $U(t)$ satisfying \fer{eq4.1'} and \fer{**}
would be a unitary group. In the non-equilibrium case $\beta_1\neq \beta_2$, one can see that \fer{**} cannot be satisfied for a unitary $U(t)$
implementing the dynamics as in \fer{eq4.1'}. For technical
reasons, we choose $U(t)$ to satisfy \fer{**} for a convenient
vector $\Omega$, rather than to be unitary (c.f. \cite{JP:NESS}).

We will show that $U(t)$ is strongly differentiable on a dense set
of vectors and we will calculate explicitly its generator,
$K:=-i\frac{\partial}{\partial
  t}U(t)|_{t=0}$. In the non-equilibrium situation $K^*\neq K$ ($U(t)$ is not unitary!) and
\fer{**} implies that $K\Omega=0$. The main effort of our analysis
is to derive enough spectral information on the operator $K$ to
enable us to show \fer{new2.10} and to identify the NESS with
\begin{equation}
\eta(A)=\scalprod{\Omega^*}{\pi(A)\Omega}, \label{weaksense}
\end{equation}
where $\Omega^*$ is a zero resonance of the operator $K^*$:
$K^*\Omega^*=0$ (in the sense of distributions) and $\Omega^* \in \cD'_{\anal}$, for an appropriate
dense set $\cD_\anal \subset \cH$, and $A$ are such that $\pi (A)
\Omega \in \cD_\anal$.

In order to obtain rather subtle spectral information on the
operator $K$, and to give a precise meaning to expression
\fer{weaksense}, we develop a new type of spectral deformation,
$K\mapsto K_\theta$, with a spectral deformation parameter
$\theta\in {\mathbb C}^2$, in combination with an application of a
Feshbach map acting on $K_\theta$.

In conclusion of this outline we present here the GNS triple
provided by the Araki-Woods construction, which forms a
mathematical framework for our analysis
(see \cite{BFS:RtE, JP:QFII, BR} for details and \cite{AW,
HHW} for original papers).
In the Araki-Woods GNS representation the (positive temperature)
Hilbert space is given by
\begin{equation}\label{eqn_6_1}
  \cH = \cH^p \otimes \cH^r,
\end{equation}
where $\cH^p = \cH_p \otimes \cH_p$ and $\cH^r = \cH^{r1} \otimes
\cH^{r2}$ with
\begin{equation}\label{eqn_6_2}
  \cH^{rj} = {\cal H}_{rj} \otimes {\cal H}_{rj}.
\end{equation}
We denote by $a^{\#}_{\ell, j}(f)$ (resp., $a^{\#}_{r, j}(f)$) the
creation and annihilation operators which act on the left (resp.,
right) factor of (\ref{eqn_6_2}). They are related to the zero
temperature creation and annihilation operators $a_j^{\#}(f)$ by
\begin{equation}\label{eq_AWrepr}
  \pi(a_j(f)) = a_{\ell j} (\sqrt{1+\rho_j} \, f) +
  a^*_{r j} (\sqrt{\rho_j} \, \bar{f})
\end{equation}
and
\begin{equation}\label{eq_AWrepr'}
  \pi'(a_j(f)) = a^*_{\ell j} (\sqrt{\rho_j} \, f) +
  a_{r j} (\sqrt{1+\rho_j} \, \bar{f})
\end{equation}
where $\rho_j \equiv \rho_j(k) = (e^{\beta_j \omega(k)} - 1)^{-1}$
with $\omega(k) = |k|$. Finally, we denote $\Omega_r :=
\Omega_{r1} \otimes \Omega_{r2}$, where $\Omega_{rj} :=
\Omega_{rj,\ell} \otimes \Omega_{rj,r}$ are the vacua in
$\cH^{rj}$. Thus, $\Omega_r$ is the vacuum in $\cH^r$.

Definition (\ref{eq2.2}) and our choice of $\omega_p$ made at the
beginning of this section imply that
\begin{equation}\label{eq_6_3}
  \Omega_0 = \Omega_p \otimes \Omega_r \quad \text{with} \quad
  \Omega_p \equiv \Omega_p^{(\beta_p)} =\frac{\sum_j e^{-\beta_pE_j/2}
  \varphi_j \otimes \varphi_j}{[\sum_j e^{-\beta_pE_j}]^{1/2}},
\end{equation}
where $E_j$ and $\varphi_j$ are the eigenvalues and
normalized eigenvectors of $H_p$.

The self-adjoint operator $L_0$ generating the free
evolution $U_0(t)$ defined in \fer{2.25a} is of the form $L_0 = L_p \otimes \id^r + \id^p
\otimes L_r$ with $L_r = \sum_{j=1}^2 L_{rj}$. The operator $L_p$
has the standard form
\begin{equation*}
  L_p = H_p \otimes \id_{p} - \id_{p} \otimes H_p,
\end{equation*}
and
\begin{equation*}
  L_{rj} = \int \omega(k) \left( a^*_{\ell,j}(k) a_{\ell,j}(k)
     - a^*_{r,j}(k) a_{r,j}(k) \right) \,d^3k.
\end{equation*}
The operator $K$ can be written as $K=L_0 + g(V - W)$ with $V=\pi(v)$
and $W=\pi'(w)$ with $w$ a non-self-adjoint operator obtained by a
simple transformation of $v$.

A standard argument shows that the spectrum of the operator $L_0$
fills the axis $\BR$ with the thresholds and eigenvalues located
at $\sigma(L_{p})$ and with $0$ an eigenvalue of multiplicity at least $\dim
H_p$ and at most $(\dim H_p)^2$ (depending on the degeneracy of the spectrum of $L_p$). A priori we do not know anything about the spectrum of the non-self-adjoint
operator $K$ besides the fact that it has an eigenvalue $0$. For all we
know its spectrum might fill in the entire complex plane! Thus understanding the evolution generated by the operator $K$ is a
subtle matter.

This paper is organized as follows. In Section 3 we give a precise
formulation of our assumptions, state the results and discuss
assumptions and results. In Section 4 we present the Hilbert space
framework and we define the vector $\Omega$ and the evolution $U(t)$ and in Section 5 we describe
the generator $K$. In Section 6 we introduce the complex
deformation $K_\theta$ of $K$ and establish the connection between
the resolvents of $K$ and $K_\theta$. In Section 7 we
establish the spectral properties of $K_\theta$ which we then use in Section 8 to express the dynamics in terms of an integral over the resolvent of $K_\theta$. In Section 9 we prove our first main result, the
existence and explicit form of the NESS, and its dynamical
stability. Section 10 contains a supplementary result on the structure of the level shift operators, used in Section 7. In Section 11 we develop a perturbation theory for NESS
and in Section 12 we prove the positivity of the entropy
production.
Finally, in Appendices \ref{appA}--\ref{appE} we collect some technical results.

\section{Assumptions and Results}
\label{Sect3}

In order to state assumption (B) below, it is practical to define
the map $\gamma: L^2({\mathbb R}^3) \to L^2({\mathbb R}\times S^2)$,
\begin{equation}\label{eq_2_8}
  (\gamma f)(u,\sigma) =
  \sqrt{|u|}\
\left\{
\begin{array}{ll}
 f(u\sigma), & u\geq 0,\\
 -\overline{f}(-u\sigma), & u<0.
\end{array}
\right.
\end{equation}
Let $j_\theta(u) = e^{\delta {\rm sgn}(u)} u + \tau$ for $\theta =
(\delta, \tau) \in \BC^2$ and $u \in \BR$ (see \fer{eqn_A.25}) and
define $({\gamma_{\theta}} f)(u, \sigma) = (\gamma f)(j_\theta(u),
\sigma)$, for $f\in L^2(\BR  \times S^2)$, $\theta\in{\mathbb R}^2$.
The maps  $\gamma$ and ${\gamma_{\theta}}$ have obvious extensions
to operator valued functions.

\begin{itemize}
\item[(A)] {\it Ultraviolet cut-off.\ } $\int \|G_j(k)\|^2 e^{a|k|^2}[1+ |k|^{-1}]d^3k
< \infty$ for some $a >0$.

\item[(B)] {\it Analyticity.\ }
{}For $j=1,2$ and every fixed $(u,\sigma)\in{\mathbb R}\times S^2$, the maps
\begin{equation}
\theta \mapsto (\gamma_\theta G_j)(u,\sigma)
\end{equation}
from ${\mathbb R}^2$ to the bounded operators on ${\cal H}_p$ have analytic continuations to
\begin{equation}\label{eq_2_11}
\Big\{ (\delta, \tau) \in \BC^2 \big| |\rIm\delta| < \delta_0,
|\tau|
           < \tau_0 \Big\} \ ,
\end{equation}
for some $\delta_0$, $\tau_0 > 0$, $\frac{\tau_0}{\cos\delta_0} \le
\frac{2\pi}{\beta}$, where $\beta=\max(\beta_1,\beta_2)$. Moreover,
\begin{equation}
\label{eqn_cond_B}
\norm{G_j}_{\delta\beta_j, \mu, \theta} := \sum_{\nu=1/2,\mu}
\left[\ \int\limits_{\BR \times S^2}
\left\|\gamma_\theta\left( \frac{\sqrt{|u|+1}}{|u|^\nu} e^{\delta\beta_j |u|/2} G_j\right)(u,\sigma)\right\|^2 du d\sigma\right]^{1/2}<\infty,
\end{equation}
for some fixed $\mu>1/2$ and where $\delta\beta_j = \beta- \beta_j$.
For future references, $\theta_0 := (\delta_0, \tau_0)$.

\item[(C)] {\it Non-Degeneracy of the Particle System.}
We have $\dim\cH_p=N<\infty$ and the Hamiltonian $H_p$ has non-degenerate spectrum $\{E_n\}_{n=0}^{N-1}$.

\item[(D)] {\it Fermi Golden Rule Condition.} We have, for $j=1,2$,
     \begin{eqnarray} \label{eqn_2.7}
       \gamma_0 := \min_{0 \le n < m \le N-1} \int\limits_{{\mathbb R}^3} \delta(|k| -E_{mn})  |G_j(k)_{nm}|^2 d^3k > 0,
     \end{eqnarray}
 where $E_{mn}=E_m-E_n$, $G_j(k)_{mn}:=\innerprod{\varphi_m}{G_j(k)\varphi_n}$, the $\varphi_n$ are normalized eigenvectors of $H_p$ corresponding to the eigenvalues
 $E_n$, and $\delta$ is the Dirac delta distribution.

\item[(E)] Either $\dim {\cal H}_p=2$, or $\dim {\cal H}_p\geq 3$ and the inverse temperatures satisfy
\begin{equation}
|\beta_1-\beta_2| \leq c, \mbox{\ \ or\ \ $|\beta_1-\beta_2|\geq C$ and $\min_j\beta_j\geq C$ },
\label{aaahhh}
\end{equation}
where $0<c,C<\infty$ are constants depending only on the interaction $G_{1,2}$.
\end{itemize}

{\it Remarks.} {1)\ } The map \fer{eq_2_8} has the following
origin. In the positive-temperature representation of the CCR (the
Araki-Woods representation on a suitable Hilbert space, see
Appendix A), the interaction term $v_j$ is
represented by $a_j(\widetilde\gamma_{\beta_j}G_j)+ a^*_j(\widetilde\gamma_{\beta_j}G_j)$, where
\begin{equation}
\widetilde\gamma_\beta := \sqrt{\frac{u}{1-e^{-\beta u}}}\ \gamma.
\label{gammatilde}
\end{equation}
\indent 2)\ The Condition (A) can be removed at expense of a
slightly more involved proof of existence of the operator
$\Gamma(z)$ in Lemma \ref{lemma4.2}, see the remark after the proof
of Lemma \ref{lemma4.2}. However, since the ultraviolet behavior of
the coupling functions is inessential in the models considered, we
choose a stronger assumption over extra steps in the corresponding
proof.

3)\ A class of interactions satisfying Conditions (A) and (B) is given by
$G_j(k)=g(|k|) G$, where $g(u)=u^{\alpha}e^{-u^2}$, with $u\geq 0$,
$\alpha=n+1/2$, $n=0,1,2,\ldots$, and $G=G^*\in{\cal B}({\cal
H}_p)$. A straightforward estimate gives that the norms
(\ref{eqn_cond_B}) have the bound
\begin{equation}
 \norm{G_j}_{\delta\beta_j,\mu, \theta} \leq C \left( 1+e^{(\delta\beta_j)^2/4
}\right) ||G||,\label{b1}
\end{equation}
provided
$\mu < \alpha + 1$, where the constant $C$ does not depend on the
inverse temperatures, nor on $\theta$ varying in any compact set (compare this with the bound
(4.13) of \cite{JP:NESS}). The restriction $\alpha=n+1/2$ with
$n=0,1,2,\ldots$ comes from the requirement of translation
analyticity (the $\tau$--component of $\theta$), which appears
also  in \cite{JP:NESS}.

4)\ The condition $\tau_0/\cos\delta_0<2\pi /\beta$ after
\fer{eq_2_11} guarantees that the square root in \fer{gammatilde} is
analytic in translations $u\mapsto u+\tau$.

5)\ What we need in our analysis is that the level shift operator
$\Lambda_0$, the $N\times N$ matrix defined in \fer{m2}, has a
spectral gap at zero which is bounded below by a strictly positive
constant independent of the temperatures. Condition (E) ensures this
property. If one can show the desired property of the gap by other
means then Condition (E) can be dropped.

Let
\begin{equation}
\sigma :=\min \left\{|\lambda-\mu|\ |\ \lambda,\mu\in\sigma(H_p), \lambda\neq\mu\right\}
\label{gap}
\end{equation}
and define
\begin{equation*}
g_0:= C\sigma^{1/2} \sin(\delta_0)\left[(1+\beta_1^{-1/2}+\beta_2^{-1/2}) \max_j \sup_{|\theta|\leq\theta_0}\|G_j\|_{\delta\beta_j,1/2,\theta}\right]^{-1},
\end{equation*}
where $C$ is a constant depending only on $\tan\delta_0$, and $\delta\beta_j=\beta-\beta_j$, $\delta\beta_p=|\beta-\beta_p|$, and set
\begin{equation}
g_1:=\min\Big(
(g_0)^{1/\alpha},[\min(\beta_1^{-1},\beta_2^{-1})]^{\frac{1}{2+\alpha}}\Big),
\label{g1}
\end{equation}
where $\alpha=\frac{\mu-1/2}{\mu+1/2}$, and $\mu>1/2$ is given in \fer{eqn_cond_B}.

The main results of this paper are given in the following theorems, where by a ``state'' on a subalgebra (which is not necessarily a $C^*$-subalgebra), we mean a positive normalized linear functional.

\begin{theorem}
\label{thm2.1} Assume conditions (A) -- (E) are obeyed for some
$0<\beta_1,\beta_2<\infty$, $\mu>1$, and let $\beta = \max(\beta_1,
\beta_2)$.

If $0<|g|<g_1$ then there is a stationary state $\eta =
\eta_{\beta_1\beta_2}$, defined on a strongly dense subalgebra ${\cal A}_1$ of $\cal A$ (see \fer{calA}), satisfying
\begin{equation} \label{eqn_2.15}
  \psi^t \to \eta,\ \ t\rightarrow\infty
\end{equation}
for any $\beta_1 \beta_2$-normal initial state $\psi$. $\eta$ is continuous in the norm of $\cA$.
The convergence is in the weak$*$ sense (i.e., pointwise for each $A\in\cA_1$).
For $A\in{\cal A}_1$, $\eta(A)$ is analytic in $g$.
\end{theorem}

{\it Remark.\ } 6)\ Our analysis shows that the NESS is actually
defined on a bigger (but somewhat less explicit) Banach space of
operators ${\cal A}_{0}\supseteq{\cal A}_1$ (see \fer{a_0}), and the
convergence to the NESS, \fer{eqn_2.15}, holds on ${\cal A}_{0}$. On
$\cA_1$ one can introduce a ``deformation norm''
$|||\cdot|||\leq\|\cdot\|$, see \fer{simnorm}, such that in this
norm, the convergence in \fer{eqn_2.15} is uniform,
$\sup_{A\in\cA_1}|\psi^t(A)-\eta(A)|/|||A|||\rightarrow 0$.
Moreover, on ${\cal A}_1$, the convergence is exponentially fast for
initial conditions $\psi$ in a dense set (in the topology of bounded
linear functionals on $\cA$) -- this set is the convex hull of
vector states with deformation analytic vectors.

\begin{theorem}
\label{theorem3.2} Assume that the
conditions of Theorem \ref{thm2.1} are satisfied and let
$\eta_{\beta_1\beta_2}$ and $g$ be as in Theorem \ref{thm2.1}. Let
$\beta_1 \ne \beta_2$, and let $g$ and $|\beta_1-\beta_2|$ be
sufficiently small (independently). If either  $G_1=G_2$ or
$\dim{\cal H}_p=2$, then $EP(\eta_{\beta_1 \beta_2}) >0$.
\end{theorem}

Our analysis gives a stronger result than the one presented in Theorem \ref{theorem3.2}. Namely, for $\mu>3/2$, we show that $EP(\eta_{\beta_1\beta_2})>0$, provided $o(g^0)O(\delta\beta)\leq \eta'$, where $\delta\beta=|\beta_1-\beta_2|$ (see Theorem \ref{thm_12_1}). Here, $\eta'$ depends on the inverse temperatures and the coupling functions and is given by
\begin{equation*}
  \eta' = \frac{2\pi}{\sqrt N} \sum_{j>i} (\gamma_j e^{\beta_1 E_{ji}} - \gamma_i)
  \frac{E_{ji}\ g_{ji}(E_{ji})^2}{e^{\beta_1E_{ji}}-1},
\end{equation*}
where $E_{ji} = E_j - E_i$, $g_{ji}(E)^2 = \int_{{\mathbb R}^3}d^3k
|\innerprod{\varphi_i}{G_1(k)\varphi_j}|^2\delta(E_{ji}-\omega)$ (see Condition (C)). The numbers $\gamma_j \ge 0$ are the
coordinates (in the basis $\{\varphi_j\otimes\varphi_j\}$ of Null$(L_p)$) of the unique vector $\zeta^*$ in the kernel of the adjoint level shift operator $\Lambda_0^*$, at the value $\beta_p=0$ (and normalized as $\sum_j\gamma_j=\sqrt N$). (The operator $\Lambda_0$ is defined in Section \ref{Sect13}.)

By general arguments one can prove that $\eta'\geq 0$ for
$\beta_1>\beta_2$, $\eta'\leq 0$ for $\beta_1<\beta_2$, and
$\eta'=0$ if $\beta_1=\beta_2$. We also show that $\eta'>0$ for
$\beta_1>\beta_2$ and $\eta'<0$ for $\beta_1<\beta_2$, for all
$\beta_1,\beta_2$, except possibly for finitely many values in any
compact set, see the remark at the end of Section \ref{Sect10}.

The dependence of $\eta'$ on $\delta\beta$ is determined by the coordinates $\gamma_j$. We compute those in the cases when $G_1=G_2$ and $\dim{\cal H}_p=2$ (see the proof of Theorem \ref{thm_12_1}, and equation \fer{eqn_11.26}, respectively).

{\it Remarks.\ }
7)\ Using Araki's theory of perturbation of KMS states, one shows that if the temperatures of both reservoirs are equal then the limit state is an equilibrium state and has zero entropy production. Non-existence of equilibrium states for $\beta_1\neq\beta_2$ has been shown in \cite{MMS1}.

8)\
For a model with fermionic reservoirs, using a sufficiently fast convergence rate in
(\ref{eqn_2.15}) (e.g. $O(t^{-\alpha})$ with $\alpha > 1$
suffices) and the fact that $\eta$ is not a normal state for $\beta_1\neq\beta_1$, it has been shown by an abstract argument that
$EP(\eta_{\beta_1\beta_2}) > 0$,
provided $|\beta_1 - \beta_2| \ge Cg$ for some $C > 0$
(see \cite{JP:NESS}). Instead of this indirect derivation we compute
$EP(\eta_{\beta_1\beta_2})$ to the leading order in $g$ and derive the results stated in the theorem.

9)\
The condition $G_1=G_2$ can be relaxed to $G_1-G_2$ being small in a suitable sense.

\section{Spectral Theory of NESS}
\label{Sect4}

In this section we outline a spectral theory of  NESS applicable
to Bose and Fermi reservoirs. Our approach follows the one
developed for the Fermi reservoirs in \cite{JP:NESS}. Fix a state,
$\omega_0$, of the form (\ref{eq2.2}) with $\omega_p (A) := {\rm
Tr}(e^{-\beta_pH_p}A) / {\rm Tr} (e^{-\beta_pH_p})$.

In this and the next section we use Condition (A) and
\begin {equation} \label{part}
(H_p + i)^{-1}
\mbox{\ is of a trace class.}
\end {equation}
In particular, the $\theta$-analyticity of the coupling
functions and the finiteness of the dimension of the particle space
are not required.

We pass to the Araki-Woods GNS representation for the unperturbed
system:
$$ (\cA , \omega_0) \to (\cH , \pi , \Omega_0)$$
where $\cH$, $\pi$ and $\Omega_0$ are a Hilbert space, a
representation of the algebra $\cA$ by bounded operators on $\cH$,
and a cyclic element in $\cH$
s.t.
\begin{equation*}
\omega_0(A) = \innerprod{ \Omega_0 }{ \pi (A) \Omega_0}.
\end{equation*}
There is also an anti-linear representation, $\pi'$, of the
algebra $\cA$ in bounded operators on the space $\cH$, s.t. $\pi'$
commutes with $\pi$ (i.e. $[\pi' (A) , \pi (B) ] = 0 \ \forall A,
B \in \cA ) $, and $ \overline{\pi'(\cA) \Omega_0 } = \cH$.

The full dynamics is implemented by a one-parameter group $U(t)$ satisfying
\begin{equation}\label{eq4.1}
U(t) B\Omega = \sigma^t(B)\Omega, \quad \forall B \in \pi(\cA),
\end{equation}
where $\Omega$ is a cyclic and separating vector for $\pi({\cal
A})$ to be specified below and where
\begin{equation}
\label{fm4}
\sigma^t(B):=e^{itL^{(\ell)}} B e^{-itL^{(\ell)}}
\end{equation}
for any $B \in \pi(\cA)$ with, recall,  $L^{(\ell)}:=L_0+g\pi(v)$
and $L_0$ is defined on the line before \eqref{2.25a} (see also the
paragraph after \eqref{eq_6_3}). Observe that \fer{eq4.1} implies
that
\begin{equation}\label{eq4.2}
U (t) \Omega = \Omega.
\end{equation}
If the state $\omega$ corresponding to $\Omega$ is stationary
(i.e. $\omega^t=\omega$) then $U (t)$ comes out to be unitary. In
our situation
we expect that there is no $\omega_0$-normal stationary state and $U(t)$
will be a non-unitary group.

We pick the vector $\Omega$ as follows. Let $\beta =
\max_{j=1,2}{\beta_j}$. We define
\begin{equation}
  \Omega := e^{-\beta\Ll/2} \Omega_0/\|e^{-\beta\Ll/2} \Omega_0\|.
\end{equation}
The facts that $\Omega$ is well defined, i.e., that $\Omega_0 \in \dom(e^{-\beta\Ll/2})$, and that $\Omega$ is cyclic and separating, are established in Proposition \ref{cyc&sep} at the
end of this section.

The family $U (t)$ is not unitary since $\omega :=
\innerprod{\Omega}{\pi(\,\cdot\,) \Omega}$ is not stationary:
\begin{eqnarray}
\scalprod{U(t)\pi(A)\Omega}{U(t)\pi(B)\Omega} &=& \scalprod{\Omega}{\sigma^t(\pi(A^*B))\Omega}=\omega^t(A^*B)\nonumber\\
&\neq&\omega(A^*B) = \scalprod{\pi(A)\Omega}{\pi(B)\Omega},
\label{notunitary}
\end{eqnarray}
for some $A,B,t$.
Let now $\psi$ be an $\omega_0$-normal state corresponding to the
vector $Q\Omega \in \cH$ (i.e. $\psi(A) = \innerprod{ Q\Omega }{
\pi(A) Q\Omega } $), where $Q \in \pi' (\cA)$ then
\begin{equation}\label{eq4.4}
\psi^t (A) =
\innerprod{ Q\Omega }{\sigma^t (\pi(A)) Q\Omega } \ .
\end{equation}
Due to Eqns \fer{mm32} and
$\sigma_0^{t}(\pi(A)):=\pi(\alpha_0^t(A))$, and due to the convergence of $\sigma_{(n)}^t(\pi(A))$ to $\sigma^t(\pi(A))$ established after \fer{mm32}, we see that the operator $Q$
commutes with $\sigma^t (\pi(A))$. Using this together with
 \fer{eq4.1} and \fer{eq4.2} we arrive at
\begin{equation}\label{eq4.5}
\psi^t (A) = \innerprod{ Q^* Q \Omega }{ U(t) \pi (A) \Omega } \ .
\end{equation}
This key formula, due to \cite{JP:NESS}, connects the long time
behaviour of $\psi^t(A)$ with spectral properties of $U
(t)$ or its generator. We explain what this means.

Assume we can show that, for a certain class of $\phi$ and $\Psi$,
and as $t\rightarrow\infty$,
\begin{equation}\label{eq4.6}
\innerprod{ \phi}{ U (t) \Psi \rangle \to \langle \phi, P\Psi } \,
\end{equation}
where $P$ is the eigenprojection on the fixed point subspace of $U
(t)$ (i.e. $U (t) P = PU (t) = P)$, which we assume for a moment
to exist. Relations \fer{eq4.5} and \fer{eq4.6} imply
\begin{equation}\label{eq4.60}
\psi^t\rightarrow\eta,\ \ \ t\rightarrow\infty,
\end{equation}
where the state $\eta$ is defined (on an appropriate set of
observables) by
\begin{equation}
\eta(A):=\scalprod{Q^*Q\Omega}{P\pi(A)\Omega}. \label{soso}
\end{equation}
We will show below that \fer{eq4.6} holds for some unbounded
projection operator $P$. To
understand the structure of this operator, we proceed as follows.

We will show that $U (t)$ is strongly
differentiable on a  dense set and we will compute its non-self-adjoint
generator, $K := -i \frac{\partial}{\partial t}|_{t=0} U(t)$,
which satisfies (see (\ref{eq4.2}))
\begin{equation}\label{eq4.7}
K \Omega = 0.
\end{equation}
\indent The operator $P$ is the eigenprojection onto the
eigenspace of $K$ associated with the eigenvalue $0$ (i.e.
$KP=PK=0$). We show that $\dim P=1$ and
\begin{equation}\label{eq4.8}
P = \ket{ \Omega } \bra{ \Omega^*}
\end{equation}
for some $\Omega^*\notin\cH$, satisfying $K^*\Omega^* = 0$ in a weak
sense ($\Omega^* \in \D'_\anal$, where $\D_\anal = \cup_{{\rm
Im}\theta > 0} \dom (U_\theta)$ with the family $U_\theta$ defined
in Section \ref{Sect6}).
Understanding the nature of the vector $\Omega^*$, which we call the
{\it NESS vector}, is a goal of our analysis.

Substituting \fer{eq4.8} into \fer{soso} and using that
$\innerprod{ Q^* Q \Omega
  }{ \Omega } = \norm{ Q \Omega }^2
= \psi (\bfone)=1$, we obtain
\begin{equation}\label{eq4.10}
\eta (A) = \innerprod{ \Omega^*}{ \pi (A) \Omega }.
\end{equation}

Since $\Omega^*\notin\cH$ the state $\eta$ is not normal but it is well defined for a
dense set of observables.  The question now is what is $\Omega^*$?
The answer, provided in subsequent sections, is that $\Omega^*$ is
a resonance of $K^*$.

In the following sections we construct a mathematical framework
which provides meaningful expressions replacing formal ones,
(\ref{eq4.6})--(\ref{eq4.10}), and with the help of which we
can prove the convergence (\ref{eq4.60}).

{\it Remark.} Evolution groups and their generators given by
conditions of the type of (\ref{eq4.2}) (or (\ref{eq4.7}))
were introduced in \cite{JP:NESS}, where
the group $U(t)$ is specified by the condition $U(t)\Omega_0 =
\Omega_0$.
where $\Omega_0$ is the unperturbed vector (``vacuum") introduced in \fer{eq_6_3} above. However, an analysis of the operator $K$ (see Sections \ref{Sect6} and \ref{Sect_7}) defined this way
requires,
instead of Condition (B), the condition obtained from Condition (B) by replacing the weight $e^{\delta\beta_j|u|/2}$ by $e^{\beta_j|u|/2}$.
This leads to an additional restriction on the temperatures of the form
\begin{equation}
g\leq  c \min_j\{e^{-\beta_j}\},\ \ \mbox{i.e.,} \ \  \min_j T_j\geq c[\ln(1/g)]^{-1}.
\label{additionalrestriction}
\end{equation}
Using in \cite{JP:NESS} the vector $\Omega$ instead of $\Omega_0$ would improve this bound to $\min_j T_j\geq cg$.

Now we proceed to the main technical result of this section - the
proof of the existence of the vector $\Omega$ and establishing its
properties mentioned and used above.

\begin{proposition}
\label{cyc&sep}
$\Omega_0 \in \dom(e^{-\beta\Ll/2})$ and the vector
$\Omega := e^{-\beta\Ll/2} \Omega_0/\|e^{-\beta\Ll/2} \Omega_0\| $
is cyclic and separating for the von Neumann algebra $\pi({\cal A})''$.
\end{proposition}

We begin with with some preliminary technical results. To do
manipulations with unbounded operators we use the dense subset,
$\cD$, of our Hilbert space, $\cH$, defined by
\begin{equation}\label{eq4.16}
\cD:=\pi(\tilde{\cA})\Omega_0,
\end{equation}
where $\tilde{\cA} = \cB(\cH_p)\otimes \mathcal{P}_1\otimes
\mathcal{P}_2$. Here $\mathcal{P}_j$ is the polynomial algebra
generated by the annihilation and creation operators, $a_{j}(f)$ and
$a_{j}^*(f)$, of the $j$-th reservoir acting on $\cH_{rj}$ with $f
\in L^2({\mathbb R}^3,e^{a'|k|^2} (1+|k|^{-1}) d^3k)$
for some $a'>0$.
$\cal D$ is a subset of
\begin{equation}
{\cal F}_0 := \left\{ \psi\in{\cal H}\ |\ P_n\psi = \psi,\ \ \mbox{for some $n<\infty$}\right\},
\label{mmm6}
\end{equation}
where $P_n$ is the spectral projection of the self-adjoint operator
$\widetilde N=\max\{N_1,N_2\}:= \int \max(\lambda, \mu)
dE_{N_1}(\lambda) \otimes dE_{N_2}(\mu)$ associated to the interval
$[0,n]$. Here, $N_j: = \int \left[ a^{\ast}_{\ell,j}(k)
a_{\ell,j}(k) + a^{\ast}_{r,j}(k) a_{r,j}(k) \right]d^3k$
is the number
operator of reservoir $j$ (see also \fer{mmm20}). ${\cal F}_0$ is commonly called the {\it
finite-particle subspace}.

Let $\Delta$ and $J$ be the modular operator and modular
conjugation associated with the pair $\{\pi(\cA)'',\ \Omega_0\}$
and let $\kappa^t$ be defined by
\begin{equation}
\kappa^t=\alpha^{\beta_{ p}t}_p \otimes \alpha^{\beta_1 t}_{r1} \otimes \alpha^{\beta_2 t}_{r2} .
\label{kappakappa}
\end{equation}
The vector $\Omega_0$ defines a $(\kappa, 1)$-KMS state and one can show that
\begin{equation}
\Delta^{-it} A \Delta^{it} = \pi\left(\kappa^t(A)\right)\ \ \forall\
A\in\cA.
\label{afterglow}
\end{equation}
Using this one computes that $\Delta = e^{-\tilde L}$, where $\tilde
L = \beta_{p} L_{p} + \beta_{r1} L_{r1} + \beta_{r2} L_{r2}$, and,
in particular, $\Delta^{i t}$ commutes with $e^{i t L_0}$. Observe
that
\begin{equation}
e^{\tau L_0}\cD = \cD\ \hbox { and }\ \Delta^{\tau}\cD = \cD\ \ \  \forall \tau
\in \BR.
\end{equation}

In the sequel, an important role is played by the operators
\begin{equation}
\Gamma(z):=e^{-z L^{(\ell)}/2} e^{z L_{0}/2} \label{mmm5}
\end{equation}
(defined as products of two, in general, unbouded operators).
We study properties of these operators in the following lemma.

\begin{lemma}
\label{lemma4.2}
The following statements hold for all $z\in{\mathbb C}$ and all
$\varphi,\ \psi\in \cD$:
\begin{equation}
\cD \subset\dom(\Gamma(z)),
\label{eq4.21}
\end{equation}
\begin{equation}
z\mapsto \Gamma(z)\varphi
\mbox{\ \ is entire,} \label{eq4.20a}
\end{equation}
\begin{equation}
e^{-z L^{(\ell)}/2}= \Gamma(z)e^{-z L_{0}/2}\mbox{\ \  on}\ \cD,
\label{eq4.22a}
\end{equation}
\begin{equation}
\left<B\Gamma(\bar{z})\psi,\Gamma(-z)\varphi\right>= \left<B\psi\
,\varphi\right>\ \forall B\in\pi({\cal A})', \label{eq4.21a}
\end{equation}
%
\begin{equation}
\left<JA\Gamma(z)\Omega_0,\varphi\right>= \left<A^*\Omega_0\
,\Gamma(z)\Delta^{1/2}\varphi\right>,
\ \forall A \in \pi(\cA)'',\ \label{eq4.22}
\end{equation}
\begin{equation}
\Gamma(z)\cD\mbox{\ \  is dense.} \label{mmm10}
\end{equation}
\end{lemma}

{\it Remark.\ } $\pi({\cal A})'$ is the von Neumann algebra $\pi'({\cal A})''$, the weak closure of $\pi'({\cal A})$.

\noindent {\it Proof of Proposition \ref{cyc&sep}. } Let
$\Gamma =\Gamma(\beta)$, where $\Gamma(z)$ is defined in \fer{mmm5}.
Since $e^{\tau L_0}\Omega_0=\Omega_0$, the property $\Omega_0\in
\dom (e^{-\beta \Ll/2})$ is equivalent to $\Omega_0 \in
\dom(\Gamma)$ (see \eqref{eq4.22a}), which is proven in Lemma
\ref{lemma4.2}, relation \fer{eq4.21}. Hence the vector $\Omega$
exists.

For the cyclicity it is enough to show that if $B\in\pi({\cal A})'$
and $B\Omega=0$ then $B=0$. Let $\Omega':=e^{-\beta
L^{(\ell)}/2}\Omega_0 = \Gamma \Omega_0$. By
(\ref{eq4.21a}), $\forall \varphi \in \cD$,
$0=\left<B\Gamma(\beta)\Omega_0,\Gamma(-\beta)\varphi\right>=
\left<B\Omega_0\ ,\varphi\right>$.
This implies $B\Omega_0=0$. Since $\Omega_0$ is separating for
$\pi({\cal A})'$ we have that $B=0$.

Now we show that $\Omega$ is separating for $\pi(\cA)''$. Let
$A\in\pi(\cA)''$ be such that $A\Omega=0$.  The relation
$A\Gamma\Omega_0=0$ and equation \fer{eq4.22} imply that
\begin{equation}
\label{eq4.24} 0= \left<JA\Gamma\Omega_0,\varphi\right>=
\left<A^*\Omega_0\ ,\Gamma\Delta^{1/2}\varphi\right>,
\end{equation}
for any $\varphi \in \D\subset \cF_0$. Now $\Delta^{1/2}\D=\D$ and
$\Gamma\D$ is dense (as is shown in \fer{mmm10}), so we have that
$\Gamma\Delta^{1/2}\D$ is dense and it follows from \fer{eq4.24}
that $A^*\Omega_0=0$. Since $\Omega_0$ is separating this implies
that $A=0$ and therefore $\Omega$ is separating. \hfill
$\blacksquare$\

\medskip

{\it Proof of Lemma \ref{lemma4.2}.}  We first show that for all
$\varphi\in\cF_0$ the following formal Dyson expansion of the
operator $\Gamma(z)$ is well defined:
\begin{equation} \label{mm53}
\tilde\Gamma(z) \varphi:=\sum_{m\geq 0}\left(\frac{-g z}{2}\right)^m\int_0^{1}d\tau_1\cdots\int_0^{\tau_{m-1}} d\tau_m
\sigma_0^{iz\tau_m/2}( V)\cdots \sigma_0^{iz\tau_1/2}( V)\varphi.
\end{equation}
The integrals on the r.h.s. are understood as strong limits of
Riemann sums. Due to the UV-cut-off Condition (A), the
transformation $\sigma_0^{iw}(V)= e^{-wL_0} V e^{wL_0}$
is well defined and strongly analytic on $\cF_0$ for $w\in\mathbb C$. Since $\varphi\in\cF_0$, there is a $\nu_0$ s.t.
$P_{\nu_0}\varphi=\varphi$ (see also \fer{mmm6}). Because each
interaction $V$ can increase the number of particles in each
reservoir by at most one, we can write the integrand of \fer{mm53}
as
$$
\sigma_0^{iz \tau_m/2}(V)P_{\nu_0+m-1}\  \sigma_0^{iz
\tau_{m-1}/2}(V)P_{\nu_0+m-2}\cdots \sigma_0^{iz
\tau_1/2}(V)P_{\nu_0}\varphi.
$$
Since $\sigma_0^{i w}(V)P_{k}$ are bounded operators, the
integrand on the r.h.s. of \fer{mm53}  belong to $\cH$. Moreover, it is strongly continuous in $\tau_1, \ldots, \tau_m$.

Let us first show that the series in \fer{mm53} converges
absolutely, for all values of $g,z\in {\mathbb C}$. We use the bound
$\| \sigma_0^{i w}(V)P_{\nu}\|\leq C({\rm Im}w) (\nu+1)^{1/2}$ which
follows in a standard way from the explicit expression of $V$. The
constant is given by
$$
C({\rm Im}w)=2e^{2{\rm Im}w \| H_p\|}\sum_{j=1,2}\left[\int_{{\mathbb R}^3}
\|G_j(k)\|^2 \frac{e^{|k|({\rm Im}w+\beta_j)}} {e^{\beta_j|k|}-1} d^3k
\right]^{1/2}.
$$
It follows that the norm of the $m-$th
term in the series of \fer{mm53} has the upper bound $
\frac{[|gz|C({\rm Im}z/2)]^m}{2^m m!}\sqrt{\frac{(\nu_0+m)!}{\nu_0!}}. $
Consequently, the series converges absolutely, for any
$g,z\in{\mathbb C}$, and for any $\varphi\in\cF_0$.

Next, we show that $\cD\subset\dom(\Gamma(z))$, and that $\tilde{\Gamma}(z)\varphi=\Gamma(z)\varphi$ for all $\varphi\in\cD$. It suffices to establish that for any $\varphi\in\cD$,
\begin{equation}
\innerprod{e^{-\overline {z} L^{(\ell)}/2}\psi}{e^{z
L_{0}/2}\varphi}=\innerprod{\psi}{\tilde{\Gamma}(z)\varphi},
\label{mmm2}
\end{equation}
for all $\psi\in{\cal H}$ s.t. $\psi=f(L^{(\ell)})\psi$ for some
$f\in C_0^\infty({\mathbb R})$ (such $\psi$ form a core for
$e^{-\overline{z}L^{(\ell)}/2}$). Indeed, this would show that $e^{z
L_{0}/2}\varphi\in \dom (e^{-z L^{(\ell)}/2})$ and therefore
$\varphi\in \dom (\Gamma(z))$. Equation \fer{mmm2} can be shown e.g.
using the analyticity of both sides in $z$, and the fact that the
equation holds for $z\in i\mathbb{R}$. Indeed, in the latter case
$\Gamma(z)$ are bounded operators and the Dyson series expansion
\eqref{mm53} is valid for them. In particular, $\cD \subset \dom(\Gamma(z))$. Moving $e^{-\overline{z}
L^{(\ell)}/2}$ in \fer{mmm2} to the right factor proves that
$\Gamma(z)\varphi=\tilde{\Gamma}(z)\varphi$, for all
$\varphi\in\cD$. This
also shows  \fer{eq4.20a}, since the series \fer{mm53} is entire in
$z$. Thus we have shown  \eqref{eq4.21} and  \eqref{eq4.20a}.

Furthermore, since $e^{z L_{0}/2}\cD=\cD$, we have by the argument
above (see \eqref{mmm2} and the argument after it) that
\eqref{eq4.22a} holds. Moreover, $\forall \varphi \in \cD,\ e^{-z
L^{(\ell)}/2}\varphi$ is analytic due to the formula $e^{-z
L^{(\ell)}/2}\varphi = \Gamma(z) e^{-z L_{0}/2} \varphi$ and
analyticity of the factors on the r.h.s. .

To prove \fer{eq4.21a} we note that, due to the Dyson expansion,  it
is true for purely imaginary $z$ and by analyticity of the l.h.s.
for all $z$.

Now, we prove \eqref{eq4.22}. Denote by $\Gamma_k (z)\varphi$ the
truncated series on the r.h.s. of \fer{mm53}, with $m\leq k$. Let
$A\in\pi({\cal A})''$. Choose a sequence of linear combinations of
(${\cal H}_p\otimes{\cal H}_p$-valued) field operators
$A_n\in\pi(\widetilde{\cal A})$ converging weakly to $A$. We use the
defining property of the operator $S = \Delta^{-1/2}J$, and the
results proven above, to obtain the following relation for all
$\varphi \in \cD$:
\begin{eqnarray}
\scalprod{JA\Gamma(z)\Omega_0}{\varphi}
&=&\lim_{k\rightarrow\infty} \scalprod{JA\Gamma_k(z)\Omega_0}{\varphi}\nonumber\\
&=&\lim_{k\rightarrow\infty} \lim_{n\rightarrow\infty} \scalprod{JA_n\Gamma_k(z)\Omega_0}
{\varphi}\nonumber\\
&=&\lim_{k\rightarrow\infty} \lim_{n\rightarrow\infty}
\scalprod{\Gamma_k(z)^* (A_n)^*\Omega_0}{\Delta^{1/2}
\varphi}\label{u1}\\
&=&\lim_{k\rightarrow\infty} \scalprod{A^*\Omega_0}{\Gamma_k(z)\Delta^{1/2}\varphi}\nonumber\\
&=&\scalprod{A^*\Omega_0}{\Gamma(z)\Delta^{1/2}\varphi}.\nonumber
\end{eqnarray}
We have used that $(A_n)^*\Omega_0\in\cD\subset\cF_0$,
$\Delta^{1/2}\psi\in \cD\subset\cF_0$, and in \fer{u1} that
$J\Delta^{1/2}A_n \Gamma_k(z)\Omega_0=\Gamma_k(z)^*(A_n)^*\Omega_0$. The
latter relation follows from the facts that $\Gamma_k(z)$ is
affiliated with the von Neumann algebra $\pi({\cal A})''$, and that
$(A_n)^*$ leaves $\cF_0$ invariant. It can also be verified
directly, using the explicit actions of $J$ and $\Delta^{1/2}$. This
shows \fer{eq4.22}.

To prove the last statement we introduce a new family of operators
$\Gamma^\&(z):= e^{z L_{0}/2} e^{-z L^{(\ell)}/2} $
related to the adjoint of $\Gamma(z)$. First we prove that $\cD
\subset \dom(\Gamma^{\&}(z))$. To this end we note that, exactly as
above, the formal expansion of $\Gamma^{\&}(z)$, which we denote by
$\tilde{\Gamma}^{\&}(z)$, converges on elements of $\cD$ and is
entire as a function of $z$. Next, we observe that, since $e^{z
L_{0}/2}\cD=\cD$,
the equation $e^{z L_{0}/2}\varphi\in \dom (e^{-z L^{(\ell)}/2})$,
proven above, implies that $\cD \subset \dom (e^{-z L^{(\ell)}/2})$.
Let $\cD_1:= \{\psi\in{\cal H}|\ \psi=f(L_0)\psi$ for some $f\in
C_0^\infty({\mathbb R})\}$. Then $\cD_1$  is a core for
$e^{-zL_0/2}$. Now we claim that for any $\varphi \in\cD$ and any
$\psi\in\cD_1$,
\begin{equation}
\innerprod{e^{\overline {z} L_{0}/2}\psi}{e^{-z
L^{(\ell)}/2}\varphi}=\innerprod{\psi}{\tilde{\Gamma}^{\&}(z)\varphi}.
\label{mmm2b}
\end{equation}
Indeed, the latter relation is true for $z$ purely imaginary and it
remains to be true for complex $z$ by analyticity of both sides. The
last relation and the fact that $\cD_1$ is a core of the operator
$e^{\overline {z} L_{0}/2}$ show that $e^{-z L^{(\ell)}/2}\varphi
\in \dom (e^{z L_{0}/2})$ and $\innerprod{\psi}{e^{z L_{0}/2}e^{-z
L^{(\ell)}/2}\varphi}=\innerprod{\psi}{\tilde{\Gamma}^{\&}(z)\varphi}.$
In other words, $\varphi \in \dom (\Gamma^{\&}(z))$ and
$\Gamma^{\&}(z)\varphi =\tilde{\Gamma}^{\&}(z)\varphi$. This proves
that $\cD \subset \dom(\Gamma^{\&}(z))$ and that $\Gamma^{\&}(z)$ is
an entire operator-function.

To prove \fer{mmm10} it suffices to show that $\Gamma(z)\cD
\supseteq \cD$. To prove the latter, let
$\psi\in \cD \subset \dom(\Gamma^{\&}(z))$. Then
$e^{zL_0/2}e^{-zL^{(\ell)}/2}\psi\in
\dom(e^{zL^{(\ell)}/2}e^{-zL_0/2})$, and the image of this vector
under $e^{zL^{(\ell)}/2}e^{-zL_0/2}$ is just $\psi$. Hence $\psi \in
\Gamma(z)\cD$ as required.

This completes the proof of Lemma \ref{lemma4.2}. \hfill
$\blacksquare$

\textit{Remark}: A sharper estimate can be obtained by using a
standard argument to estimate the norms of the integrands (to keep
notation simple we set $\varphi = \Omega_0$),
\begin{equation}
\parallel\sigma_0^{i\tau_m}( V) \cdots \sigma_0^{i\tau_1}(
V)\Omega_0\parallel = \omega_0(\alpha_0^{-i\tau_1}( v)\cdots
\alpha_0^{-i\tau_m}( v) \alpha_0^{i\tau_m}( v) \cdots
\alpha_0^{i\tau_1}( v))^{1/2},
\end{equation}
by using Wick's theorem, the expression for the imaginary-time
two-point functions, in the same way as
it is done, for instance, in \cite{BFS:RtE}, Thm IV.3.

\section{Generator $K$ and interpolating family $K_{(s)}$}
\label{Sect5}

In this section we find an explicit form and some properties of
the generator $K$ of the one-parameter group $U(t)$ introduced in
the preceding section (cf.~\cite{JP:NESS}), and of the family $K_{(s)}$ which interpolates $K$ to a selfadjoint operator.

Let $\omega_0$ be the state of the algebra $\cA$ fixed at the
beginning of the Section \ref{Sect4} and let $J$ and $\Delta$ be
the Tomita-Takesaki modular conjugation and modular operator
associated with the couple $(\cA,\omega_0)$. We have the following
standard relations:
\begin{equation}
J \pi(A) J  =  \pi'(A), \label{eqn_5.2}
\end{equation}
$J\Delta^{1/2}\pi(A) \Omega_0
 =  \pi(A^*) \Omega_0$, $\Delta^{1/2} \Omega_0
 = \Omega_0$ and
$\Delta^{-it}\pi(A)\Delta^{it} = \pi(\kappa^{t}(A))$, where
$\kappa $ is the automorphism of the algebra $\cA$ defined in
\fer{kappakappa}. The last three equations imply
\begin{equation}
J\pi(A) \Omega_0
 =  \pi(\kappa^{i/2}(A^*)) \Omega_0. \label{eqn_5.3} \\
\end{equation}

Finally, we recall that $\beta = \max(\beta_1, \beta_2)$ and that
$\Ll$ is the self-adjoint operator defined as $\Ll := L_0 + gV$
where $V := \pi(v)$.

\begin{theorem}\label{thm5.1}
Assume that (A) and (\ref{part}) hold. The semigroup $U(t)$,
defined in (\ref{eq4.1})--(\ref{eq4.2}), is differentiable
on the domain $\dom (\Ll)
\cap \pi(\cA)\Omega$, and the generator $K = -i
\frac{\partial}{\partial t} \Big|_{t=0} U (t)$ is given on this domain by the
expression
\begin{equation}\label{eq5.2}
K = L_0 + g(V - V'_{-i/2}),
\end{equation}
where $V = \pi(v)$ and $V'_s = \pi'(\gamma^{\overline{s}}(v))$ with $\gamma^s
:=\alpha_0^{-\beta s} \circ \kappa^s$ (due to condition (A)
the operator $V'_{-i/2}$ is well defined). Furthermore, the domain
$\dom(\Ll) \cap \pi(\cA) \Omega$ is dense in $\cH$.
\end{theorem}

{\it Remark.} The imaginary part of the generator $K$ is not
semi-bounded. Therefore, the group $U(t)$, densely defined on
$\pi(\cA)\Omega$, does not extend to a group of bounded operators.

{\it Proof of Theorem \ref{thm5.1}.\ } Due to Condition (A), we have
$e^{-\beta L_0/2}\Ll\Omega_0 = g e^{-\beta L_0/2}\pi(v) \Omega_0 = g
\pi(\sigma_0^{i\beta/2}(v)) \Omega_0 \in\cD$. It thus follows from
Lemma \ref{lemma4.2} that $\Ll\Omega_0 \in \dom(e^{-\beta\Ll/2})$.
The latter fact implies that $e^{i t\Ll}\Omega$ is differentiable at
$t=0$ and therefore $\Omega \in \dom(\Ll)$. Now, let $B \in
\pi(\cA)$ be such that $B \Omega \in \dom (\Ll)$. Taking into
account equation \fer{fm4} we see that $\sigma^t(B)$ and therefore,
due to \fer{eq4.1}, also $U(t)B\Omega$, are differentiable in $t$ at
$t=0$. Indeed, let $\frac{1}{t}(e^{i t\Ll}Be^{-i t\Ll} -B)\Omega =
F+G$, where $F:= e^{i t\Ll}B\frac{1}{t}(e^{-i t\Ll} -1)\Omega$ and
$G:=\frac{1}{t}(e^{i t\Ll}-1)B\Omega$. Clearly, $F\rightarrow -i
BL^{(\ell)}\Omega$ and $G\rightarrow i L^{(\ell)}B\Omega$. Now,
differentiating equation \fer{eq4.1} we find
\begin{equation} \label{eqn_5.7a}
K B \Omega = \Ll B \Omega -B \Ll \Omega.
\end{equation}
Now we compute the last term on the r.h.s. of this expression. To
this end we use the following relations:
\begin{equation}\label{eqn_5.2a}
  \pi(A) \Omega_0 = \pi'(\kappa^{i/2}(A^*)) \Omega_0
\end{equation}
and, for $z = i t$,
\begin{equation}
\label{eqn_5.2b}
e^{z\Ll}\pi'(A)e^{-z\Ll} = \pi'(\alpha_0^{i\overline{z}}(A)).
\end{equation}
Eqn (\ref{eqn_5.2a}) follows from relations (\ref{eqn_5.2}) and
(\ref{eqn_5.3}). Equality (\ref{eqn_5.2b}) is proven by using the
Kato-Trotter product formula.

Now, we claim that $\forall \psi \in \cD$
\begin{equation}\label{eqn_5.8a}
 \langle \psi, B e^{z\Ll} \pi(v)
  \Omega_0\rangle =   \langle [\pi'(\alpha_0^{i \overline{z}} \circ \kappa^{i/2}(v))]^{*}\psi, B
    e^{z\Ll} \Omega_0\rangle.
\end{equation}
Observe that the vectors $e^{z\Ll} \varphi=\Gamma(-2z)e^{zL_0}\varphi$, $\varphi \in \cD$, are entire in $z$, by Lemma \ref{lemma4.2}.
Hence both sides of Eqn \fer{eqn_5.8a} are entire in $z$.
Therefore it suffices to prove (\ref{eqn_5.8a}) for
$z = i t$. Eqn \fer{eqn_5.8a} with $z = i t$ follows from the
relations (\ref{eqn_5.2a}) and (\ref{eqn_5.2b}) proven above.
Thus (\ref{eqn_5.8a})
is demonstrated.

Now, let $\Omega' := e^{- \beta \Ll/2} \Omega_0$. Recall that $\Ll
\Omega' =  g e^{- \beta \Ll/2} \pi(v) \Omega_0$. Then
(\ref{eqn_5.8a}) with $z=\beta/2$ and the definition of the
transformation $\gamma^{s}$ imply that for all $\psi\in {\cal D}$
\begin{eqnarray}\label{eqn_5.8b}
\langle \psi, B \Ll \Omega'\rangle & = & g \langle \psi, B e^{-
\beta \Ll/2} \pi(v)
  \Omega_0\rangle \nonumber\\
 &=& g \langle [\pi'(\gamma^{i/2}(v))]^{*}\psi, B \Omega'\rangle.
   \end{eqnarray}
This relation and the fact that $\cal D$ is a core for
$\pi'(\gamma^{i/2}(v))^*$ (which is a linear combination of creation
and annihilation operators) show that $B \Omega' \in
\dom(\pi'(\gamma^{i/2}(v))$ and
\begin{equation}\label{eqn_5.8c}
  B \Ll \Omega =  \pi'(\gamma^{ i/2}(v))
B \Omega.
   \end{equation}
Since the r.h.s. of (\ref{eqn_5.8c}) is $V'_{-i/2} B \Omega$,
this equation together with (\ref{eqn_5.7a}) implies
(\ref{eq5.2}).
Finally, $\dom(\Ll) \cap \pi(\cA) \Omega$ contains the set
$\cD$ and is therefore dense in $\cH$. \hfill \qed

Observe that
\begin{equation}\label{eq5.9}
  \kappa^t|_{\beta_j = \beta_p = \beta} = \alpha_0^{\beta t}
  \qquad \text{and} \qquad
  \gamma^t|_{\beta_j = \beta_p = \beta} = {\rm id}.
\end{equation}
This implies that
\begin{equation}
  K|_{\beta_j = \beta_p = \beta} = L,
\label{el}
\end{equation}
where $L := L_0 + gV - gV'$, with $V' := \pi'(v)$, is the standard
self-adjoint Liouville operator.
In what follows we write $K = L_0 + gI$,
where
\begin{equation*}
I = V -V'_{-i/2}.
\end{equation*}

The operator $K$ is non-self-adjoint for $\delta\beta\neq 0$, and
the perturbation $I$ is not relatively bounded w.r.t. the
unperturbed operator $L_0$. To study the evolution generated by $K$
we use the family of operators
\begin{equation}\label{eq7.5}
K_{(s)} := L_0 +g(V- V'_s) \ ,
\end{equation}
where, recall, $V = \pi (v)$, and
$V'_s := \pi' (\gamma^{\overline s} (v))$. This family
interpolates between the
operator $K$,
\begin{equation}\label{eq7.8}
K = K_{(-i/2)} \ ,
\end{equation}
(see Eqn (\ref{eq5.2})) and the self-adjoint operators
$K_{(s)}$ with real $s$.
Under condition (A) on $v$, $V'_{(s)}$ and $K_{(s)}$ are
well defined on the dense domain $\dom(\Ll) \cap \pi(\cA) \Omega$ for all $s$
in the strip
\begin{equation}\label{eq7.7}
S_\vep := \left\{ t \in \BC \big| |\rIm t| < {\textstyle \frac12} + \vep \right\},
\end{equation}
for $\varepsilon > 0$, and are strongly analytic there (recall that
$\pi'$ is anti-linear).

\section{Spectral Deformation of $K$ and $K_{(s)}$}
\label{Sect6}

Since the operator $K$ is not self-adjoint it is not a simple
matter to derive long-time properties of the dynamics $e^{iKt}$
from spectral properties of $K$. As a result we bypass
establishing the connection of the dynamics to the spectrum of $K$
and instead connect it to certain spectral properties of a complex
deformation, $K_{\theta}$, of this operator.
To do this we use the interpolating family $K_{(s)\theta}$, which is the complex deformation of the
family $K_{(s)}$, \fer{eq7.5}.
In this section we
define complex deformations $K_{\theta}$ and $K_{(s)\theta}$
and in the next
section we establish their spectral characteristics which are relevant for us.

In order to carry
out the spectral analysis of the operator $K$, which we begin in
this section, we use the specifics of the Araki-Woods
representation in an essential way. They were not used in an essential way for the developments up to this section.

As a complex deformation we choose a combination of the complex
dilation used in \cite{BFS:RtE} and complex translation due to
\cite{JP:QFII} (see \cite{BFS:RtE}, Section V.2 for a sketch of
the relevant ideas). This complex deformation was used in \cite{MMS1} in the spectral analysis for a general class of Liouville type operators.

First we define the group of dilations. Let $\hat{U}_{d, \delta}$
be the second quantization of the one-parameter group
\begin{equation*}
  u_{d, \delta} : f(k) \to e^{3\delta/2} f(e^{\delta}k)
\end{equation*}
of dilations on $L^2(\BR^n)$. This group acts on creation and
annihilation operators $a_r^{\#}(f)$ on the Fock space, $\cH_r$,
according to the rule
\begin{equation} \label{eqn_6_3}
  \hat{U}_{d, \delta} a_r^{\#}(f) \hat{U}_{d, \delta}^{-1} =
  a_r^{\#}(u_{d, \delta}f), \qquad \hat{U}_{d, \delta}
  \Omega_{rj} = \Omega_{rj}.
\end{equation}
We lift this group to the positive-temperature Hilbert space,
 (\ref{eqn_6_1}), according to the formula
\begin{equation} \label{eqn_6_4}
  U_{d, \delta} =  \id_{p} \otimes \id_{p} \otimes
  \hat{U}_{d, \delta} \otimes \hat{U}_{d, -\delta} \otimes
\hat{U}_{d, \delta} \otimes \hat{U}_{d, -\delta}.
\end{equation}
Note that we could dilate each reservoir by a different amount.
However, this does not give us any advantage, so to keep notation
 simple we use the same dilation parameter for both
reservoirs.

We record for future reference how the group $U_{d, \delta}$ acts on
the Liouville operator $L_0$ and the positive-temperature photon
number operator $N := \sum_{j=1}^2N_j$, where
\begin{equation}
  N_j: = \int \left[ a^{\ast}_{\ell,j}(k) a_{\ell,j}(k) + a^{\ast}_{r,j}(k) a_{r,j}(k) \right] \,
  d^3k,
\label{mmm20}
\end{equation}
and the operators $a_{\{\ell,r\},j}^{\#}(k)$ were introduced after \fer{eqn_6_2}.
We have (below we do not display the identity operators):
\begin{equation} \label{eqn6_6}
  U_{d, \delta} L_{rj} U_{d, \delta}^{-1} = \cosh(\delta) L_{rj}
  + \sinh(\delta) \Lambda_j,
\end{equation}
where $\Lambda_j$ is the positive operator on the $j$th reservoir
Hilbert space given by
\begin{equation}
  \Lambda_j = \int \omega(k) \left( a^{\ast}_{\ell,j}(k) a_{\ell,j}(k) + a^{\ast}_{r,j}(k) a_{r,j}(k) \right)
  \, d^3k,
\end{equation}
and
\begin{equation} \label{eqn6_8}
  U_{d, \delta} N_j U_{d, \delta}^{-1} = N_j.
\end{equation}

Now we define a one-parameter group of translations. It can be
defined as one-parameter group arising from transformations of the
underlying physical space similarly to the dilation group. This is
done in Appendix \ref{App_Repr}. We define here the translation
group by means of the selfadjoint generator $T := \sum_{j=1}^2 T_j$, where
\begin{equation} \label{eqn_6_12}
  T_j = \int \left[ a^{\ast}_{\ell,j}(k) \gamma a_{\ell,j}(k)
       + a^{\ast}_{r,j}(k)\gamma a_{r,j}(k) \right] \,
       d^3k.
\end{equation}
The operator $\gamma = i (\hat{k} \cdot \nabla + \nabla \cdot \hat{k})$,
with $\hat{k} = k/|k|$, is a symmetric, but not a self-adjoint operator. Nevertheless, the operators $T_j, j=1,2$, {\it are} self-adjoint \cite{MMS1}. Thus the operator $T$ is self-adjoint as well. We define the
one-parameter group of translations as
\begin{equation} \label{eqn_6_13}
  U_{t, \tau} := \bfone_p \otimes \bfone_p \otimes
  e^{i\tau T}.
\end{equation}
Eqns.~(\ref{eqn_6_12}) - (\ref{eqn_6_13}) imply the following
expressions for the action of this group on the Liouville
operators:
\begin{equation} \label{eqn6_11}
  U_{t, \tau} L_{rj} U_{t, \tau}^{-1} = L_{rj} + \tau N_j,
\end{equation}
and  $U_{t,\tau}N_j U_{t,\tau}^{-1}=N_j$.
Observe that neither the dilation nor the translation group
affects the particle vectors.

Now we want to apply the product of these transformations to the
full operator $K = L_0 + gI$. Since the dilation and translation
transformations do not commute we have to choose the order in
which we apply them. Since the operator $\Lambda = \sum_{j}
\Lambda_j$ is not analytic under the translations while the
operator $N$ is analytic under dilations we apply first the
translation and then the dilation transformation. We define the
combined translation-dilation transformation as
\begin{equation}
  U_{\theta} = U_{d, \delta} U_{t, \tau}
\end{equation}
where $\theta = (\delta, \tau)$. Note that $U_\theta$ leaves the finite-particle space $\cF_0$, as well as $\dom(\Lambda)\cap \dom(N)$ invariant, for $\theta\in{\mathbb R}^2$.

In what follows we will use the
notation $|\theta| = (|\delta|, |\tau|)$, $\rm{Im} \theta =
(\rm{Im}\delta, \rm{Im}\tau)$, and similarly for $\rm{Re}\theta$,
and
\begin{equation}
  \rm{Im} \theta > 0 \quad \iff \quad \rm{Im}\delta > 0 \wedge
  \rm{Im}\tau > 0.
\end{equation}

Now we are ready to define a complex deformation of the operator
$K$. On the set $\dom(\Lambda)\cap \dom(N)$ we define for $\theta \in
\BR^2$
\begin{equation}
  K_{\theta } := U_{\theta } K U_{\theta}^{-1}.
\end{equation}
Recalling the decomposition $K = L_0 + gI$, where $L_0:=L_p+L_r$, $L_r :=
\sum_{j=1}^2 L_{rj}$ and $I = V - V'_{-i/2}$, we have
\begin{equation} \label{eq6.18}
K_{\theta} = L_{0,\theta } + gI_{\theta},
\end{equation}
where the families $L_{0, \theta}$ and $I_{\theta}$ are defined
accordingly. Due to Eqns.~(\ref{eqn6_6}), (\ref{eqn6_8}) and
(\ref{eqn6_11}) we have:
\begin{equation}
\label{eqn6_16}
L_{0, \theta} = L_p + \cosh(\delta) L_r +\sinh(\delta)\Lambda +
\tau N,
\end{equation}
where $\theta = (\delta, \tau)$, and $\Lambda = \sum_{j=1}^2
\Lambda_j$. An explicit expression for the family $I_\theta$ is
given in Appendix \ref{b2} (see Eqns \fer{itheta} and \fer{A2.7}).

Similarly, we define the family $K_{(s)\theta}:= U_\theta
K_{(s)}U_{\theta}^{-1}$ (recall \fer{eq7.5}).

The operator families above are well defined for real $\theta$. Our
task is to define them as analytic families on the strips
\begin{equation} \label{eqn_6.17}
S_{\theta_0}^\pm = \left\{\theta \in \BC^2 | 0<\pm \rm{Im} \theta
< \theta_0 \right\}
\end{equation}
where $\theta_0 = (\delta_0, \tau_0) > 0$ is the same as in
Condition (B). Recall that the inequality $\pm\rm{Im}\theta <
\theta_0$ is equivalent to the following inequalities:
$\pm\rm{Im}\delta < \delta_0$ and $\pm\rm{Im}\tau < \tau_0$. (The fact that analyticity in a neighbourhood of a fixed $\theta\in S^\pm_{\theta_0}$ implies analyticity in the corresponding strip in which Re$\theta$ is not constraint follows from the explicit formulas \fer{eqn6_16}, \fer{itheta} and \fer{A2.7}.)
The
analytic continuations (if they exist) are denoted by the same
symbols.

We define the family $K_\theta$ for $\theta \in \{ \theta \in \BC^2
\big| |\rIm \theta| < \theta_0 \}$ by the explicit expressions
(\ref{eq6.18}), (\ref{eqn6_16}), (\ref{itheta}) and \fer{A2.7}.
Clearly, $\dom(\Lambda)\cap \dom(N) \subset \dom(L_{0\theta})$ and
on this domain the family $L_{0\theta}$ is manifestly strongly
analytic in $\theta \in \{ \theta \in \BC^2 \big| | {\rm Im} \theta
| < \theta_0 \}$. It is shown in Appendix \ref{App_Repr} that for
$|{\rm Im}\theta| < \theta_0$ we have $\dom(\Lambda^{1/2}) \subset
\dom(I_\theta)$ and $I_{\theta} f$ is analytic $\forall f \in
\dom(\Lambda^{1/2})$. Here Condition (B) of Section \ref{Sect3} is
used. Hence the family $K_\theta$ for $\theta \in \{ \theta \in
\BC^2 \big| |\rIm \theta| < \theta_0 \}$ is bounded from
$\dom(\Lambda) \cap \dom (N)$ to $\cH$ (and $K_\theta f$ is analytic
in $\theta\in \{ \theta \in \BC^2 \big| |\rIm \theta| < \theta_0
\}$, $\forall f\in\dom(\Lambda)\cap\dom(N)$). Moreover, for $|{\rm
Im}\ \theta| > 0$ the operators $K_\theta$ are closable on the
domain $\dom(\Lambda)\cap \dom(N)$, since their adjoints are defined
on this dense domain. We denote their closure by the same symbols.

However, $\{ K_\theta  \ |\ {\rm Im}\theta<\theta_0\}$  is not an analytic family in the sense of Kato. The problem here is the lack of coercivity -- the
perturbation $I$ is not bounded relatively to the unperturbed
operator $L_0$. To compensate for this we have chosen the
deformation $U_{\theta}$ in such a way that the operator
$M_{\theta} := {\rm Im} L_{0, \theta}$ is coercive for
$\rm{Im}\theta > 0$ , i.e., the perturbation $I_{\theta}$ as well
as ${\rm Re} L_{0, \theta}$ are bounded relative to this operator.
The problem here is that $M_{\theta} \to 0$ as $\rm{Im}\theta \to
0$ so we have to proceed carefully.

Everything said about $K_\theta$ applies also to the family
$K_{(s)\theta}$.

The next result gives some analyticity properties and some global spectral properties of $K_{(s)\theta}$.
\begin{theorem}[\cite{MMS1}]
\label{thm_6.1}
Assume that Condition (B) holds and let $\theta_0 = (\delta_0,
\tau_0)$ be as in that condition. Take an
\begin{equation}
a>\frac{g^2}{\sin({\rm Im}\delta)}C_0^2\left(\sum_{j=1}^2\|G_j\|_{\delta\beta_j,1/2,\theta}\right)^2,
\label{condition on a}
\end{equation}
where $C_0:=C (1+\beta_1^{-1/2}+\beta_2^{-1/2})$,
and where $C$ is a constant depending only on $\tan\delta_0$. Then we have:
\begin{enumerate}
\item[(i)] $\left\{ z \in \BC | \rIm z \le -a \right\}\subseteq
\rho(K_{(s)\theta})$ (resolvent set) if either $s\in \BR$ and $\theta\in
\overline{S_{\theta_0}^+}$, or if $s\in S_\epsilon$ and $\theta\in
S_{\theta_0}^+$;
\item[(ii)] Let $C_{a,b}$ be the truncated cone
\begin{equation*}
C_{a,b} := \left\{ z\in \BC \ | \ \rIm z > -{\textstyle \frac a2}, \ | \rRe z | <
2[(\sin b)^{-1}+a/4] (\rIm z +a) +\|L_p\|+1 \right\}.
\end{equation*}
Take $s\in S_\varepsilon$, $\theta\in S_{\theta_0}^+$, and take
$a$ as in \fer{condition on a}. Then $\sigma (K_{\theta}) \subset C_{a,\rIm \delta}$, and for $z\in \BC \backslash C_{a, \rIm \delta}$ we have
\begin{equation}
\| (K_{\theta}-z)^{-1} \| \le (\dist (z,C_{a, \rIm \delta}))^{-1}.
\label{eq6.3}
\end{equation}
\item [(iii)] The family $K_{(s)\theta}$, $s\in
S_\epsilon$, $\theta\in \overline{S_{\theta_0}^+}$, is analytic of
type A (in the sense of Kato) in $\theta\in S_{\theta_0}^+$, for
all $s\in S_\epsilon$, and in $s\in S_\epsilon$, for all
$\theta\in S_{\theta_0}^+$;
\item [(iv)] Let $s\in{\mathbb R}$. For any $u$
and $v$ which are $U_{\theta}$-analytic in a strip $\left\{ \theta
\in \BC^2 | 0 \le {\rm Im} \theta < \theta_1 \right\}$, for some
$\theta_1 = (\delta_1 ,\tau_0)$, $\delta_1\in
[0,\min\{\pi/3,\theta_0\})$, the following relation holds:
\begin{equation}
\label{eq6.21}
\left< u, (K_{(s)}-z)^{-1} v \right> = \left<
u_{\overline{\theta}}, (K_{(s) \theta} - z)^{-1} v_{\theta}
\right>,
\end{equation}
where $u_{\theta} = U_{\theta} u$, etc., $\rIm z \le -a$ and $0 < \rIm\theta < \theta_1/2$.
\end{enumerate}
\end{theorem}
{\it Proof.\ }
Statements (i), (iii) and (iv) are special cases of Theorem 5.1 in \cite{MMS1}, with the exception of the assertion about analyticity of $s\mapsto K_{(s)\theta}$ in (iii). This assertion is easily proven by noticing that $\partial_s(K_{(s)\theta}-z)^{-1} = -(K_{(s)\theta}-z)^{-1}(\partial_s V'_s) (K_{(s)\theta}-z)^{-1}$. Statement (ii) is the content of Proposition 5.2 in \cite{MMS1}. \hfill $\blacksquare$

\section{Spectral Analysis of $K_{\theta}$}
\label{Sect_7}

In what follows, given a self-adjoint operator $A$ and $a \in \BR $
we use the notation $\chi_{A \le a}$ for the spectral projection of
$A$ associated to the set $\set{\lambda \in \BR }{\lambda \le a}$
and similarly for $\chi_{A \ge a}$ and $\chi_{A=a}$, etc. Fix a
$\theta$ satisfying $0 < \rIm \theta < \theta_0$.

\begin{theorem}
\label{thm6.1} Assume Conditions (A) -- (D). Let $\theta_0$ be the
same as in Condition (B), let $g_1$ be as in Theorem \ref{thm2.1}
and let $\theta_0 > {\rm Im}\theta = (\delta' , \tau' )
> 0$. If $0<|g|<g_1$ and $\tau' > |g|^{2+\alpha}$
then
\begin{description}
\item{(a)} $0$ is an isolated and simple eigenvalue of
$K_{\theta}$; \item{(b)} $\sigma (K_{\theta}) \backslash \{ 0\}
\subset \{ z \in \BC^+ | {\rm Im}z \ge
\min(cg^2,\frac12 \tau')\}$,
\end{description}
for some $c>0$, independent of $\theta$.
\end{theorem}

Theorem \ref{thm6.1} is proven at the end of this section. Together with
Theorem \ref{thm_6.1} it gives the following picture for the spectrum of $K_\theta$.

\bigskip
\centerline{{\epsfbox{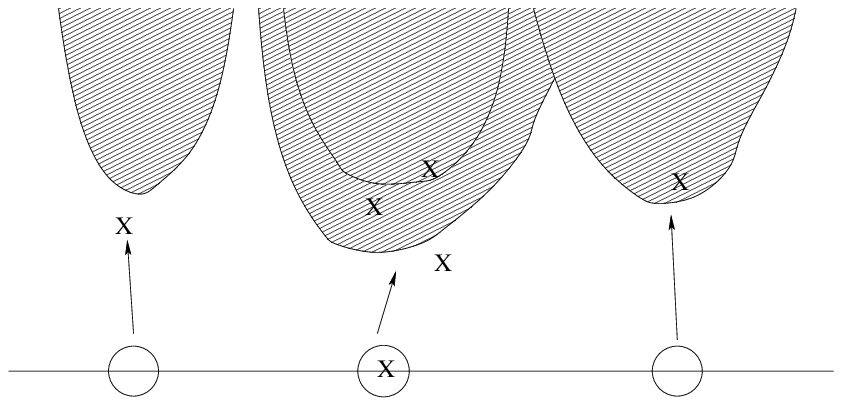}}}
\centerline{Spectrum of $K_{\theta}$ for ${\rm Im}\theta>0$}
\bigskip
\noindent

The motion of resonances bifurcating out of the eigenvalues of $L_0$ is governed, to second order in the coupling constant $g$, by {\it level shift operators}, see \cite{MMS1,Mlso} for a discussion closest to the situation at hand. Let $e$ be an eigenvalue of $L_p$ and let $\Lambda_e$ be the
level shift operator acting on ${\rm Ran} \chi_{L_p = e}$, defined by
\begin{equation}
\Lambda_e := - P_e I (L_0-e+i0)^{-1} I P_e,
\label{m2}
\end{equation}
where $P_e = \chi_{L_p = e} \otimes \chi_{L_r = 0}$. The notation $+i0$ in \fer{m2} stands for the limit of $i\epsilon$ as $\epsilon\downarrow 0$. The following result summarizes properties of the level shift operators which are essential for the proof of Theorem \ref{thm6.1}.
\begin{theorem}
\label{thmlso} Assume Condition (C). The level shift operators
$\Lambda_e$, \fer{m2}, satisfy
\begin{eqnarray}
\label{eq6.4} \qquad &&\sigma (\Lambda_e) \subset \overline{\BC^+} \\
\label{eq6.5} \qquad &&\sigma (\Lambda_e) \cap \BR = \left\{
\begin{array}{ll}
\emptyset &{\hbox{if }} e\ne 0\\
\{ 0 \}   &{\hbox{if }} e=0
\end{array} \right. \\
\label{eq6.6} \qquad &&\dim \Ker (\Lambda_{e=0} ) =  1 \ .
\end{eqnarray}
Furthermore, there is a $\gamma_0>0$ which does not depend on the inverse temperatures, s.t. ${\rm Im}\Lambda_e\geq\gamma_0$, for all $e\neq 0$. Moreover, if Condition (D) is satisfied, then there is a $\delta_0>0$ which does not depend on the inverse temperatures, s.t. ${\rm Im}(\sigma(\Lambda_0)\backslash\{0\})\geq \delta_0$.
\end{theorem}
We prove this theorem is in Section \ref{Sect13}.

{\it Proof of Theorem \ref{thm6.1}.\ }
That $0$ is an eigenvalue of $K_{\theta}$ follows
readily from the equations $K \Omega = 0$ and the fact that
$\Omega$ is $U_{\theta}$-analytic in the strip (\ref{eqn_6.17}), as we show in Lemma \ref{lemma7.2} below. So we have
\begin{equation}\label{eq6.1}
K_{\theta} \Omega_{\theta} = 0,
\end{equation}
where $\Omega_{\theta} := U_{\theta} \Omega$.
\begin{lemma}
\label{lemma7.2}
$\Omega$ is $U_\theta$-analytic, for
$\theta=(\delta,\tau)\in S^\pm_{\theta_0}$, see (\ref{eqn_6.17}).
\end{lemma}

{\it Proof of Lemma \ref{lemma7.2}.\ }
This follows from the Dyson series expansion \fer{mm53} given in Lemma \ref{lemma4.2},  and the analyticity condition (B).
\hfill $\blacksquare$

Let $\rho_0 \in (0,\sigma/2)$, where $\sigma$ is given in \fer{gap}, and consider the half-space
\begin{equation}\label{eq6.7}
S = \left\{ z\in \BC \big| \rIm z <{\textstyle{\frac14}}\rho_0 \sin (\rIm \delta )\rho_0\right\}.
\end{equation}
We decompose this region into the strips $S_e = \left\{ z\in S |\, |\rRe z -e | \le \rho_0 \right\}$, where $e \in \sigma (L_p)$, and the complement.
The following result is a special case of Theorem 6.1 of \cite{MMS1}.
\begin{theorem}
\label{spec1} Assume that condition (B) holds and that there are
constants $\gamma_0>0$ and $\delta_0>0$ satisfying ${\rm
Im}\Lambda_e\geq\gamma_0$ for all $0\neq e\in \sigma(L_p)$ and ${\rm
Im}(\sigma(\Lambda_0)\backslash\{0\})\geq \delta_0$. Take
$0<|g|<\sqrt{\rho_0}g_0$ and let $\alpha=(\mu-1/2)/(\mu+1/2)$, where
$\mu>1/2$ is given in Condition (B). Then

1. We have $\sigma(K_\theta)\cap S\ \subset\ \bigcup_{e\in\sigma(L_p)}S_e$.

2. Choose $\rho_0=|g|^{2-2\alpha}$. There is a $C>0$ s.t. if $0<|g|< C (\gamma_0)^{1/\alpha}$, then, for all $e\neq 0$,
\begin{equation}
\sigma(K_\theta)\cap S_e \ \subset \ \{ z\in {\mathbb C} \ |\ {\rm Im}z\geq {\textstyle \frac12} g^2\gamma_0\},
\label{mf1}
\end{equation}

3. Choose $\rho_0=|g|^{2-2\alpha}$. There is a $C>0$ s.t. if $0<|g|<C \min[(\delta_0)^{1/\alpha},(\tau')^{\frac{1}{2+\alpha}}]$ then
\begin{equation}
\sigma(K_\theta)\cap S_0 \ \subset \ \{z_0\}\cup\{z\in{\mathbb C}\ |\ {\rm Im}z\geq {\textstyle \frac 12}\min(g^2\delta_0,\tau')\},
\label{mf2}
\end{equation}
where $z_0$ is a simple isolated eigenvalue of $K_\theta$, satisfying $|z_0|=O(|g|^{2+\alpha})$.
\end{theorem}
For a coupling constant satisfying
\begin{equation}
0<|g|<\min\left[(g_0)^{1/\alpha}, (\gamma_0)^{1/\alpha}, (\delta_0)^{1/\alpha},(\tau')^{\frac{1}{2+\alpha}}\right]
\label{mf3}
\end{equation}
all three parts of Theorem \ref{spec1} apply. Thus $K_\theta$ has a simple isolated eigenvalue $\{z_0\}$ in a neighbourhood $O(|g|^{2+\alpha})$ of the origin, and $\sigma(K_\theta)\backslash\{ z_0 \} \subset \{z\in{\mathbb C}_+\ | \ {\rm Im}z \geq  \min(cg^2,{\textstyle\frac12}\tau')\}$, where $c={\textstyle\frac12}\min(\gamma_0,\delta_0)$. In order to complete the proof of Theorem \ref{thm6.1} we only need to remark that $z_0=0$ since zero is an eigenvalue of $K_\theta$, see \fer{eq6.1}. (One can also give a dynamical argument to prove that $z_0=0$, see the remark after \fer{eq8.5}.) \hfill $\blacksquare$

\section{Resolvent representation and pole approximation}\label{Sect7}

In order to study the long-time behaviour of the evolution
$U(t)=e^{iKt}$ (defined on the domain $\pi(\cA)\Omega$ by \fer{eq4.1})
we relate it to an object which we understand relatively
well, namely the resolvent $(K_{\theta} - z)^{-1}$ of the deformation $K_{\theta}$, defined in Section~\ref{Sect6}.
The main result of this section is

\begin{theorem}\label{thm7.1}
Assume that Conditions (A), (B) and \fer{part} hold.
Let $\phi$ and $\Psi$ be $U_\theta$-analytic vectors, and let
$\Psi=\pi(A)\Omega$ for some $A\in{\cal A}$. We have the following
representation
\begin{equation}\label{eq7.1}
\innerprod{ \phi}{ e^{iKt} \Psi } = -\frac1{2\pi i}
\oint\limits_\Gamma \innerprod{ \phi_{\btheta}}{ (K_{\theta}-z)^{-1}
\Psi_\theta } e^{izt} dz \ ,
\end{equation}
where $\Psi_\theta = U_\theta \Psi$ (similarly for $\phi$ or any
other vector), $\theta \in S^+_{\theta_0}$, and $\Gamma$ is the
path
\begin{multline}
\Gamma:=\{z=\lambda- i \tau'/3, |\lambda|\leq C\}\cup \{
z=\lambda - i 2\tau'/3+i \lambda \tau' / 3 C , \lambda\geq C\}\\
\cup
\{ z=\lambda - i2\tau'/3-i
\lambda \tau' / 3 C , \lambda\leq -C\}
\end{multline}
for a sufficiently large constant $C$. The integral on the r.h.s. of \fer{eq7.1} is well defined in virtue of Theorem \ref{thm_6.1}, \fer{eq6.3}, and the estimate $|e^{izt}|\leq e^{-\lambda\tau'/3C}$ on the infinite branches of $\Gamma$.
\end{theorem}

{\it Proof.\ } In the proof below, the vectors $\phi$ and $\Psi$ are
as in the theorem. To prove the equality in (\ref{eq7.1}) we use the
family $K_{(s)}$ of operators defined in \fer{eq7.5}-\fer{eq7.8}
(here we use Conditions (A) and \fer{part}). Note that
$\Omega\in\dom(e^{\nu N})$ for any $\nu
> 0$, as follows from the relation $\Omega_{0}\in\dom(e^{\nu N}
e^{-\beta L^{(\ell)}/2})$ which is shown in the same way as the
relation $\Omega_{0}\in\dom(e^{-\beta L^{(\ell)}/2})$, see Lemma
\ref{lemma4.2}.

Next, we define the operator $e^{iK_{(s)}t}$ as follows:
$e^{iK_{(s)}t}\Omega$ is
given by a Dyson expansion, where the part $-gV_s'$ of $K_{(s)}$ is
treated as a perturbation. The fact that the Dyson series converges is
easily seen from the relation $\Omega\in\dom(e^{\nu N})$,
$\nu >0$, shown above. Moreover, it is clear that
this series defines a vector which is analytic in $s\in S_\varepsilon$,
i.e., $s\mapsto e^{iK_{(s)}t}\Omega$ is analytic for $s\in S_\varepsilon$.
We define the action of $e^{iK_{(s)}t}$ on vectors
$\pi(A)\Omega$, $A\in \cA$ (which form a dense set), by
\begin{equation}
\label{eq7.11}
e^{iK_{(s)} t} \pi (A) \Omega = \sigma^t(\pi(A)) e^{iK_{(s)}t}\Omega.
\end{equation}
Consequently, the map $s\mapsto e^{iK_{(s)}t}\pi(A)\Omega$ is analytic for
$s\in S_\varepsilon$.
For $s=\frac{-i}{2}$ this definition gives $e^{iKt}$.

Since $K_{(s)}$ is self-adjoint for $s$ real we derive from
 Stone's formula
\begin{equation}\label{eq7.13}
\innerprod{\phi}{e^{i K_{(s)} t} \Psi} = -\frac1{2\pi i}
\oint\limits_{\BR-i}
\innerprod{\phi}{ (K_{(s)}-z)^{-1} \Psi} e^{izt} dz,\ \ \ s\in{\mathbb R}.\\
\end{equation}
Next, using $e^{izt} = \frac1{it}
\frac{\partial}{\partial (\rRe z)} e^{izt} $ and integrating by
parts we can represent the r.h.s. of (\ref{eq7.13}) as
\begin{equation*}
RHS (\ref{eq7.13}) = -\frac1{2\pi i} \int\limits_{\BR-i}
\innerprod{ \phi}{ (K_{(s)}-z)^{-2} \Psi }
e^{izt} d z.
\end{equation*}
Now we perform the spectral deformation, Theorem \ref{thm_6.1},
(iv),  to obtain for $\theta_1/2>{\rm Im}\theta> 0$ (where
$\theta_1$ is given in Theorem \ref{thm_6.1}, (iv); here we use
Conditions (B))
\begin{equation}
\label{eq7.4}
RHS (\ref{eq7.13}) = -\frac1{2\pi i} \frac{1}{it} \int\limits_{\BR-i}
\innerprod{ \phi_{\btheta }}{ (K_{(s)\theta}-z)^{-2} \Psi_\theta }
e^{izt} dz.
\end{equation}
The integral converges since due to \fer{eq6.3} we have $\big\| (K_{(s)\theta}-z)^{-n} \| \le C_n \langle \rRe z\rangle^{-n}$ for $z\in \BR - i\vep$,
where $\langle x \rangle := \sqrt{1+x^2}$.

In (\ref{eq7.4}) we deform the contour of integration from $\BR-i$ to
$\Gamma$ which is fine due to \fer{eq6.3}, and we integrate by parts in the opposite direction of above, to obtain
\begin{equation}
 \innerprod{\phi}{e^{i K_{(s)} t} \Psi} =
-\frac1{2\pi i} \oint\limits_{\Gamma} \innerprod{
\phi_{\btheta}}{ (K_{(s)\theta}-z )^{-1} \Psi_\theta} e^{izt} dz
\ . \label{eq7.14}
\end{equation}
Both sides of this expression are
well defined and analytic for $s\in S_\vep$ (see Theorem~\ref{thm_6.1}, \fer{eq6.3}, and after equation \fer{eq7.11}). Since
they are equal for real $s$ they are
equal for all $s$ in their domain of analyticity. Taking $s =
-\frac{i}{2}$ in this formula gives (\ref{eq7.1}). \hfill \qed

It is shown in Section~\ref{Sect_7} that the operator family
$K_\theta$ has a simple isolated eigenvalue at $0$ and the rest of
its spectrum is located in a truncated cone in $\{ z \in \BC^+ |
\rIm z > \frac13 \tau' \}$, where $\tau' = \rIm \tau $. In the
integral on the r.h.s. of formula (\ref{eq7.1}) we deform the
contour of integration to
\begin{equation}
\Gamma':=\{z=\lambda+ {\textstyle \frac13} i \tau', |\lambda|\leq C\}\cup \{
z=\lambda +i \lambda \tau' / 3 C , \lambda\geq C\} \cup \{ z=\lambda -i
\lambda \tau' / 3 C , \lambda\leq -C\}
\label{gamma'}
\end{equation}
where $C$ is sufficiently large. Picking up the residue from the
simple eigenvalue $0$ of $K_{\theta}$ we derive from (\ref{eq7.1})
\begin{equation}\label{eq8.1}
\innerprod{ \phi }{ e^{iKt} \Psi \rangle = \langle \phi_\btheta ,
P_{\theta} \Psi_\theta } + O(\|\phi_\btheta\|\
\|\Psi_\theta\|e^{-\tau' t/3}) \ ,
\end{equation}
where $ P_{\theta}=\frac{-1}{2\pi i}\oint(K_\theta-z)^{-1}dz$ is the
eigenprojection of $K_{\theta}$ corresponding to the simple and
isolated eigenvalue $0$ and the remainder bound is coming from the
term  $\frac{1}{2\pi} \oint_{\Gamma'} \innerprod{\phi_\btheta}
{(K_\theta-z)^{-1}\Psi_\theta} e^{izt}dz $. The contour integral is
over a small circle around the origin and the path $\Gamma'$  is
defined in \fer{gamma'}. This is the only place where we use that
$0$ is an isolated eigenvalue of $K_{\theta}$ -- the fact we show
using complex translation in addition to complex dilation.

$P_\theta$ is a rank one projection which is analytic in $\theta\in S_{\theta_0}^+$. One proves using a standard argument that it satisfies $K_\theta P_\theta = P_\theta K_\theta = 0$. Hence,
$P_\theta$ can be written as
\begin{equation}\label{eqn_10_2}
P_\theta = \ket{\Omega_\theta} \bra{\Omega^*_\btheta},
\end{equation}
where $\Omega_\theta$ and $\Omega^*_\btheta$ are zero eigenvectors
of $K_\theta$ and its adjoint operator
\begin{equation}
(K_\theta)^*=(K^*)_\btheta=:K^*_\btheta,
\label{eqn_10_2.5}
\end{equation}
i.e. we have
\begin{equation}
\label{eqn_10_3}
K_\theta \Omega_\theta = 0\
{\hbox{ and }}\
K^*_\btheta \Omega^*_\btheta = 0,
\end{equation}
with the normalization $\langle \Omega^{*}_{\thetabar} ,
\Omega_\theta \rangle = 1$.
Since $P_\theta$ and $\Omega_\theta$ are analytic in $\theta\in S_{\theta_0}^+$, then so is $\Omega_\thetabar^*$ in the variable $\thetabar\in S_{\theta_0}^-$.
(The possibility of the normalization, $\langle\Omega^*_\btheta, \Omega_\theta\rangle \neq 0$, follows also from
results of Section~\ref{Sect_11}. Analyticity of $\Omega_\thetabar^*$ in $\thetabar\in S_{\theta_0}^-$ can also be shown directly by using the analyticity and spectral properties of $K^*_\thetabar$, see Section \ref{Sect_7}). Equation \fer{eqn_10_2} implies
\begin{equation}\label{eq8.5}
\innerprod{ \phi_\btheta }{ P_{\theta}\Psi_\theta } = \innerprod{
\phi }{ \Omega } \innerprod{ \Omega^*_\btheta }{ \Psi_\theta } \ .
\end{equation}

The resonance vector $\Omega^*$ appearing in \eqref{weaksense} and
\eqref{eq4.8} is defined as
\begin{equation}\label{eq8.5a}
\innerprod{ \phi }{  \Omega^*} = \innerprod{ \phi_{-\btheta}}{
\Omega^*_\theta} \ .
\end{equation}

\indent {\it Remark.\ } We present here another proof of the
relation $z_0=0$, where $z_0$ is the simple isolated eigenvalue of
$K_\theta$ given in Theorem \ref{spec1}. Starting with the
information on the spectrum of $K_\theta$ given in Theorem
\ref{spec1} and proceeding with a contour deformation as above we
find that $\scalprod{\phi}{e^{iKt}\psi} =
e^{iz_0t}\scalprod{\phi_{\overline\theta}}{P_\theta \psi_\theta} +
O(e^{(|\rIm z_0|-\tau'/3)t})$ instead of \fer{eq8.1}. Applying this
formula to $\phi=\psi=\Omega$ and using \fer{eq4.1} we obtain
$\scalprod{\Omega}{\Omega} = e^{iz_0t}\scalprod{\Omega}{\Omega}
+O(e^{(|\rIm z_0|-\tau'/3)t})$. Since
$|z_0|=O(|g|^{2+\alpha})\lless\tau'$ the error term tends to zero as
$t\rightarrow\infty$, so by taking
$\lim_{T\rightarrow\infty}\int_0^T dt$ on both sides we see that
$z_0$ must be zero.

\section{ Proof of  Theorem \ref{thm2.1}}
\label{Sect8}

Let $\psi$ be any $\omega_0$-normal state on $\cA$. For the following reasoning, we may assume without loss of generality that $\psi(A)=\innerprod{\Omega_\psi}{\pi(A)\Omega_\psi}$, for some $\Omega_\psi\in {\cal H}$. Since $\Omega$ is cyclic for $\pi({\cal A})'$, $\psi$ can be approximated as follows. For any $\epsilon >0$ there is a $Q\in \pi({\cal A})'$ s.t., for all $A\in\cA$,
\begin{equation}
|\psi(A)-\innerprod{Q\Omega}{\pi(A)Q\Omega}|<\epsilon\|A\|.
\end{equation}
Applying this to $\psi^t(A)=\scalprod{\Omega_\psi}{\sigma^t(\pi(A))\Omega_\psi}$, pulling $Q$ through $\sigma^t(\pi(A))$ and taking into account \fer{eq4.1'}, $U(t)=e^{itK}$ and $U(t)\Omega=0$, we obtain
\begin{equation}
|\psi^t(A)-\innerprod{Q^*Q\Omega}{e^{itK}\pi(A)\Omega}|<\epsilon\|A\|,
\label{yep0}
\end{equation}
uniformly in $t\in{\mathbb R}$. In order to examine the long time behaviour of $\innerprod{Q^*Q\Omega}{e^{itK}\pi(A)\Omega}$, via \fer{eq8.1}, we first approximate the vector $Q^*Q\Omega\in{\cal H}$ by a family of $U_\theta$-analytic vectors $\chi_\epsilon$, s.t. $\|\chi_\epsilon- Q^*Q\Omega\|<\epsilon$. We have
\begin{equation}
|\innerprod{Q^*Q\Omega-\chi_\epsilon}{e^{itK}\pi(A)\Omega}|\leq\epsilon \|e^{itK}\pi(A)\Omega\| =\epsilon \|\sigma^t(\pi(A))\Omega\| \leq \epsilon \|A\|.
\label{yep1}
\end{equation}
It follows from \fer{yep0}, \fer{yep1}, \fer{eq8.1} and \fer{eq8.5} that
\begin{equation}
\left|\psi^t(A)-\scalprod{\chi_\epsilon}{\Omega}\scalprod{\Omega_{\overline\theta}^*}
{(\pi(A)\Omega)_\theta}\right| \leq 2\epsilon \|A\| +
C\|\chi_{\epsilon,\btheta}\|\ \|(\pi(A)\Omega)_\theta\|
e^{-\tau't/3}. \label{--1}
\end{equation}

Since $\innerprod{Q^*Q\Omega}{\Omega}\rightarrow 1$ as $\epsilon\rightarrow 0$, we have  $\innerprod{\chi_\epsilon}{\Omega}=1+o(\epsilon^0)$, where $o(\epsilon^0)$ denotes a quantity that vanishes in the limit $\epsilon\rightarrow 0$. Thus
\begin{equation*}
|\psi^t(A)-\innerprod{\chi_\epsilon}{\Omega}\innerprod{\Omega_\btheta^*}{(\pi(A)\Omega)_\theta}|
\geq |\psi^t(A)-\innerprod{\Omega_\btheta^*}{(\pi(A)\Omega)_\theta}| -o(\epsilon^0)\ |\innerprod{\Omega_\btheta^*}{(\pi(A)\Omega)_\theta}|.
\end{equation*}
Combining this estimate with \fer{--1}
we arrive at
\begin{eqnarray}
\lefteqn{
|\psi^t(A) -\innerprod{\Omega_\btheta^*}{(\pi(A)\Omega)_\theta}|}\nonumber\\
&& \leq o(\epsilon^0) \left( \|A\| +|\innerprod{\Omega_\btheta^*}{(\pi(A)\Omega)_\theta}|\right)
+C\|\chi_{\epsilon,\overline\theta}\|\ \|(\pi(A)\Omega)_\theta\| e^{-\tau't/3},
\label{yep5}
\end{eqnarray}
where $o(\epsilon^0)$ is independent of $A$ and $\theta$. Let $\eta$
be the state  on $\widehat \cA$
given by
\begin{equation}
\label{eq8.7} \eta (A) := \innerprod{ \Omega^*_\btheta }{ (\pi (A)
\Omega )_\theta}.
\end{equation}
Note that $\eta(A)$ is independent of the deformation parameter
$\theta$, if $0 < \rIm \theta < \theta_0 $, and $0<g<g_0$ ($g_0$
depends on $\sin(\rIm\delta)$, see the equation after \eqref{gap}).
Taking in \eqref{yep5}  first $t\rightarrow\infty$ and then
$\epsilon\rightarrow 0$ yields
\begin{equation}
\lim_{t\rightarrow\infty} \psi^t(A)= \eta(A),
\label{yep6}
\end{equation}
Equation \fer{yep6} shows that $|
\innerprod{\Omega_\btheta^*}{(\pi(A)\Omega)_\theta}|\leq \|A\|$. We
re-inject this inequality into r.h.s. of \fer{yep5} to arrive at
\begin{equation}
\lim_{t\rightarrow\infty} {\sup_{A\in{\widehat{\cA}}}} \frac{ |\psi^t(A)-  \eta(A)|}{|||A|||}=0,
\label{yep7}
\end{equation}
where
\begin{equation}
\widehat\cA =\{A\in\cA\ |\ \pi(A)\Omega \mbox{\ is $U_\theta$-analytic for $|\theta|<\theta_0$}\},
\label{eq8.10}
\end{equation}
and where $|||\cdot|||$ is the norm on $\widehat\cA$ defined by
\begin{equation}
|||A|||= \|A\|+\sup_{|\theta|<\theta_0}\|(\pi(A)\Omega)_\theta\|, \mbox{\ \ for $A\in\widehat \cA$}.
\label{simnorm}
\end{equation}

Observe that $\bfone\in\widehat \cA$, and that the normalization
$\innerprod{\Omega^*_\btheta}{\Omega_\theta} = 1$ implies
$\eta(\id) = 1$. $\widehat \cA$ is a linear subspace of $\cA$, but not an algebra. \\
\indent
We show in Appendix \ref{AppD}, Proposition \ref{propc1},  that ${\cal A}_1$
(defined in \fer{eqn_4_9}) is strongly dense in ${\cal A}$
(defined in \fer{calA}), and that any $A\in \cA_1$ has the property that $\pi(A)\Omega$ is $U_\theta$-analytic, for $\theta$ in a neighbourhood of $\theta=0$. Hence ${\cal A}_1\subseteq \widehat \cA \subseteq{\cal A}$,
and therefore $\widehat \cA$ is strongly dense in $\cA$. \\
\indent
We have thus shown that for any $\omega_0$-normal state $\psi$,
$
\psi^t \rightarrow \eta$ as $t\rightarrow\infty$,
where the convergence is understood in the $|||\cdot|||$-topology of linear functionals on $\widehat \cA$. \\
\indent It is clear from \fer{yep7} that $|\eta (A) | \le \| A \|$,
for $A\in\widehat \cA$, hence $\eta$ extends to a bounded positive
linear functional on the Banach space of observables
\begin{equation}
\mbox{$\cA_0:=$ $\|\cdot\|$-closure of $\widehat\cA$,}
\label{a_0}
\end{equation}
normalized as $\eta(\bfone)=1$. Standard perturbation theory shows that $\eta(A)$, $A\in\widehat\cA$, is analytic in the coupling constant $g$.

Observe that we can rewrite the state $\eta(A)$ also in the form
\begin{equation}\label{eq8.14}
\eta (A) = \rTr \big(\pi (A)_\theta P_\theta \big)
\end{equation}
where $\pi (A)_\theta := U_\theta \pi (A) U^{-1}_\theta $.
Formally one can undo the rotation in (\ref{eq8.7}) to obtain
(\ref{eq4.10}) with $\Omega^* := U^{-1}_\btheta \Omega^*_\btheta$.
However, in the non-equilibrium situation $\Omega^* \not\in \cH$!
The set $\cA_1$ is exactly the set on which (\ref{eq4.10}) makes
sense. Thus we gave a rigorous meaning to (\ref{eq4.10}) and the
NESS vector $\Omega^*$.

\section{Proof of Theorem \ref{thmlso}}
\label{Sect13}
Our task is to show that the spectrum of \fer{m2} lies in the
upper complex half plane $\{\rIm z >0\}$ if $e\neq 0$; and that it
has a simple eigenvalue at zero and all the other eigenvalues lie
in the upper complex half plane if $e=0$. While this analysis is
standard in the case when $I$ is a selfadjoint operator (then the
imaginary part of \fer{m2} is just $P_e I \delta(L_0-e) I P_e$,
manifestly a non-negative operator; see e.g. \cite{BFS:RtE}), it
needs some more thought in our case, where the interaction is
non-selfadjoint.  Let
\begin{equation*}
V_j = \pi(v_j),\ \ \ \mbox{and}\ \ \ V'_j = \pi'(v_j).
\end{equation*}
The main ingredient of the proof is
\begin{proposition}
\label{aepsilon} Assume Condition (A). We have
\begin{equation}
 \Lambda_e = \left(e^{-\beta_p
    H_p/2}\otimes\bbbone_p\right)
    \left[\sum_{j=1,2}\left(e^{\beta_j
    H_p/2}\otimes \bbbone_p\right) \Lambda_{je} \left(e^{-\beta_j
    H_p/2}\otimes \bbbone_p\right) \right]\left(e^{\beta_p
    H_p/2}\otimes\bbbone_p\right),
\label{m3}
\end{equation}
where, setting $\res:=(L_0-e+i0)^{-1}$,
\begin{equation}
\label{eq_8_3}
\Lambda_{je} = - P_e (V_j - V'_j)\res (V_j - V'_j) P_e.
\end{equation}
Notice that \fer{m3} shows that the spectrum of $\Lambda_e$ is
independent of $\beta_p$.
\end{proposition}

The importance of \fer{m3} is that it relates $\Lambda_e$ to the operators $\Lambda_{je}$ whose spectral
characteristics are well known. Indeed, $\Lambda_{je}$ are the
level shift operators corresponding to the reservoir $j$ coupled
to the particle system, studied in \cite{BFS:RtE,Mlso}.

Before proceeding to the proof we examine some consequences of this
proposition. We assume Conditions (C) and (D) in addition to Condition (A).

\indent {\bf The case $e\neq 0$.\ } Let us assume that the nonzero eigenvalues of $L_p$ are simple, i.e. $E_i-E_j=E_m-E_n \Leftrightarrow i=m,j=n$. For a treatment of the more general case where $E_i-E_j=E_m-E_n$, for $(i,j)\neq (m,n)$, with simple $E_j$, we refer to \cite{MaMueThesis}. Since $P_e$ is of rank one $\Lambda_e$ is just
a complex number, namely the sum of $\Lambda_{1e}+\Lambda_{2e}$
(the dependence on $\beta_1, \beta_2$ disappears).  Under condition
(\ref{eqn_2.7}), one has $\rIm \Lambda_{e}\geq \gamma_0$, where $\gamma_0$ is a strictly
positive constant which is independent of the inverse temperatures, see \cite{BFS:RtE}). This shows (\ref{eq6.4})
and (\ref{eq6.5})
for $e\neq 0$.

\indent {\bf The case $e=0$.\ } Zero is necessarily a degenerate
eigenvalue of $L_0$, so the above reasoning does not apply. In
particular, $\Lambda_{10}$ and $\Lambda_{20}$ do not commute. It is shown in \cite{Mlso, MaMueThesis} that
\begin{equation}
\label{eqn_9.4}
\Lambda_{j0} = i\rIm \Lambda_{j0} =: i\Gamma_{j0},
\end{equation}
where $\rIm \Lambda_{j0} :=
\frac{1}{2i}(\Lambda_{j0}-\Lambda_{j0}^*)$. We use here implicitly
Condition (C) on the non-degeneracy of $H_p$; if the small system
has degenerate energy levels then $\Lambda_{j0}$ are not purely
imaginary \cite{Mlso}. One shows as in \cite{BFS:RtE, MaMueThesis,
Mlso} that $\Gamma_{j0}$ are real matrices having strictly negative
off-diagonal entries, $(\Gamma_{j0})_{mn}<0$, for $m\neq n$, and
satisfying
\begin{equation}
\Gamma_{j0}\Omega^{(\beta_j)}_p=0, \label{8.4a}
\end{equation}
where $\Omega^{(\beta_j)}_p$ is the particle Gibbs state at
temperature $\beta_j$. Hence, since
\begin{equation*}
\left(e^{(\beta_p-\beta_j)H_p/2}\otimes\bbbone_p\right)
\Omega^{(\beta_p)}_p=\sqrt{\frac{ {\rm tr} e^{-\beta_j H_p/2}}{{\rm tr}
e^{-\beta_pH_p/2}}} \ \Omega^{(\beta_j)}_p,
\end{equation*}
we see that
\begin{equation*}
\Gamma_{0}\Omega^{(\beta_p)}_p= \sum_{j=1,2}
\left(e^{(-\beta_p+\beta_j)H_p/2}\otimes\bbbone_p\right) \sqrt{\frac{
{\rm tr}
  e^{-\beta_j H_p/2}}{{\rm tr} e^{-\beta_pH_p/2}}}\ \Gamma_{j0}
\Omega^{(\beta_j)}_p =0,
\end{equation*}
where $\Gamma_0:=-i\Lambda_0$.
Thus, $\Omega^{(\beta_p)}_p$ is an eigenvector of the real matrix
$\Gamma_0$ with eigenvalue zero. Notice that the vector
$\Omega^{(\beta_p)}_p$ has strictly positive components, $[{\rm
tr}e^{-\beta_p H_p/2}]^{-1} e^{-\beta_p E_n/2}$, in the
orthonormal basis $\{\varphi_n\otimes\varphi_n\}$ of $\rRan
\chi_{L_p=0}$ (where $H_p\varphi_n=E_n\varphi_n , \| \varphi_n \|
= 1$). Moreover, the off-diagonal elements of the real matrix (which is not symmetric for $\beta_1\neq\beta_2$)
$\Gamma_0$ are given by
\begin{eqnarray}
\lefteqn{(\Gamma_0)_{m,n}=
\sum_{j=1,2}
\left[\left(e^{(-\beta_p+\beta_j)H_p/2}\otimes\bbbone_p\right)
\Gamma_{j0}
\left(e^{-(-\beta_p+\beta_j)H_p/2}\otimes\bbbone_p\right)\right]_{mn}}\nonumber\\
&&=-\pi\sum_{j=1,2}
\frac{E^2_{mn} e^{-(\beta_p-\beta_j)E_{mn}/2}}{|\sinh(\beta_j E_{mn}/2)|}
\int_{S^2}d\sigma\ |[G_j(E_{mn},\sigma)]_{nm}|^2
\label{fgrc}
\end{eqnarray}
for $m>n$ (and similarly for $m<n$, see also \cite{BFS:RtE,
MaMueThesis}). Hence Condition (D) implies that
$(\Gamma_0)_{m,n}<0$. A standard {\it Perron-Frobenius argument}
shows that zero is a simple eigenvalue of $\Gamma_0$, and that
$\sigma(\Gamma_0)\backslash\{0\} \subset {\mathbb C}^+$. This shows
equations (\ref{eq6.5}) -- (\ref{eq6.6}) for $e=0$.  It is shown in
\cite{BFS:RtE} that the gap at the bottom of the spectrum of
$\Gamma_{j0}$, $j=1,2$ has a lower bound which is independent of the
inverse temperatures.

We now prove the existence of $\delta_0$, assuming that Condition
(E) is satisfied. If $\dim {\cal H}_p=2$ then one eigenvalue of
$\Lambda_0$ is zero and the other equals the trace of $\Lambda_0$.
Expressions \fer{m3} and \fer{eqn_9.4} show that ${\rm
Tr}(\Lambda_0) = i[ {\rm Tr}(\Gamma_{10}) +{\rm Tr}(\Gamma_{20})]$.
Thus the spectral gap of $\Lambda_0$ is the sum of the gaps of
$\Gamma_{01}$ and $\Gamma_{02}$, which have lower bounds uniform in
the inverse temperatures.

Next take $\dim {\cal H}_p\geq 3$. For $\delta\beta=|\beta_1-\beta_2|=0$ the matrix $\Lambda_0$ has the same spectrum as $\sum_j \Lambda_{j0}$, see \fer{m3}. An application of the minimax principle demonstrates that the spectral gap of the latter operator has to be at least as large as the maximum of the gaps of $\Gamma_{j0}$, $j=1,2$. For small values of $\delta\beta$ (c.f. \fer{aaahhh}), the existence of $\delta_0$ follows by perturbation theory.

Finally we consider the case where $\delta\beta$ and $\beta_1, \beta_2$ are large (see \fer{aaahhh}). Let us take $\beta_2=\beta_1+\delta\beta$. As is easily seen from \fer{m3} we have
\begin{equation}
\sigma(\Lambda_0) = i\sigma\left(\left[ \Gamma_{10} + (e^{\delta\beta H_p/2}\otimes\bbbone) \Gamma_{20} (e^{-\delta\beta H_p/2}\otimes\bbbone)\right]\right).
\label{aahh2}
\end{equation}
Using the explicit expression for the matrix elements of $\Gamma_{20}$ in the basis $\varphi_j\otimes\varphi_j$ (which can be read off of relation \fer{fgrc} for off-diagonal terms, and is easy to obtain for the diagonal ones), one verifies that the matrix $(e^{\delta\beta H_p/2}\otimes\bbbone) \Gamma_{20} (e^{-\delta\beta H_p/2}\otimes\bbbone)$ converges to a lower-triangular matrix $Q(\beta_1)$, in the limit $\delta\beta\rightarrow\infty$ (uniformly in $\beta_1$), and furthermore, that $Q(\beta_1)\rightarrow D$ as $\beta_1\rightarrow\infty$, where $D$ is a diagonal matrix with non-negative entries. The minimax principle implies that all but one eigenvalues of $\Gamma_{10}+D$ are greater than, or equal to the gap of $\Gamma_{10}$. From perturbation theory we know that for $\delta\beta$ and $\beta_1$ sufficiently large (independently of each other), all but one eigenvalues of the operator $\Gamma_{10} + (e^{\delta\beta H_p/2}\otimes\bbbone) \Gamma_{20} (e^{-\delta\beta H_p/2}\otimes\bbbone)$ must have real part greater than, or equal to half of the gap of $\Gamma_{10}$. The existence of $\delta_0$ now follows from \fer{aahh2}. \hfill $\blacksquare$

{\it Proof of Proposition \ref{aepsilon}. \ } Let $V =
\sum_{j=1}^2 V_j $ and $V' = \sum_{j=1}^2 V'_j$. By the definition
of the operator $I$, $I := V - V'_{-i/2}$, and the relation
$V'_{-i/2} = e^{\tilde{L}/2} V' e^{-\tilde{L}/2}$ we have
\begin{equation*}
  I = V - e^{\tilde{L}/2} V' e^{-\tilde{L}/2},
\end{equation*}
where $\tilde{L} := \delta\beta_p L_p + \delta\beta_1 L_{r1} +
\delta\beta_2 L_{r2}$ with $\delta\beta_p = \beta_p - \beta$ and
$\delta\beta_j = \beta_j - \beta$. Using that $P_e
e^{-\delta\beta_pL_p/2}=P_e e^{-\delta\beta_pe/2}$ we decompose
\begin{eqnarray}
\Lambda_e &=& P_e V\res V P_e + P_e V'\res V'P_e \nonumber\\
&& -P_e V \res e^{\tilde{L}/2} V'P_e e^{-\delta\beta_p e/2} -P_e
V' e^{-\tilde{L}/2}\res V P_e e^{\delta\beta_p e/2}. \label{m4}
\end{eqnarray}
Notice that $V$ and $\res$ commute with $\bbbone_p\otimes
e^{\delta\beta_pH_p/2}$. Using this and the relation
\begin{equation}\label{eqn_8_7}
  (\id_p \otimes e^{\delta\beta_p H_p/2}) P_e =
  e^{-\delta\beta_p e/2} (e^{\delta\beta_p H_p/2} \otimes \id_p)
  P_e,
\end{equation}
we obtain
\begin{eqnarray*}
P_e V\res VP_e &=& e^{\delta\beta_pe/2}
P_e\left(e^{-\delta\beta_pH_p/2}\otimes
  e^{\delta\beta_pH_p/2}\right)  V\res VP_e \\
&=&e^{\delta\beta_pe/2}P_e
\left(e^{-\delta\beta_pH_p/2}\otimes\bbbone_p\right) V\res V
  \left(\bbbone_p\otimes e^{\delta\beta_pH_p/2}\right)P_e\\
&=&\left(e^{-\delta\beta_pH_p/2}\otimes\bbbone_p\right) P_eV\res
VP_e
  \left(e^{\delta\beta_pH_p/2}\otimes\bbbone_p\right).
\end{eqnarray*}
Next, using (\ref{eqn_8_7}) again, we find
\begin{eqnarray*}
\lefteqn{
P_eV\res e^{\tilde{L}/2} V'P_e e^{-\delta\beta_pe/2}}\\
&=& P_eV\res\left(e^{\delta\beta_pH_p/2}\otimes e^{-\delta\beta_p
H_p/2}\right)
e^{(\delta\beta_1L_{r1}+\delta\beta_2L_{r2})/2} V'P_e e^{-\delta\beta_pe/2}\\
&=& P_e\left(\bbbone_p\otimes e^{-\delta\beta_p H_p/2}\right)
V\res e^{(\delta\beta_1L_{r1} +\delta\beta_2L_{r2})/2} V' \left(
e^{\delta\beta_p H_p/2}\otimes\bbbone_p\right)
P_e e^{-\delta\beta_pe/2}\\
&=&\left(e^{-\delta\beta_pH_p/2}\otimes\bbbone_p\right) P_e V\res
e^{(\delta\beta_1L_{r1} +\delta\beta_2L_{r2})/2}
V'P_e\left(e^{\delta\beta_pH_p/2}\otimes\bbbone_p\right).
\end{eqnarray*}
Treating the other two terms in \fer{m4} in a similar way, we
arrive at
\begin{eqnarray}
\lefteqn{ -\Lambda_e =
\left(e^{-\delta\beta_pH_p/2}\otimes\bbbone_p\right) P_e [ V\res V
+
  V'\res V'}\nonumber\\
&& -V\res e^{(\delta\beta_1 L_{r1} +\delta\beta_2L_{r2})/2} V'
-V'\res
  e^{(-\delta\beta_1L_{r1} -\delta\beta_2L_{r2})/2} V] P_e
\left(e^{\delta\beta_pH_p/2}\otimes\bbbone_p\right). \label{m5}
\end{eqnarray}
We now examine the term in $[\cdots]$. Write $V=V_1+V_2$,
$V'=V'_1+V'_2$, where, recall, $V_j=\pi(v_j)$, $V'_j=\pi'(v_j)$.
Notice that we have $(\bbbone_p\otimes\bbbone_p\otimes
P_\Omega\otimes\bbbone_2) V_1 (\bbbone_p\otimes\bbbone_p \otimes
P_\Omega\otimes\bbbone_2)=0$, and similarly for $V_2$, from which
it follows that the expression $P_e[\cdots]P_e$ in \fer{m5} splits
into a sum
\begin{equation*}
P_e\sum_{j=1,2}\left[ V_j\res V_j +V'\res V'_j -V_j\res
  e^{(\delta\beta_1L_{r1} +\delta\beta_2L_{r2})/2}V'_j -V'_j\res
  e^{(-\delta\beta_1L_{r1}-\delta\beta_2L_{r2})/2}V_j\right]P_e.
\end{equation*}
We consider the $j=1$ term. Using that $L_{r2}$ commutes with
$V_1, V'_1$ and that $L_{r2}P_e=0$, we see that
\begin{eqnarray*}
P_eV_1\res e^{\delta\beta_1L_{r1}+\delta\beta_2L_{r2}}V'_1P_e
&=&P_e V_1\res e^{\delta\beta_1L_{r1}/2} V'_1 P_e=
P_e V_1\res e^{\delta\beta_1L_0/2} e^{-\delta\beta_1 L_p/2} V_1'P_e\\
&=&P_e\left(\bbbone_p\otimes e^{\delta\beta_1H_p/2}\right) V_1\res
e^{\delta\beta_1L_0/2}
V_1'\left( e^{-\delta\beta_p H_p/2}\otimes\bbbone_p\right) P_e\\
&=&\left(e^{\delta\beta_1 H_p/2}\otimes\bbbone_p\right)P_e V_1\res
e^{\delta\beta_1(L_0-e)/2} V'_1P_e \left( e^{-\delta\beta_1
H_p/2}\otimes\bbbone_p\right).
\end{eqnarray*}
All other terms in \fer{m5} for $j=1$, as well as the terms for
$j=2$, are treated similarly and one arrives at
\begin{eqnarray*}
\lefteqn{- \Lambda_e =
\left(e^{-\beta_pH_p/2}\otimes\bbbone_p\right) \sum_{j=1,2}\left(
  e^{\beta_j H_p/2}\otimes\bbbone\right)P_e[ V_j\res V_j +V'_j\res V'_j}\\
&& -V_j\res e^{\delta\beta_j(L_0-e)/2} V'_j -V_j' \res
e^{-\delta\beta_j(L_0-e)/2} V_j]P_e \left(e^{-\beta_j
H_p/2}\otimes\bbbone_p\right)\left(e^{\beta_pH_p/2}\otimes\bbbone_p\right),
\end{eqnarray*}
where we used $\delta\beta_p - \delta\beta_j = \beta_p - \beta_j$.
Hence, $- \Lambda_e = RHS(\ref{m3}) +
(e^{-\delta_pH_p}\otimes\bbbone_p) R
(e^{\delta_pH_p}\otimes\bbbone_p)$, where
\begin{eqnarray*}
\lefteqn{ R=\sum_{j=1,2}\left(e^{\beta_j
H_p/2}\otimes\bbbone\right)P_e [
V_j\res (1-e^{\delta\beta_j(L_0-e)/2}) V_j'}\\
&&+ V_j'\res (1-e^{-\delta\beta_j(L_0-e)/2})V_j]P_e
\left(e^{-\beta_j
    H_p/2}\otimes\bbbone_p\right).
\end{eqnarray*}
Since $L_0$ implements the free dynamics, we have that
$e^{izL_0}V_je^{-izL_0}$ commutes with $V_j'$, for $z\in \BC$.
Using this, writing $\frac{\bfone - e^{\delta \beta_j (L_0 - e)/2
}}{L_0-e\pm i0} = -\int_0^{\delta \beta_j / 2}  \ ds \,  e^{s(L_0
- e)} $ and using that $P_eL_0=eP_e$, we see that
\begin{equation*}
P_eV_j\frac{\bfone - e^{\delta\beta_j(L_0-e)/2}}{L_0-e-i0}
V_j'P_e= -P_e V_j' \frac{\bfone
-e^{-\delta\beta_j(L_0-e)/2}}{L_0-e+i0} V_jP_e.
\end{equation*}
Consequently, $R=0$.
This concludes the proof of Proposition \ref{aepsilon}.\hfill \qed

\section{Perturbation Theory for NESS} \label{Sect_11}

In this section we develop a perturbation theory for the NESS $
\eta(A) := \innerprod{\Omega^*_\btheta}{\pi(A)_\theta
\Omega_\theta}$. The vectors $\Omega^*_\btheta$ and $\Omega_\theta$
are the zero eigenvectors of the operators $K^*_\btheta$ and
$K_\theta$ respectively, see \fer{eqn_10_2.5} and \fer{eqn_10_3}. We
derive perturbation expansions for $\Omega^*_\btheta$ and
$\Omega_\theta$, see \fer{eqn_11.12} and \fer{eqn_12_13} below,
using the Feshbach maps introduced in \cite{BFS:QED, BFS:RG}, and
extended in \cite{BCFS:SmoothFeshbach}. We review the definitions
and some properties of these maps referring the reader to
\cite{BFS:RG,BCFS:SmoothFeshbach} for more detail. For simplicity we
present here the original version, \cite{BFS:QED, BFS:RG}, though
the refined one, \cite{BCFS:SmoothFeshbach}, the smooth Feshbach
map, is easier to use from a technical point of view.

Let $X$ be
a Banach space and let $P$ be a projection on $X$. Define $\oP := \id
- P$ and let $H_\oP := \oP H \oP$ and $R_\oP (H) := \oP H_\oP^{-1}
\oP$ if $H_\oP$ is invertible on ${\rm Ran}\oP$. We define the
Feshbach map $F_P$ by the relation $F_P(H) := P \left(H - H R_\oP(H)H\right)P$
on the domain
\begin{eqnarray}
\lefteqn{
  \dom(F_P) = }\nonumber \\
&&\!\!\!\!\!\! \{ H : X \to X | H_\oP \text{ is invertible}, {\rm Ran}P \subseteq \dom(H),
  {\rm Ran}R_\oP(H) \subseteq \dom(PH\oP) \}.
\nonumber
\end{eqnarray}
A key property of the maps $F_P$ is given in the following
statement proven in \cite{BFS:RG}:

\begin{theorem}[Isospectrality Theorem]
 \label{thm_7_4}
\begin{enumerate}
  \item[(i)] $0 \in \sigma(H) \iff 0 \in \sigma(F_P(H))$,
  \item[(ii)] $H \psi = 0 \iff F_P(H)\varphi = 0$
  with $\varphi = P\psi$ (``$\Rightarrow$") and $\psi=(\id -
  R_\oP(H)H)\varphi$ (``$\Leftarrow$").
\end{enumerate}
\end{theorem}
Let $P_{e\rho}$ be defined as
\begin{equation} \label{eqn7_23}
  P_{e\rho} := \chi_{L_p=e} \otimes \chi_{M_{\theta} \le \rho},
\end{equation}
where $\chi_{L_p = e}$ is the eigenprojection for the operator $L_p$
corresponding to an eigenvalue $e\in\sigma(L_p)$ and
$\chi_{M_{\theta} \le \rho}$ is the spectral projection for the
self-adjoint operator $M_{\theta} := {\rm Im} L_{0, \theta}$
corresponding to the spectral interval $[0,\rho]$ (remember that
$M_{\theta}$ is a positive operator).

The following result is proven in \cite{MMS1}, Lemma 6.3.
\begin{lemma}
\label{lem_7_5} Assume Condition (B). Take $\rho_0\in (0,\sigma/2)$
and let $|g|<\sqrt{\rho_0}\, g_0$, where $g_0$ is given after
\eqref{gap}. If $z \in S_e$ then $K_{\theta z} := K_\theta - z \in
\dom(F_{P_{e\rho_0}})$, and the operator $K^{(1)}_{\theta z} :=
F_{P_{e\rho_0}}(K_{\theta z})$ acting on ${\rm Ran}P_{e\rho_0}$ is
of the form
\begin{equation}\label{eqn_7_26}
  K^{(1)}_{\theta z} = (e-z) \id + L_{r\theta} + g^2 \Lambda_e +
  O(\epsilon (g, \rho_0)).
\end{equation}
The remainder is estimated in operator norm, $\|O(\epsilon (g, \rho_0))\|\leq C \epsilon(\g,\rho_0)$, where $\g=|g| \max_j\sup_{|\theta|<\theta_0} \|G_j\|_{\delta\beta_j,1/2,\theta}$, with a $C$ independent of $\delta\beta_j$, $\theta$,  and where we have set
\begin{equation}
\epsilon (g, \rho) := |g|\rho^{\mu} +
|g|^{3}\rho^{-1/2} +|g|^{2}\rho^{2\mu - 1}.
\end{equation}
\end{lemma}
To unify the following analysis we write $\sOmega_{\stheta}$ for
either $\Omega_\theta$ or $\Omega^*_\btheta$. Correspondingly,
$\sI_{\stheta}$, $L_{0\theta^\#}$ and $K^{\#}_{\stheta}$ stand for
either $I_\theta$ or $I^*_\btheta$, for either $L_{0\theta}$ or
$L_{0\thetabar}$, and for either $K_\theta$ or $K^*_{\btheta}$,
respectively. We use the shorthand $P_0\equiv P_{0\rho_0}$ and
$\overline R_0(A):=\overline P_0A^{-1}_{\overline P_0}\overline
P_0$, where $A_P:=PAP$. We will assume that
\begin{equation}
\label{eqn_12_11} \tau' \gg g^{2+\alpha},\ \ \ \alpha=\frac{\mu-1/2}{\mu+1/2}.
\end{equation}

Theorem~\ref{thm_7_4} and Lemma~\ref{lem_7_5} imply that
$K^{\#}_{\stheta} \in \dom(F_{P_0})$, that
\begin{equation}
\label{eqn_11.9}
  F_{P_0} (K^{\#}_{\stheta}) P_0 \sOmega_{\stheta} = 0,
\end{equation}
and that the original eigenvector $\sOmega_{\stheta}$ can be reconstructed as
\begin{equation}\label{eqn_11.10}
\sOmega_{\stheta} = \Big[ \id - g \oR_0(K^{\#}_{\stheta})
\sI_{\stheta} \Big] P_0 \sOmega_{\stheta}.
\end{equation}
We expand $\oR_0(K^{\#}_{\stheta})$ in this expression into a
Neumann series,
\begin{equation}\label{eqn_11.11}
\sOmega_{\stheta} = \sum_{n=0}^{N-1} g^n (-\oR_0(L_{0\sstheta})
\sI_{\stheta})^n P_0 \sOmega_{\stheta} + O\left((\g\rho_0^{-1/2})^N\right),
\end{equation}
for any $N \ge 1$, provided that $O((\g\rho_0^{-1/2})) =
o_g(1)$. The remainder term in (\ref{eqn_11.11}) is obtained
by using a standard estimate on the $N$th term of the convergent
Neumann series. Indeed, writing
\begin{eqnarray}
  \lefteqn{
    \left[\oR_0 (L_{0\sstheta}) \sI_{\stheta} \right]^N}  \label{eqn_11.11b} \\
  & = & (M_{\theta} + \rho_0)^{-1/2} \left[
      \frac{{M_\theta} + \rho_0}{L_{0\sstheta} \oP_0} \oP_0 (M_{\theta}+\rho_0)^{-1/2}
      \sI_{\stheta} (M_{\theta}+\rho_0)^{-1/2} \right]^N (M_{\theta}+\rho_0)^{1/2} \nonumber
\end{eqnarray}
and using the estimates
\begin{equation}\label{eqn_11.7}
  |g|  \norm{ (M_{\theta} + \rho_0)^{-1/2} \sI_{\stheta} (M_{\theta} +
  \rho_0)^{-1/2}} \le \g\rho_{0}^{-1/2}
\end{equation}
(see also Lemma 5.3 of \cite{MMS1}) and $\norm{\frac{M_0+\rho_0}{L_{0\sstheta}\overline P_0}\overline
P_0}\leq C$, we obtain
\begin{equation}\label{eqn_11.11a}
|g|^N \norm{\left[\oR_0 (L_{0\sstheta}) \sI_{\stheta} \right]^N P_0
\sOmega_{\stheta}} \le \rho_0^{-1/2}
(\g \rho_0^{-1/2})^N \rho_0^{1/2} \ .
\end{equation}
Observe that since $\g \rho_0^{-1} = o_g(1)$ we have
\begin{equation}\label{eqn_11.12}
\sOmega_{\stheta} = \sum_{n=0,1} g^n (- \oR_0 (L_{0\sstheta})
\sI_{\stheta})^n P_0 \sOmega_{\stheta} + o(g).
\end{equation}

Let $K^{\#(1)} := F_{P_0} (\sK_{\stheta})$. As in
(\ref{eqn_7_26}) it can be written as
\begin{equation*}
  K^{\#(1)} = K^{\#(1)}_0 + W^{\#},
\end{equation*}
where $K^{\#(1)}_0 := L_{r\sstheta} + g^2 \Lambda^{\#}$ with
$\Lambda^{\#}$ either $\Lambda_0$ or $\Lambda^*_0$, where
$\Lambda_0$ is given by (\ref{m2}) with $e =0$. We assume that $\delta\beta$ varies
in the set $|\delta\beta|\leq c$, for some $c>0$, so that
$\g$ can be replaced by $g$. Take $\rho_0=g^{2-2\alpha}$ with $\alpha=\frac{\mu -1/2}{\mu+1/2}$, then Lemma \ref{lem_7_5} gives
\begin{equation}
W^{\#} =O(g^{2+\alpha}).
\label{mm2}
\end{equation}
 By Theorem~\ref{thm_7_4},
\begin{equation}\label{eqn_12_10}
  K^{\#(1)} P_0 \sOmega_{\stheta} = 0.
\end{equation}
Let $Q_0^{\#} := \chi_{K^{\#(1)}_0 = 0} = \chi_{\Lambda^{\#}=0}
\otimes \chi_{L_r=0}$, where $\chi_{\Lambda^\#=0}$ is the Riesz
projection onto the kernel of $\Lambda^\#$. The operator
$K_0^{\#(1)}$ is normal with the simple eigenvalue $0$ separated
from the rest of the spectrum by a gap $\ge c \min(\tau', g^2)$ for
some $c>0$. The operator $K^{\#(1)}$ has also the simple eigenvalue
$0$, which, by the Kato-Rellich theorem, \eqref{mm2} and the
assumption \eqref{eqn_12_11}, is separated from the rest of the
spectrum by a gap $\ge c' \min(\tau', g^2)$ for some $c'>0$.
%
Hence, we conclude that $K^{\#(1)} \in \dom(F_{Q_0^\#})$. Therefore,
by Theorem~\ref{thm_7_4},
\begin{equation}
\label{eqn_12_12}
P_0 \sOmega_{\stheta} = \left( \id - R_{\overline{Q_0^\#}}
(K^{\#(1)}) W^{\#} \right) Q_0^\#\sOmega_{\stheta}.
\end{equation}
Since $\|R_{\overline{Q_0^\#}}(K^{\#(1)})\| \le C [\min(\tau', g^2
)]^{-1}$ we have the absolutely convergent perturbation expansion
\begin{equation}
\label{eqn_12_13}
P_0 \sOmega_{\stheta} = C^{\#} \sum_{n=0}^{\infty} \left( -
R_{\overline{Q_0^\#}} (K_0^{\#(1)}) W^{\#} \right)^n \szeta \otimes
\Omega_r.
\end{equation}
Here,
$\szeta \in \rRan \chi_{L_p = 0}$ is the unique vector in the kernel of $\Lambda^{\#}$, normalized as
\begin{equation}
\label{eqn_12_14}
\Lambda^{\#} \szeta = 0, \ \|\zeta\|=1,\ \scalprod{\zeta^*}{\zeta}=1.
\end{equation}
(Recall that $\Lambda^\#$ is either $\Lambda_0$ or $\Lambda_0^*$, so
$\zeta^\#$ is either a null vector, $\zeta$, of $\Lambda_0$ or a
null vector, $\zeta^*$, of $\Lambda_0^*$.) Letting $\zeta^{\#\#}$
equal $\zeta$ if $\#=*$ and letting it equal $\zeta^*$ otherwise,
the constant in \fer{eqn_12_13} takes the form $C^{\#}
=\innerprod{\zeta^{\#\#}\otimes\Omega_r}{\sOmega_{\stheta}}$.

The overlap $\innerprod{\zeta^*}{\zeta}$ can be chosen strictly positive since the $\zeta^{\#}$ are the Perron-Frobenius eigenvectors of $\Lambda^{\#}$ (i.e., their components can be chosen non-negative), and every component of $\zeta$ is strictly positive (see below). The last relation in \fer{eqn_12_14} is then achieved by scaling $\zeta^*$ properly. The normalization
$\innerprod{\Omega^*_\theta}{\Omega_\theta} = 1$ together with \fer{eqn_11.11}, \fer{mm2} \fer{eqn_12_13} and  \fer{eqn_12_14} implies that
\begin{equation}\label{eqn_11_16}
\overline{C^*} C =1+o(g).
\end{equation}

If the condition \fer{eqn_12_11}, $\tau' \gg g^{2+\alpha}$, does not
hold then we have to apply the Feshbach map iteratively and use a
corresponding perturbation theory for eigenvectors. We omit here
this analysis and refer the reader to \cite{BFS:QED, BFS:RG,
BCFS:SmoothFeshbach} for general references on such a RG
perturbation theory and we will present elsewhere the RG
perturbation theory in our specific case.

In Section~\ref{Sect13} we have shown that ${\rm Null}\Lambda =
\BC\Omega_p$ and consequently $\zeta = \Omega_p$ and the vector
$\Omega_0 = \zeta \otimes \Omega_{r} = \Omega_p \otimes \Omega_r$
is our unperturbed state introduced in Section~\ref{Sect6}. Recall
that $\Omega_p = \Omega^{\beta_p}_p$ is the particle Gibbs state
at temperature $\beta_p$.

Expressions \fer{m3}--\fer{eq_8_3} for $\Lambda_0$ imply the
following relation among vectors $\zeta^*$ corresponding to
different particle temperatures
\begin{equation}
\label{11.19} \zeta^*=\sqrt{\frac{{\rm Tr\,}{\rm e}^{-\beta_p
H_p}}{N}}(e^{\beta_pH_p/2}\otimes\bbbone_p)\zeta^*|_{\beta_p=0}\ .
\end{equation}
In view of \fer{11.19}, it suffices to consider $\beta_p=0$.

In general there is no simple expression for the eigenvector $\zeta^*$.
However, there are three cases where such an expression can be
obtained. We expand $\zeta^*|_{\beta_p=0}$ in the basis
$\varphi_j\otimes\varphi_j$ as
\begin{equation}
\zeta^*|_{\beta_p=0}=\sum_j\gamma_j\varphi_j\otimes\varphi_j.
 \label{11.15a}
\end{equation} Here $\varphi_j$ and $E_j$ are defined after Eqn \eqref{eq_6_3}. Note that the
normalization condition and Eqn \eqref{eq_6_3} imply that $\sum_j
\gamma_j = \sqrt{N}$. We have the following results:
\begin{enumerate}
\item[(i)] If $\beta_1=\beta_2=\beta$ and $\beta_p=0$, then
\begin{equation}
\gamma_j=\frac{\sqrt{N}}{{\rm Tr\, }e^{-\beta H_p}} \ e^{-\beta E_j}.
 \label{11.15}
\end{equation}
\item[(ii)] If $\beta_1$ is fixed and $\beta_2 \to 0$, and $\beta_p=0$,  then
\begin{equation}\label{eqn_11_11}
\gamma_j =
\frac{1}{\sqrt{N}} + O(\beta_2).
\end{equation}
\item[(iii)] If $N = {\rm dim} \cH_p = 2$ and $\beta_p=0$, then
\begin{equation}\label{eqn_11.26}
\gamma_1 =\frac{\sqrt 2 \alpha(E)}{\alpha(E)+1}\ \mbox{\ \ and\ \ }
\gamma_2 =\frac{\sqrt 2}{\alpha(E)+1},
\end{equation}
where
\begin{equation}\label{eqn_11.26a}
\alpha (E)= 1 + \frac{\sum_{j=1,2} g_j(E)} {\sum_{j=1,2}
g_j(E) \rho_j (E)}.
\end{equation}
Here, we use the notation $E :=  E_2 - E_1$, $\rho_j(E) :=
\frac{1}{e^{\beta_j E} -1 }$ and
\begin{equation*}
g_j(E) :=  \int\limits_{|k| = |E|} \left[ |G_j(k)_{12}|^2 +
|G_j(k)_{21}|^2 \right] \, dk \ .
\end{equation*}
\end{enumerate}
Equation \fer{11.15} follows from \fer{m3}, \fer{eqn_9.4} and
\fer{8.4a}. The expressions \fer{eqn_11.26} -- \fer{eqn_11.26a}
come simply from solving a two-dimensional eigenvalue problem.
Equation~(\ref{eqn_11_11}) follows from a straightforward
perturbation theory in $\beta_2$. See Appendix \ref{appE} for an outline of the proofs.

\section{Entropy production rate for $\eta$}
\label{Sect10}

In this section we prove Theorem~\ref{theorem3.2}.
Recall that the stationary state $\eta$ mentioned in this theorem
is, in fact, given in (\ref{eq8.7}). To analyze the entropy
production, $EP(\eta)$, in this state $\eta$ we use expression
(\ref{eq3.9}),
\begin{equation}\label{eqn_12_1}
  EP(\eta) = (\beta_1 - \beta_2) \eta(\phi_1),
\end{equation}
which relates it to the heat flow, $\eta(\phi_1)$, in the state
$\eta$. Recall that
\begin{equation}\label{eqn_12_2}
  \phi_1 = i g [v_1, H_{r1}]=ig \eta\big( a_1(\omega G_1)
   - a^*_1(\omega G_1)\big).
\end{equation}
If $\beta_1 = \beta_2$, then $\eta(\phi_1) = 0$, c.f.
\cite{JP:EP}. We want to show here that
$ \eta(\phi_1) > 0 \quad \text{if} \quad \beta_1 > \beta_2$.
Our proof is based on

\begin{theorem}
\label{thm_12_1}
Set $\betamax=\max(\beta_1,\beta_2)$ and let $\Omega^*_0 = \zeta^* \otimes
\Omega_r$ with $\Lambda^* \zeta^* = 0$ (see (\ref{eqn_12_14})) and with
the vector $\Omega_0$ defined in \fer{eq_6_3}.
Under the conditions of Theorem~\ref{thm2.1}, we have
\begin{equation}\label{eqn_12_4}
  \eta(\phi_1) = g^2\eta' + o(g^2)
  O\left(\delta\beta\right),
\end{equation}
where, recall, $\delta\beta:=\beta_1-\beta_2$ and
\begin{equation}\label{eqn_12_5}
\eta' = 2 {\rm Re} \innerprod{\Omega^*_0}{\pi(v_1) L_{r1}
i(L_0 +i0)^{-1} e^{-\betamax L_0/2}\pi(v_1)\Omega_0}.
\end{equation}
Moreover, we have the  explicit expression
\begin{equation}\label{eqn_12_6}
  \eta' = \frac{2\pi}{\sqrt N} \sum_{j>i} (\gamma_j e^{\beta_1 E_{ji}} - \gamma_i)
  \frac{E_{ji}\ g_{ji}(E_{ji})^2}{e^{\beta_1E_{ji}}-1},
\end{equation}
where $E_{ji} = E_j - E_i$, $g_{ji}(E)^2 = \int_{{\mathbb R}^3}d^3 k
|\innerprod{\varphi_j}{G_1(k)\varphi_i}|^2 \delta(E_{ji}-\omega)$.
The numbers $\gamma_j \ge 0$ are the components of the vector
$\zeta^*$, see \eqref{11.15a}, normalized as in \fer{eqn_12_14}, at
$\beta_p = 0$.
Observe that by \fer{11.15}, $\eta'=0$ for $\beta_1=\beta_2$.
\end{theorem}

The following result shows that $\eta'$ is strictly positive for small nonzero temperature differences.

{\theorem \label{thm_12_2} If $\delta\beta=\beta_1-\beta_2>0$ is
small and either $\dim {\cal H}_p=2$ or the coupling functions
\fer{intop} satisfy $G=G_1=G_2$, then the linear in $\delta\beta$
part of $\eta'$ is $>0$. Moreover, in the latter case, this part is
\begin{equation}
\frac{\delta\beta}{2} \frac{Z_p(\beta_p)}{Z_p(\beta_1+\beta_p/2)} \sum_{j>k} \frac{ E_{jk}^2 \ g_{jk}(E_{jk})^2}{e^{\beta_1 E_j}-e^{\beta_1 E_k}} >0,
\label{_+_+}
\end{equation}
where $Z_p(\beta)=\mbox{\rm tr\,} e^{-\beta H_p}$ is the particle partition function.
}

{\it Remarks.\ }  1. In the general case, if $G_1$ is close to $G_2$
one deduces strict positivity of $\eta'$ in the linear term in
$\delta\beta$ by a perturbation argument.

2. For an expression of $\eta'$ in the case $\dim {\cal H}_p=2$
similar to \eqref{_+_+}, see the
proof of Theorem \ref{thm_12_2}, given below.\\

{\bf Proof of Theorem \ref{thm_12_1}.\ }
To simplify the exposition we restrict ourselves to the case $\tau'
\gg g^{2+\alpha}$, for $\alpha =\frac{\mu-1/2}{\mu+1/2}$, and
$\mu>3/2$. Pick $\rho_0=g^{2-2\alpha}$.  First we prove an estimate
on $\eta(\phi_1)$ which is rougher than (\ref{eqn_12_4}) and then we
explain how to obtain (\ref{eqn_12_4}). Recall that
\begin{equation}\label{eqn_13_6}
  \eta(\phi_1) =  \innerprod{\Omega^*_\btheta}{\pi(\phi_1)_\theta
  \Omega_\theta}.
\end{equation}
Let $\sOmega_0 = \szeta \otimes \Omega_r$ and take a number $M$ so large that
$(g^{2+\alpha}[\min(\tau', g^2)]^{-1} )^M = o(g)$. Substituting
expansions (\ref{eqn_11.11}) and (\ref{eqn_12_13}) into the r.h.s.
of this expression, using that $(\g \rho_0^{-1/2})^2
= o(g)$, for $\rho_0 = g^{2-2\alpha}$, $\alpha > \frac{1}{2}$, and
using that $\phi_1$ is proportional to $g$, we find
\begin{equation}\label{eqn_13_7}
  \eta(\phi_1) = \eta_0 + \eta_1 + {\rm Rem} + o(g^2),
\end{equation}
where
\begin{eqnarray}
  \eta_0 & := & \innerprod{\Omega^*_0}{\pi(\phi_1)\Omega_0}, \\
  \eta_1 & := & -g \innerprod{\Omega^*_0}{\left[I_\theta \oR_0(L_{0\theta})
    \pi(\phi_1)_\theta + \pi(\phi_1)_\theta \oR_0(L_{0\theta}) I_\theta)\right]
    \Omega_0},
    \label{eqn_13_8}
\end{eqnarray}
\begin{eqnarray}
  {\rm Rem} & = & \sum_{m+n \ge 1}^M \sum_{k,l = 0}^1 g^{k+l}
  \Big<(-\oR_0(L_{0\btheta}) I^*_\btheta)^k (-R_{\overline{Q_0^*}}(K_0^{\#(1)})W^*)^m
  \Omega^*_0, \nonumber \\
  && \pi(\phi_1)_\theta (-\oR_0(L_{0\theta}) I_\theta)^l
(-R_{\overline{Q}_0}(K_0^{(1)})W)^n
  \Omega_0 \Big>. \label{eq_12_11}
\end{eqnarray}
Here we replaced on the r.h.s the factor $\overline{C^*}C=1+o(g)$ by
$1$ (see (\ref{eqn_12_13}) and (\ref{eqn_11_16})). Using the
pull-through procedure and elementary estimates of the resulting
integrals we obtain that
\begin{equation}\label{eqn_13_9}
  {\rm Rem} = o(g^2).
\end{equation}
Since $\phi_1$ is linear in creation and annihilation operators, see \fer{eqn_12_1}, we have
\begin{equation}\label{eqn_13_10}
  \eta_0 = 0.
\end{equation}

It remains to compute $\eta_1$. Using
\begin{equation}\label{eq12.15}
(\oP_{\rho_0} - \id) \sI_{\stheta} \sOmega_0 = O(\rho_0)
\end{equation}
(\cite{MMS1} Lemma 5.3), removing the spectral deformation and using
that $\pi(\phi_1) =g [\pi(v_1), iL_{r1}]$, we obtain $\eta_1 = g^2
\eta' + o(g^2 )$, where
\begin{equation}\label{eq12.16}
\eta' = -\innerprod{\Omega_0^*}{\left[\pi(v_1) iL_{r1} (L_0 +
i0)^{-1} I - I (L_0 +i0)^{-1} iL_{r1} \pi(v_1) \right]\Omega_0}.
\end{equation}
Next, note that the contribution of the $v_2$-part of $I$ to
$\eta'$ is zero since the resulting expression is linear in
creation and annihilation operators for the first and second
reservoirs separately. The contribution of the $\pi (v_1)$-part of
$I$ is also zero by the symmetry of (\ref{eq12.16}). Hence we have
$\eta' = A-B$, where
\begin{eqnarray*}
A &=& \langle \Omega^*_0 , \pi(v_1) iL_{r1} (L_0 + i0)^{-1} \pi'
\big(\gamma^{i/2} (v_1)\big) \Omega_0 \rangle,\\
B &=& \langle \Omega^*_0 , \pi' \big(\gamma^{i/2} (v_1)\big) iL_{r1}
(L_0 + i0)^{-1} \pi (v_1 ) \Omega_0 \rangle .
\end{eqnarray*}
Using that $\pi' \big(\gamma^{i/2} (v_1) \big) \Omega_0 = J e^{(\beta L_0 -\tilde L)/2} \pi
(v_1) \Omega_0$, where $\tilde L$ is given after \fer{afterglow}, and the fact that $J e^{-\tilde L/2}\pi(v_1)\Omega_0=\pi(v_1^*)\Omega_0=\pi(v_1)\Omega_0$,
we transform
\begin{equation*}
A = \langle \Omega^*_0 , \pi (v_1 ) iL_{r1} (L_0 + i0)^{-1}
e^{-\beta L_0/2 } \pi (v_1 ) \Omega_0 \rangle \ .
\end{equation*}
We use the relations $\pi' \big(\gamma^{i/2} (v_1)\big) = J e^{(\beta L_0 - \tilde L )/2} \pi
(v_1) e^{-(\beta L_0 - \tilde L)/2}  J$
and
\begin{equation*}
J iL_{r1} (L_0 + i0)^{-1} \pi (v_1) \Omega_0 = - iL_{r1} (L_0 +
i0)^{-1} e^{-\tilde L/2} \pi (v_1 ) \Omega_0 \ ,
\end{equation*}
see also (\ref{eqn_5.2}), to find that
\begin{equation*}
B = - \langle \Omega^*_0 , Je^{(\beta L_0 + \tilde L)/2} \pi (v_1)
iL_{r1} (L_0 + i0)^{-1} e^{-\beta L_0 / 2} \pi (v_1) \Omega_0
\rangle \ .
\end{equation*}
Finally, since  $\langle Ju , Jv \rangle = \overline{\langle u,v
\rangle} $, $J \Omega^*_0 = \Omega^*_0 $, $L_p\Omega_0^*=0$ and
$(\beta L_0-\overline L)\Omega^*_0=0$  we obtain
\begin{equation*}
\overline B = - \langle \Omega^*_0 ,
\pi (v_1 ) iL_{r1} (L_0 + i0)^{-1} e^{-\beta L_0 / 2} \pi (v_1)
\Omega_0 \rangle =-A.
\end{equation*}
Since $\eta'=A-B$ this gives \eqref{eqn_12_5}.

Collecting estimates (\ref{eqn_13_7}), (\ref{eqn_13_9}),
(\ref{eqn_13_10}) and $\eta_1=g^2\eta'+o(g^2)$ we find that
\begin{equation}
\label{eqn_13_20}
\eta(\phi_1) = g^2 \eta' + o(g^2)
\end{equation}
where $\eta'$ is given by (\ref{eqn_12_5}).  This proves the rough
version of \fer{eqn_12_4}--\fer{eqn_12_5}.

Before we refine estimate (\ref{eqn_13_20}) let us show (\ref{eqn_12_6}).
We expand the vectors $\Omega_0^*$
and $\Omega_0$ in the basis $\varphi_{jj}\otimes\Omega_{r}$,
$\varphi_{jj}=\varphi_j\otimes\varphi_j$,
\begin{equation}
\Omega_0 = \sum_{j=1}^n
\alpha_j \varphi_{jj}\otimes\Omega_{r}, \ \ \ \Omega_0^* =  \sum_{j=1}^n \tilde{\gamma}_j
\varphi_{jj}\otimes\Omega_{r}
\label{eqn_12.21}
\end{equation}
with $\alpha_j \ge 0$, $\sum_j \alpha_j^2 =1$ and $\tilde{\gamma}_j
\ge 0$, $\sum_j \alpha_j\tilde{\gamma}_j = 1$. Plugging the expressions in \fer{eqn_12.21} into the r.h.s. of (\ref{eqn_12_5}), using
(\ref{eq_AWrepr}) in order to express $\pi(v_1)$ in terms of
creation and annihilation operators, $a^\#_{\ell 1}$ and
$a^\#_{r1}$,
\begin{equation*}
\pi(v_1) = a_{\ell 1} (\sqrt{1+\rho_1}\,G_{1\ell}) +
a_{r1}(\sqrt{\rho_1}\,G_{1\ell}) + h.c.,
\end{equation*}
where $\rho_1 = (e^{\beta_1\omega} - 1)^{-1}$, pulling through the
annihilation operators to the right and using that $\pi(v_1)$ (or
$G_{1\ell}$) acts only on the first (left) factor in $\varphi_{jj}
= \varphi_j \otimes \varphi_j$, we obtain
\begin{eqnarray*}
\eta' & = & -2 \sum_j \alpha_j \tilde{\gamma}_j {\rm Im} \int \Big
\{ (1+\rho_1) \omega \innerprod{\varphi_j}{G^*_1 (H_p - E_j +
\omega+i0 )^{-1}
e^{-\beta(H_p - E_j + \omega)/2}G_1 \varphi_j} \\
& & -\rho_1  \omega \innerprod{\varphi_j}{G_1^* (H_p - E_j -
\omega+i0)^{-1} e^{-\beta(H_p - E_j - \omega)/2}G_1^* \varphi_j} \Big
\}d^3k.
\end{eqnarray*}
Inserting the partition of unity $\id = \sum_j \ket{\varphi_j}
\bra{\varphi_j}$ into the inner products on the r.h.s. we obtain
furthermore
\begin{equation*}
\eta'=2\pi\sum_{i,j} \alpha_i \tilde{\gamma}_i \int \omega
|\scalprod{\varphi_i}{G_1\varphi_j}|^2\left\{(1+\rho_1)\delta(E_{ji}+\omega) -\rho_1\delta(E_{ji}-\omega)\right\}d^3k.
\end{equation*}
Interchanging the labels in the sum of the first term and noticing that in the resulting expression the integrals vanish unless $E_{ji}>0$, i.e. $E_j>E_i$, or $j>i$, we arrive at
\begin{equation*}
\eta' = 2\pi\sum_{j>i} (\alpha_j \tilde{\gamma}_j e^{\beta_1E_{ji}} -
\alpha_i \tilde{\gamma}_i) E_{ji}\  g_{ji}(E_{ji})^2 \ (e^{\beta_1
E_{ji}}-1)^{-1} ,
\end{equation*}
where $g_{ji}(E)^2 := \int_{{\mathbb R}^3} d^3k\, |\innerprod{\varphi_j}{G_1(k)
\varphi_i }|^2\delta(E_{ji}-\omega)$. Observing that, due to \fer{11.19}, $\alpha_i
\tilde{\gamma}_i = N^{-1/2}\tilde{\gamma}_i |_{\beta_p = 0} \equiv
N^{-1/2} \gamma_i$, we arrive at (\ref{eqn_12_6}).

Since $\eta'=O(\delta\beta)$ (the energy flow vanishes if $\beta_1=\beta_2$) estimate (\ref{eqn_13_20}) is ineffective if $\delta\beta$ is
so small that $\delta\beta\, g^2 = o(g^2)$. However, with a
little bit more work \fer{eqn_13_20} can be upgraded to estimate
(\ref{eqn_12_4}). We sketch a proof of this estimate without going
into much detail.  We begin with some notation.

Consider the
{\it self-adjoint} Liouville operator for equal reservoir temperatures
$\beta_2 = \beta_1 = \beta$,
\begin{equation}
\tilde{L} = L_0 + g \tilde{I} ,
\end{equation}
where
\begin{equation}
\tilde{I} = \tilde{\pi} (v) - \tilde{\pi'} (v)
  \end{equation}
with $\widetilde\pi=\pi|_{\beta_1=\beta_2=\beta}$ and similarly for
$\widetilde\pi'$. Define $ \widetilde\Omega
=\frac{e^{-\beta\widetilde L^{(l)}/2}\Omega_0}{\|e^{-\beta\widetilde
L^{(l)}/2}\Omega_0\|}, $ where $\widetilde
L^{(l)}=L_0+\widetilde\pi(v)$. Since the entropy production does not
depend on $\beta_p$ we set from now on $\beta_p=\beta_1$. The
operator $K|_{\beta_p=\beta_1=\beta_2=\beta}=\tilde{L}$ is
selfadjoint, \fer{el}, and hence
$\widetilde\Omega_{\overline{\theta}}^*=\widetilde\Omega_{\overline{\theta}}$.
{}From (\ref{eqn_13_6}) we obtain
\begin{eqnarray}
\eta(\phi_1) & = &
\innerprod{(\Omega^*_{\btheta} - \widetilde{\Omega}_\btheta)}{\pi(\phi_1)_\theta \Omega_\theta }
+ \innerprod{\widetilde{\Omega}_\btheta}{\pi(\phi_1)_\theta (\Omega_\theta -
\widetilde{\Omega}_\theta)} \nonumber \\
&& + \innerprod{\widetilde{\Omega}_\btheta}{\pi(\phi_1)_\theta \widetilde{\Omega}_\theta}. \label{eqn_13.15}
\end{eqnarray}
We consider the last term first.
Recall that $\phi_1 = g[v, iH_{r1}] = g[v_1, iH_{r1}]$ and
therefore $\pi(\phi_1) = \widetilde{\pi} (\phi_1)$. It follows that
\begin{equation} \label{eqn_13.16}
\langle \widetilde{\Omega}_\btheta , \pi(\phi_1)_\theta
\widetilde{\Omega}_\theta \rangle =
\langle \widetilde{\Omega}_\btheta , \widetilde{\pi}(\phi_1)_\theta
\widetilde{\Omega}_\theta \rangle =
\langle \widetilde{\Omega}, \widetilde{\pi} (\phi_1) \widetilde{\Omega}
\rangle.
\end{equation}
Note that the r.h.s. of (\ref{eqn_13.16}) describes the heat flow into
reservoir $r1$ for the equal temperature system. Since the heat
flows vanish individually in the equal temperature case we have shown
that the last term in (\ref{eqn_13.15}) vanishes.

To estimate the first two terms on the r.h.s. we use as before
expansions (\ref{eqn_11.11}) and (\ref{eqn_12_13}) for
$\sOmega_{\stheta} = \Omega_\theta, \Omega^*_\btheta$ and similar
expansions (obtained by setting $\beta_2 = \beta_1 = \beta$ in
(\ref{eqn_11.11}) and (\ref{eqn_12_13})) for
$\widetilde{\Omega}_\theta$. As a result we obtain an expression for \fer{eqn_13.15} of
the type (\ref{eqn_13_7}) -- (\ref{eq_12_11}) but with some
of the powers in $\rm Rem$, \fer{eq_12_11}, replaced by the differences, e.g.
$(-\oR_0(L_{0\theta}) I_\theta)^l -
(-\oR_0(L_{0\theta}) \widetilde{I}_\theta )^l$
or $
(R_{\overline{Q}_0}(K^{(1)}_0)W)^n -
(R_{\overline{Q}_0}(K^{(1)}_0)W)^n |_{\beta_2 = \beta_1 = \beta}
$.
These differences are estimated by using a telescopic expansion,
e.g.,
\begin{eqnarray}
\lefteqn{
(-\oR_0(L_{0\theta})I_\theta)^l -
(-\oR_0(L_{0\theta})\widetilde{I}_\theta)^l } \nonumber \\
& = & \sum_{j=1}^l (-\oR_0(L_{0\theta})I_\theta)^{j-1}
(-\oR_0(L_{0\theta})) (I_\theta - \widetilde{I}_\theta)
(-\oR_0(L_{0\theta})\widetilde{I}_\theta)^{l-j}, \label{eqn_13_30}
\end{eqnarray}
and then estimating the first type of the differences in norm
while for the second type we do first the pull-through and
contraction procedure and then estimate the resulting integrals.
As a result we have
\begin{eqnarray}
\eta(\phi_1)
& = & -g \innerprod{\Omega_0^* - \Omega_0}{\pi(\phi_1)_\theta
\oR_0(L_{0,\theta}) I_\theta \Omega_0} \nonumber \\
&   & -g \innerprod{\oR_0(L_{0,\btheta})
\left[ I^*_\btheta \Omega_0^* - \widetilde{I}_\btheta \Omega_0 \right] }
{\pi(\phi_1)_\theta  \Omega_0} \nonumber \\
&   & -g \innerprod{\Omega_0}{\pi(\phi_1)_\theta \oR_0(L_{0,\theta})
\left[I_\theta  - \widetilde{I}_\theta \right] \Omega_0} \nonumber \\
&   & + o(g^2) O(\delta\beta) \nonumber \\
& = & \eta_1 + \eta_2 + o(g^2) O(\delta\beta) \label{eqn_12_35}
\end{eqnarray}
where $\eta_1$ is given in (\ref{eqn_13_8}) and
\begin{equation} \label{eqn_12.32}
\eta_2 = g \innerprod{\Omega_0}
{\left[\widetilde{I}_\theta
\oR_0(L_{0,\theta}) \pi(\phi_1)_\theta +\pi(\phi_1)_\theta \oR_0(L_{0,\theta})
\widetilde{I}_\theta \right]\Omega_0}.
\end{equation}
Since the contribution of the $v_2$-component of $\widetilde
I_{\theta}$ is zero we can omit the tilde $(\sim )$ in
(\ref{eqn_12.32}). Thus the expression for $\eta_2$ coincides up to
the sign and the substitution $\Omega^*_0 \to  \Omega_0$ with the
expression (\ref{eqn_13_8}) for $\eta_1$, i.e.
\begin{equation}
\eta_1+\eta_2=-g
\innerprod{\Omega_0^*-\Omega_0}
{\left[I_\theta
\oR_0(L_{0,\theta}) \pi(\phi_1)_\theta +\pi(\phi_1)_\theta \oR_0(L_{0,\theta})
I_\theta \right]\Omega_0}.
\label{mm4}
\end{equation}
We proceed with the r.h.s. of \fer{mm4} exactly as we did above with $\eta_1$ alone in equation \fer{eq12.15}, and we arrive at
\begin{equation*}
\eta(\phi_1)=2g^2{\rm Re} \innerprod{\Omega_0^*-\Omega_0}{\pi(v_1)iL_{r1}(L_0+i0)^{-1}e^{-\betamax L_0/2}\pi(v_1)\Omega_0}
+o(g^2)O(\delta\beta).
\end{equation*}
\fer{eqn_12_4} follows by noticing that
\begin{equation*}
{\rm Re} \innerprod{\Omega_0}{\pi(v_1)iL_{r1}(L_0+i0)^{-1}e^{-\betamax L_0/2}\pi(v_1)\Omega_0}=0.\mbox{\ \ \ \ \qed}
\end{equation*}

{\bf Proof of Theorem \ref{thm_12_2}.\ } Consider first the case
$\dim {\cal H}_p=2$. Using \fer{eqn_12_6}, \fer{eqn_11.26},
\fer{eqn_11.26a} we obtain in this case
$$
\eta' = \frac{2\pi}{\alpha(E_{21})+1} ( e^{\beta_1 E_{21}} - \alpha
(E_{21}))
  \frac{E_{21}\ g_{21}(E_{21})^2}{e^{\beta_1E_{21}}-1}.
$$
Next we expand $e^{\beta_1 E_{21}} - \alpha (E_{21})$ around $\delta
\beta = \beta_1-\beta_2=0$ to verify that in the two-dimensional
case the linear in $\delta\beta$ term of $\eta'$ is strictly
positive.

Now we consider the case $G_1=G_2$. We want to control the
components $\gamma_j$, appearing in the expression \fer{eqn_12_6}
for $\eta'$. To this end we employ basic analytic perturbation
theory (in $\delta\beta$) for the matrix family $M(\delta\beta):=
\Lambda_0^*(\delta\beta)|_{\beta_p=0}$, where we consider $\beta_1$
to be fixed. Write $\Gamma_j$ instead of $\Gamma_{j0}$, see
\fer{eqn_9.4}. According to Proposition \ref{aepsilon} we have
$M(\delta\beta)=M_0+\delta\beta M_1+O(\delta\beta^2)$, where
\begin{eqnarray}
M_0 &=& -i(e^{-\beta_1H_p/2}\otimes\bbbone)\left\{\Gamma_1(\beta_1)+\Gamma_2(\beta_1)\right\} (e^{\beta_1H_p/2}\otimes\bbbone),\label{M_0}\\
M_1 &=& -i(e^{-\beta_1H_p/2}\otimes\bbbone)\left\{{\textstyle\frac12}[H_p\otimes\bbbone, \Gamma_2(\beta_1)] -(\partial_\beta \Gamma_2)(\beta_1)\right\} (e^{\beta_1H_p/2}\otimes\bbbone).\ \ \ \ \ \ \ \ \label{M_1}
\end{eqnarray}
Let $\zeta^*=\sum_{j \geq 0} (\delta\beta)^j \zeta^*_j$. The normalization $\scalprod{\zeta^*}{\zeta}=1$ (where $\zeta=\Omega_p(\beta_p)$ is the particle Gibbs state) implies that $\scalprod{\zeta_0^*}{\zeta}=1$ and $\scalprod{\zeta_j^*}{\zeta}=0$, for $j\geq 1$. Solving the zero-oder eigenvalue equation $M_0\zeta^*_0=0$ gives
\begin{equation}
\zeta_0^*=\sum_j\gamma_j^{(0)}\varphi_j\otimes\varphi_j,\ \ \
\gamma_j^{(0)}=\frac{Z_p(\beta_p)}{Z_p(\beta_1+\beta_p/2)} e^{-\beta_1 E_j}.
\end{equation}
The first-order eigenvalue equation reads $M_1\zeta^*_0+M_0\zeta^*_1=0$, which implies
\begin{equation}
[\Gamma_1(\beta_1)+\Gamma_2(\beta_1)] (e^{\beta_1H_p/2}\otimes\bbbone)\zeta^*_1 = \partial_\beta|_{\beta_1} \Gamma_2(\beta) (e^{\beta H_p/2}\otimes\bbbone)\zeta_0^*.
\label{---}
\end{equation}
We use here that $(e^{\beta_1 H_p/2}\otimes\bbbone)\zeta_0^*$ is in the kernel of $H_p$. Let $\Psi(\beta):=\sum_j e^{-\beta E_j/2}\varphi_j\otimes\varphi_j$. Since $\Gamma_2(\beta)\Psi(\beta)=0$ we have $(\partial_\beta\Gamma_2)(\beta)\Psi(\beta)=-\Gamma_2(\beta)(\partial_\beta\Psi)(\beta)$, so
\begin{eqnarray}
\partial_\beta|_{\beta_1} \Gamma_2(\beta)(e^{\beta H_p/2}\otimes\bbbone)\zeta_0^* &=& C(\beta_1) \partial_\beta|_{\beta_1} \Gamma_2(\beta)\Psi(2\beta_1-\beta) \nonumber\\
&=&C(\beta_1)\Gamma_2(\beta_1)(H_p\otimes\bbbone)\Psi(\beta_1),
\label{--}
\end{eqnarray}
where $C(\beta_1)=Z_p(\beta_p)/Z_p(\beta_1+\beta_p/2)$. The r.h.s. of \fer{--} is a vector in the orthogonal complement of ker$\Gamma_2(\beta_1)={\mathbb C}\Psi(\beta_1)$. Using this fact and \fer{--} we solve \fer{---} for $\zeta_1^*$:
\begin{eqnarray}
\zeta_1^*&=& C(\beta_1) (e^{-\beta_1H_p/2}\otimes\bbbone) [\Gamma_1(\beta_1)+\Gamma_2(\beta_1)]^{-1} \Gamma_2(\beta_1)(H_p\otimes\bbbone)\Psi(\beta_1)\label{++}\\
&&+C' (e^{-\beta_1H_p/2}\otimes\bbbone)\Psi(\beta_1),
\label{+++}
\end{eqnarray}
where the constant $C'$ is determined by the normalization condition $\scalprod{\zeta_1^*}{\zeta}=0$. From expression \fer{eqn_12_6} it is clear that the term \fer{+++} does not contribute to the value of $\eta'$ (this is the same as saying that $\eta'=0$ for $\delta\beta=0$).

Under the assumption $G_1=G_2=G$ we have $\Gamma_1(\beta_1)=\Gamma_2(\beta_1)$ and the r.h.s. of \fer{++} simplifies to an easy expression, which, when used in \fer{eqn_12_6}, yields \fer{_+_+}.
\hfill \qed
\ \\

{\bf Proof of Theorem \ref{theorem3.2}.\ }
If $G_1=G_2$ or if the dimension of the particle system is $2$ then
we have $\eta'=\overline{\eta}\delta\beta +O((\delta\beta)^2)$ with
$\overline\eta>0$ independent of $\delta\beta$. This follows from
Theorem \ref{thm_12_2}. Hence for $g$ and $\delta\beta$ both small,
but independently of each other, we have, by \fer{eqn_12_4},
$EP(\eta_{\beta_1\beta_2})>0$, which is the statement of Theorem
\ref{theorem3.2}.
\hfill $\blacksquare$

\medskip

{\it Remark.\ } A stronger statement, mentioned in the second
paragraph after Theorem \ref{theorem3.2}, can be proved as follows.
By an abstract result of \cite{JP:EP}, $EP(\eta) \ge 0$. Therefore,
due to (\ref{eqn_12_1}), $\eta(\phi_1) \ge 0$ for $\beta_1 \ge
\beta_2$. Hence, due to (\ref{eqn_12_4}), for $g$ sufficiently small
(depending on $\delta\beta$ in general),
\begin{equation}\label{eqn_12_42}
  \eta' \ge 0 \quad \text{for} \quad \beta_1 \ge \beta_2.
\end{equation}

Next, the $\gamma_j$ are analytic in $\beta_1$ and $\beta_2$
separately, away from $\beta_1=0$, $\beta_2=0$. To show this, we
proceed as follows. {}From the explicit form of the level shift
operator $\Lambda_0^*$, given in Proposition \ref{aepsilon} and
equation \fer{fgrc} and similar expressions for diagonal elements,
we know that $\Lambda_0^*$ is analytic separately in $\beta_1$ and
$\beta_2$, everywhere except for $\beta_1=0, \beta_2=0$. We also
know that for each $\beta_1, \beta_2$ nonzero, $\Lambda_0^*$ has a
simple eigenvalue at zero (since $\Lambda_0$ does). It follows from
the Kato-Rellich theorem that we can find a zero-eigenvalue
eigenvector $\zeta_1^*$, which is analytic in $\beta_1, \beta_2$.
  Next, we have to normalize that vector s.t. its overlap with
$\Omega_p$ is unity. This yields $\zeta^* = \langle \zeta_1^*, \Omega_p\rangle^{-1} \zeta_1^*$.
Since $\langle \zeta_1^*, \Omega_p\rangle$ cannot vanish, we have that
 $\zeta^*$ is analytic.
  Finally, the $\gamma_j$ of Theorem \ref{thm_12_1} are the components of $\zeta^*$,
with $\beta_p=0$. From \fer{11.19} we see that analyticity of the
components of $\zeta^*$ in $\beta_1, \beta_2$ is true regardless of
what $\beta_p$ is. Hence, the $\gamma_j$ are analytic separately in $\beta_1,
\beta_2$, for $\beta_1\neq 0$, $\beta_2\neq 0$.

Analyticity of the $\gamma_j$ and expression (\ref{eqn_12_6}) show that
\begin{equation} \label{eqn_12_43}
\text{$\eta'$ is analytic separately in $\beta_1$ and $\beta_2$, for $\beta_1\neq 0$, $\beta_2\neq 0$.}
\end{equation}
Equations (\ref{eqn_11_11}) and (\ref{eqn_12_6}) imply that
\begin{equation}
\label{eqn_12_44} \eta' > 0 \quad \text{if $\beta_1$ is fixed and
$\beta_2$ is sufficiently small}.
\end{equation}
Relations (\ref{eqn_12_42}) -- (\ref{eqn_12_44}) imply $
\eta(\phi_1) > 0 \quad \text{if} \quad \beta_1 > \beta_2$ for almost
all values of $(\beta_1,\beta_2)\in(0,\infty)\times(0,\infty)$, in
the sense that for fixed $\beta_1\in(0,\infty)$, $\eta' $ can vanish
only for finitely many values of $\beta_2$ in any bounded subset of
$(0,\infty)$. The same holds for $\beta_1$ and $\beta_2$
interchanged.

\bigskip

{\bf Acknowledgements.} The authors are grateful to V. Bach, G.
Elliott, J. Fr\"ohlich, and especially C.-A. Pillet for useful
discussions, and to V. Jak\u si\'c and C.-A. Pillet, for pointing
out numerous inaccuracies and typos and the desirability of more
details in the original manuscript. Part of this work was done while
the first author was visiting the University of Toronto, the third
author, ETH, Z\"urich and the second and third authors visited ESI
Vienna. During work on this paper the second author was at the
University of Toronto on a DAAD fellowship. The authors are grateful
to these places for hospitality.

\appendix
\section{Proof of existence of dynamics}
\label{appA} In this appendix we prove the existence of dynamics
\fer{dyn1}. Recall the definition of  the operator
$L^{(\ell)}:=L_0+g\pi(v)$ and the one-parameter group
$\sigma^{t}(B):=e^{itL^{(\ell)}} B e^{-itL^{(\ell)}}$,  $B \in
\pi(\cA)''$.
\begin{proposition}
\label{propa1} Assume the operators $v_n \in \cA$ satisfy
\fer{eq2.13}. Then for any state $\psi$ normal w.r.to $\omega_0$ the
integrands on the r.h.s. of \fer{dyn1} are continuous functions, the
series is absolutely convergent, the limit exists and equals
\begin{equation}
\psi^t(A)= {\rm Tr} (\rho \sigma^{t}(\pi(A))),
\label{michael1}
\end{equation}
where $\rho$ is the positive trace-class operator defined by $\psi(A)={\rm Tr}(\rho\pi(A))$. In particular, $\psi^t(A))$ is independent of the approximating operators.
\end{proposition}

{\bf Proof.\ } Let $v_n \in \cA$ be an approximating
sequence for the operator $v$ satisfying \fer{eq2.13}. We define the selfadjoint operators $L^{(\ell)}_n:=L_0+g\pi(v_n)$ on the dense domain ${\cal
D}(L_0)$.  Let the one parameter group $\sigma^{t}_{(n)}$ on $\pi({\cal A})$ be given by
\begin{equation*}
\sigma^{t}_{(n)}(B):=e^{itL^{(\ell)}_n} B e^{-itL^{(\ell)}_n}.
\end{equation*}
Set $ \sigma_0^{t}(\pi(A)):=\pi(\alpha_0^t(A))$ and let
$\psi$ be an $\omega_0$-normal state on
$\cA$, i.e.
\begin{equation}
\psi(A)= {\rm Tr} (\rho  \pi(A))
\end{equation}
for some positive, trace-class operator $\rho$ on $\cH$ of trace 1.
Then using the definition $V_n = \pi(v_n)$ we find
\begin{equation}
\label{dyn}
\psi([\alpha_0^{t_m}(v_{n}),\cdots
[\alpha_0^{t_1}(v_{n}),\alpha_0^{t}(A)]\cdots]) = {\rm Tr} (\rho
[\sigma_0^{t_m}( V_n),\cdots [\sigma_0^{t_1}( V_n), \sigma_0^{t}(A)]\cdots]).
\end{equation}
Clearly the r.h.s. is continuous in $t_{1},\ldots, t_{m}$ and
therefore the integrals in \fer{dyn1} are well defined. Since the
r.h.s. of \eqref{dyn} is bounded by $(c\|V_n\|)^m
\|\sigma_0^{t}(A)\|$, the series on the r.h.s. of \fer{dyn1}
converges absolutely. In fact, using the Araki-Dyson series
\begin{eqnarray}
\label{mm32}
\sigma^{t}_{(n)}(\pi(A))&=&\sum_{m=0}^\infty(ig)^m\int_0^tdt_1\cdots
\int_0^{t_{m-1}}
dt_m  \ [\sigma_0^{t_m}(\pi(v_n)),
\cdots\nonumber\\
&& \cdots [\sigma_0^{t_1}(\pi(v_n)),\sigma_0^{t}(\pi(A))]\cdots],
\end{eqnarray}
one can easily see that the series in \fer{dyn1} is nothing but the Araki-Dyson
expansion of the function ${\rm Tr} (\rho \sigma^{t}_{(n)}(\pi(A)))$. Thus
we have shown that the r.h.s. of \fer{dyn1} is equal to $\lim_{n
\rightarrow \infty} {\rm Tr} (\rho \sigma^{t}_{(n)}(\pi(A)))$.

Now, $V_n$ converges to $V$ strongly on the dense set ${\cal C}:=\cH_p\otimes \cH_p\otimes{\rm Span}
\{\pi(A)\Omega_0\}$, where $A$ ranges over all polynomials in creation and annihilation operators $a_j^*(f)$, $j=1,2$, with $f\in L^{2}_0$. This follows from \fer{eq2.13} and the
relation
\begin{equation}
\|(V_n - V)\pi(A)\Omega_{0}\|^2 = \omega_{0}(A^{*}(v_{n}^* -
v^*)(v_n - v)A).
\end{equation}
Hence $L^{(\ell)}_n$ converges to $L^{(\ell)}$ strongly on $\cal C$.
The set $\cal C$ is a core for both $L^{(\ell)}_n$ and $L^{(\ell)}$.
(This can be seen by using the GJN commutator theorem, \cite{FM:TI},
Theorem 3.1, by taking $Y=\Lambda+N+\bbbone$ for the comparison
operator in this theorem, and by noticing that $\cal C$ is a core
for $Y$. The latter fact follows from \cite{RSII}, Corollary 2 to
Nelson's analytic vector theorem X.39.) It follows from Theorem
VIII.25 of \cite{RSI} that $L^{(\ell)}_n$ converges to $L^{(\ell)}$
in the strong resolvent sense as $n \rightarrow \infty$. In
particular, $e^{itL_n^{(\ell)}}\rightarrow e^{itL^{(\ell)}}$
strongly, so ${\rm Tr} (\rho \sigma^{t}_{(n)}(\pi(A)))\rightarrow
{\rm Tr} (\rho \sigma^{t}(\pi(A)))$ which, in particular, shows
\fer{michael1}. \hfill $\qed$

\section{Positive Temperature Representation}
\label{App_Repr}

\subsection{Gluing} \label{AppC}

\renewcommand{\GL}{U}
In this appendix, we represent the Hilbert space ${\cal H}$ in a
form which is well suited for a definition of the translation
transformation. This representation is due to \cite{JP:QFII}.

Consider the Fock space
\begin{equation} \label{eqn_A.21}
\cF:= \cF(L^2(X\times\{1,2\})),\ \ \ X= \BR\times S^2
\end{equation}
and denote $x=(u,\sigma)\in X$.
The vacuum in $\cF$ is denoted by $\tilde{\Omega}_r$. The smeared-out
creation operator $a^*(F)$, $F\in L^2(X\times\{1,2\})$ is given by
\begin{equation*}
a^*(F)=\sum_\alpha\int_X F(x,\alpha) a^*(x,\alpha)
\end{equation*}
and analogously for annihilation operators. The CCR read
\begin{equation*}
[a(x,\alpha),a^*(x',\alpha')]=\delta_{\alpha,\alpha'}\delta(x-x').
\end{equation*}
Following \cite{JP:QFII}, we introduce the unitary map
\begin{equation}\label{eqn_A.20}
  \GL : \left[ \cF(L^2(\R^3)) \otimes \cF(L^2(\R^3)) \right]
    \otimes \left[ \cF(L^2(\R^3)) \otimes \cF(L^2(\R^3)) \right]
    \to \cF(L^2(X \times \{1,2\}))
\end{equation}
defined by
\begin{equation}
  \GL \left(\left[\Omega_{r1} \otimes \Omega_{r1}\right] \otimes
      \left[\Omega_{r2} \otimes \Omega_{r2}\right] \right) :=
\tilde{\Omega}_{r}
\end{equation}
and
\begin{eqnarray}
  \GL \Big( \left[a^*(f_1) \otimes \id + \id \otimes a^*(g_1)\right]
      \otimes \id \otimes \id  \hphantom{+ \GL^{-1}\Big)} && \nonumber \\
      + \id \otimes \id \otimes
      \left[a^*(f_2) \otimes \id + \id \otimes a^*(g_2)\right]
  \Big) \GL^{-1} & :=  & a^*(f\oplus g),
\end{eqnarray}
where, for $x=(u,\sigma)\in X$,
\begin{equation}
  \left[ f\oplus g \right](u, \sigma, \alpha) :=
  \begin{cases}
    u \, f_\alpha(u\sigma), & u \ge 0, \\
    u \, g_\alpha(-u\sigma), & u < 0.
  \end{cases}
\end{equation}
This map is extended to the Hilbert space $\cH = \cH^p \otimes
\cF$ in the obvious way. We keep the same notation for its
extension. \\
\indent The operators $L_{r1} \otimes \id_{r2} + \id_{r1} \otimes
L_{r2}$ and $N_{r1}\otimes \id_{r2} + \id_{r1} \otimes N_{r2}$ are
mapped under $\GL$ to the (total) free field Liouvillian and
number operator given by
\begin{eqnarray*}
L_f&=&\d\Gamma(u)=\sum_\alpha\int_X a^*(x,\alpha) u a(x,\alpha),\\
N&=&\d\Gamma(\bbbone)=\sum_\alpha\int_X a^*(x,\alpha) a(x,\alpha).
\end{eqnarray*}
Moreover, the interaction takes the form
\begin{equation} \label{eqn_A_2_6}
\GL I \GL^{-1} = a^*(F_1) + a(F_2),
\end{equation}
where $F_{1,2}\in L^2(X\times\{1,2\}, \cB(\cH_p\otimes\cH_p))$ are
explicitly given by ($x = (u,\sigma) \in X = \R \times S^2$)
\begin{eqnarray}
\lefteqn{ F_1(u,\sigma, \alpha)= \sqrt{\frac{u}{1-e^{-\beta_\alpha
u}}} }\label{eqn_def_F1} \\
&& \times \left\{
\begin{array}{ll}
\sqrt{u}\left( G_\alpha(u\sigma)\otimes \bbbone_p
  -e^{-\beta_\alpha u/2} e^{\delta\beta_\alpha u/2}\ \bbbone_p \otimes
  \alpha_p^{i\, \delta\beta_p/2}(\overline{G_\alpha}^*(u\sigma))\right), & u>0\\
-\sqrt{-u}\left( G_\alpha^*(-u\sigma)\otimes\bbbone_p
  -e^{-\beta_\alpha u/2} e^{\delta\beta_\alpha u/2}\ \bbbone_p\otimes\alpha_p^{i\, \delta\beta_p/2}(\overline{G_\alpha}(-u\sigma))\right), & u<0
\end{array}
\right. \nonumber
\end{eqnarray}
and
\begin{eqnarray}
\lefteqn{
F_2(u,\sigma,\alpha)= \sqrt{\frac{u}{1-e^{-\beta_\alpha u}}} } \label{eqn_def_F2}\\
&& \times \left\{
\begin{array}{ll}
\sqrt{u}\left( G_\alpha(u\sigma)\otimes \bbbone_p
  -e^{-\beta_\alpha u/2}e^{-\delta\beta_\alpha u/2}\ \bbbone_p \otimes \alpha_p^{-i\, \delta\beta_p/2}(\overline{G_\alpha}^*(u\sigma))\right), & u>0\\
-\sqrt{-u}\left( G_\alpha^*(-u\sigma)\otimes\bbbone_p
  -e^{-\beta_\alpha u/2}e^{-\delta\beta_\alpha u/2}\ \bbbone_p\otimes \alpha_p^{-i\, \delta\beta_p/2}(\overline{G_\alpha}(-u\sigma))\right), & u<0
\end{array}
\right. \nonumber
\end{eqnarray}
where $\delta\beta_\alpha=\beta_\alpha-\beta$,
$\delta\beta_p=\beta_p-\beta$, and $\beta=\max(\beta_1,\beta_2)$. \\
\indent Thus the operator $\tilde{K} := \GL K \GL^{-1}$ can be
written as
\begin{equation*}
  \tilde{K} = \tilde{L}_0 + g\tilde{I}
\end{equation*}
where $\tilde{I} = \GL I \GL^{-1}$ is given in (\ref{eqn_A_2_6})
and $\tilde{L}_0 := \GL L_0 \GL^{-1}$ is of the form
\begin{equation*}
  \tilde{L}_0 = L_p \otimes \id_f + \id_p \otimes L_f.
\end{equation*}

\subsection{Complex Deformation}
\label{b2}

Now we express the complex deformation operators $U_\theta$
introduced in Section~\ref{Sect6} in the glued
Hilbert space. For a function $F \in L^2\left( X \times
\{1,2\}\right)$ and $\theta = (\delta, \tau)$, $x=(u,\sigma)\in X$, define
\begin{equation}
\label{eqn_complex_def}
  \left[\tilde{u}_\theta F \right](u, \sigma,\alpha) = e^{\frac{1}{2}\delta{\rm sgn}(u)}
F(j_\theta(u), \sigma, \alpha),
\end{equation}
where
\begin{equation} \label{eqn_A.25}
 j_\theta(u) =  e^{\delta{\rm sgn}(u)} u + \tau,
\end{equation}
and $\rm sgn$ is the sign function, $\mbox{sgn}(u)=1$ if $u\geq
0$, $\mbox{sgn}(-u)=-\mbox{sgn}(u)$. Next, we lift the operator
family $\tilde{u}_\theta$ from $L^2(X \times \{1,2\})$ to the
operator family, $\tilde{U}_\theta$, on ${\cal
H}^p\otimes\cF(L^2(X\times\{1,2\}))$ in a standard way (cf.
(\ref{eqn_6_3})). The family $\tilde{U}_\theta$ is related to the
family $U_\theta$ introduced in Section~\ref{Sect6} as
\begin{equation*}
  U_\theta = \GL^{-1} \tilde{U}_\theta \GL.
\end{equation*}

The operator $\tilde{K}:=UKU^{-1}$ becomes after spectral deformation
\begin{equation}
\tilde{K}_\theta:= \tilde{U}_\theta K \tilde{U}_\theta^{-1}
=\tilde{L}_{0,\theta}+g \tilde{I}_\theta \label{ktheta}
\end{equation}
where
\begin{eqnarray}
\tilde{L}_{0,\theta}
&=&L_p+\cosh\delta \ L_f+\sinh\delta \ \Lambda_f +\tau N, \label{m9}\\
\Lambda_f &=&\d\Gamma(|u|)=\sum_\alpha\int_X
a^*(x,\alpha)|u|a(x,\alpha),
\nonumber\\
\tilde{I}_\theta &=& a^*(F_{1,\theta})+a(F_{2,\theta}) \mbox{\ \ \
\ with \ \ $F_{j,\theta} = \tilde{u}_\theta F_j$.} \label{itheta}
\end{eqnarray}

This spectral deformation can be translated to the original space
$\cH$ as
\begin{equation}
\label{A2.6} K_\theta:=U^{-1} \tilde K_\theta U =
L_{0,\theta} + g I_\theta
\end{equation}
where $L_{0,\theta}:=U^{-1} \tilde L_{0,\theta} U$ is given by
\fer{eqn6_16} and
\begin{equation}
\label{A2.7}
I_\theta = U^{-1} \tilde I_\theta U.
\end{equation}

\section{The $C^*$-algebra $\cA_1$}
\label{AppD}

\begin{proposition}\label{propc1}
Let the algebras $\cA_1$ and $\cA$ be defined by (\ref{eqn_4_9}) with $D^{\rm anal}$ given in the proof below,
and (\ref{calA}), respectively.  Then $\cA_1$ is strongly dense in
$\cA$.
\end{proposition}

{\bf Proof.\ } Let $L^2_1=\{f \mbox{ a.e. continuous},\
\|f\|^2_{L^2_1}:=\int |f(k)|^2 (|k|^{-1} + 1 ) d^3k <\infty\}$.
Recall the real linear map ${\widetilde\gamma_\beta}$ defined in
\fer{gammatilde}. Using that for $x\geq 0$, $\max(1/x,1)\leq
\frac{1+e^{-x}}{1-e^{-x}}\leq 4\max(1/x,1)$, we obtain
\begin{equation}
c || f ||^2_{L^2_1} \leq \|{\widetilde\gamma_\beta}f\|^2_{L^2}
=\int_{S^2}d\sigma\int_0^\infty du\ \frac{1+e^{-\beta
u}}{1-e^{-\beta u}}\ u^2\ |f(u\sigma)|^2 \leq C\|f\|^2_{L^2_1},
\label{aa1}
\end{equation}
so ${\widetilde\gamma_\beta}$ is bounded and invertible (on $R:=
\widetilde\gamma_\beta L^2_1 \subset L^2$), and ${\widetilde\gamma_\beta}^{-1}$ is a
(real linear) bounded map. We have the equivalence
\begin{eqnarray}
g\in R & \Leftrightarrow & g(u,\sigma)=-e^{\beta u/2}
\overline{g}(-u,\sigma) \ \ \mbox{for a.e. $u\in \BR$} \nonumber\\
&& \text {and} \quad \int\limits_{\R} du \int\limits_{S^2} d\sigma
|g(u,\sigma)|^2 < \infty. \label{a1}
\end{eqnarray}
Let $R_0:=\{g\in R,\ e^{bu^2} g\in L^2$ for some $b\geq 0\}\subset
R$. The set $R_0$ is dense in $R$ and \fer{a1} implies that
\begin{equation}
R_0=\{e^{\beta u/4}h|: e^{bu^2} h\in L^2 \mbox{\ for some $b>0$
and\ } h(u,\sigma)=-\overline{h}(-u,\sigma)\}. \label{a2}
\end{equation}
Given $g=e^{\beta u/4}h\in R_0$, define
$h_\epsilon:=G_\epsilon*h$, the convolution in the variable $u$ of
$h$ with the Gaussian $G_\epsilon(u):=\epsilon^{-1}G(u/\epsilon)$,
where $G(u)=\pi^{-1/2}e^{-u^2}$ and $\epsilon>0$. $h_\epsilon$ is
continuous (actually analytic), satisfies $e^{bu^2} h_\epsilon\in
L^2$, and since $G_\epsilon(\cdot)$ is real valued and odd:
\begin{equation*}
h_\epsilon(u,\sigma)= -\overline{h}_\epsilon(-u,\sigma).
\end{equation*}
Therefore, $g_\epsilon:=e^{\beta u/4}h_\epsilon\in R_0$. Since
$h_\epsilon\rightarrow h$ in $L^2$, we conclude that
$g_\epsilon\rightarrow g$ in $L^2$ as $\epsilon \to 0$. Clearly,
$g_\epsilon$ extends to an entire function $z\mapsto
g_\epsilon(z,\sigma )$. Define the set
\begin{equation*}
R^\anal: =\{e^{\beta u/4} h_\epsilon|\ h\mbox{\ satisfies the
conditions on r.h.s. of \fer{a2},\ } \epsilon>0\}.
\end{equation*}
$R^\anal$ is a subset of $R$ that is dense in $R$. Since
${\widetilde\gamma_\beta}^{-1}$ is bounded, then
$\Danal:={\widetilde\gamma_\beta}^{-1}(R^\anal )$ is dense in $L^2_1$. Since
$L^2_1$ is dense in $L^2$, we conclude that $\Danal$ is also dense
in $L^2$.

Define $\cA_1$ as in (\ref{eqn_4_9}) with $\D^\anal$ given above.
Since $\D^\anal$ is dense in $L^2$, $\cA_1$ is strongly dense in
$\cA$ (defined by (\ref{calA})). Next we have for real $\theta$
\begin{eqnarray}\label{eqb4}
&&U_\theta \pi \big(A_p \otimes W (f_1) \otimes W(f_2) \big)
\Omega
\nonumber \\
&=& \pi_p (A_p ) \otimes W \big( (\widetilde\gamma_\beta f_1 )_\theta \big)
\otimes W \big( (\widetilde\gamma_\beta f_2 )_\theta \big) \Omega_\theta
\end{eqnarray}
where the map $g\to g_\theta$ is defined by $g_\theta (u,\sigma ):=
g(j_\theta (u) , \sigma )$ with the function $j_\theta$ defined in
(A2.2) and where we understand the Weyl operators on the r.h.s. as
acting on the (glued) GNS space and
$\Omega_\theta$ given in the same representation. Using that $\theta
\to (\widetilde\gamma_\beta f_j)_\theta$ are analytic for $f_j \in
\D^\anal$ and $(\widetilde\gamma_\beta f_j)_\theta \in L^2 (\BR
\times S^2)$ as long as $| \tan (\rIm \delta ) | < \frac{b}{1+b}$
and expanding $W\big( (\widetilde\gamma_\beta f_j )_\theta)$ into
the Taylor series and $\Omega_\theta$ into the Dayson-Araki\ one,
one can show the r.h.s. of (\ref{eqb4}) has an analytic continuation
in $\theta$ into the neighbourhood of $\BR$ given by
\begin{equation*}
\big\{ \theta \in \BC^2 \big| |\tan (\rIm \delta) | < \frac{b}{1+b}
{\hbox{\quad and\quad }} |\rIm \tau | < \beta^{-1} \big\}.
\end{equation*}
\hfill \qed

\section{The vectors $\zeta^*$}
\label{appE}

In this appendix, we outline the calculation of the vectors $\zeta^*$ in the special cases mentioned in \fer{11.19}-\fer{eqn_11.26a}. As mentioned after \fer{11.19}, we may restrict our attention to $\beta_p=0$.

\begin{itemize}
\item[(i)] $\beta_1=\beta_2=\beta$. Proposition \ref{aepsilon} gives
$$
\Lambda_0^*= \left[e^{-\beta H_p/2}\otimes\bbbone\right] (\Lambda_{10}^* +\Lambda_{20}^*)\left[e^{\beta H_p/2}\otimes\bbbone\right].
$$
Therefore, $\left[e^{\beta H_p/2}\otimes\bbbone\right]\zeta^*$ must be in the kernel of $\Lambda_{10}^* +\Lambda_{20}^*$, which is spanned by $\Omega_p^{\beta}$. Thus
$$
\zeta^*\propto \left[e^{-\beta H_p/2}\otimes\bbbone\right]\sum_{j}e^{-\beta E_j/2}\varphi_j\otimes\varphi_j=\sum_{j}e^{-\beta E_j}\varphi_j\otimes\varphi_j.
$$
The normalization is given by setting $\langle\Omega_p^{(\beta_p=0)},\zeta^*\rangle =1$. This yields the expression \fer{11.15}.

\item[(ii)] Proposition \ref{aepsilon} and equation \fer{eqn_9.4} imply that
\begin{equation}
\Lambda^*_0 = i\Gamma_{10} +O(1)
\label{-5}
\end{equation}
where the operator $O(1)$ is bounded as $\beta_2\rightarrow 0$. The
matrix elements of $\Gamma_{10}=\Gamma_{10}(\beta_2)$ in the basis
$\{\varphi_n\otimes\varphi_n\}$ are (see \cite{BFS:RtE}, Eqns
(B21)-(B22))
$$
(\Gamma_{10})_{m,n} = \delta_{m,n}\sum_{k=0, k\neq m} e^{\beta_2 E_{mk}/2}\eta_{mk} -(1-\delta_{mn})\eta_{mn},
$$
where $\delta_{mn}$ is the Kronecker symbol, and where
$$
\eta_{mn} = 2\pi E_{mn}^2\frac{e^{\beta_2 |E_{mn}|/2}}{e^{\beta_2|E_{mn}|}-1}\int_{S^2}d\sigma |G_2(|E_{mn}|,\sigma)_{mn}|^2.
$$
We expand $\Gamma_{10}$ in $\beta_2$:
\begin{equation}
\Gamma_{10} = \frac{1}{\beta_2} \Gamma' + O(1),
\label{-6}
\end{equation}
where $O(1)$ is bounded as $\beta_2\rightarrow 0$, and where
\begin{eqnarray*}
(\Gamma')_{m,n}&=& 2\pi  \delta_{m,n} \sum_{k=0, k\neq m} |E_{mk}|\int_{S^2}d\sigma |G_2(|E_{mk}|,\sigma)_{m,k}|^2 \\
&& -2\pi (1-\delta_{mn})|E_{mn}| \int_{S^2}d\sigma |G_2(|E_{mn}|,\sigma)_{m,n}|^2.
\end{eqnarray*}
It is obvious that the vector with constant coordinates $[1,1,\ldots,1]^T$ is in the kernel of $\Gamma'$, and, by the Perron-Frobenius theorem, that zero is a simple eigenvalue of $\Gamma'$.

Hence, we see from \fer{-5} and \fer{-6}, by perturbation theory, that the vector in the kernel of $\Lambda^*_0$ is of the form $\zeta^* \propto [1,1,\ldots,1]^T +O(\beta_2)$, as $\beta_2\rightarrow 0$. The proper normalization $\langle\Omega_p^{(\beta_p=0)},\zeta^*\rangle=1$ yields \fer{eqn_11_11}.

\item[(iii)] $\dim \cH_p=2$. Let $E=E_2-E_1>0$. We use Proposition \ref{aepsilon} and formula \fer{fgrc} to obtain the following expression for the level shift operator $\Lambda_0$:
\begin{equation*}
\Lambda_0 = 2\pi iE^2
\left[
\begin{array}{cc}
a & -a\\
-b & b
\end{array}
\right],
\end{equation*}
where
\begin{eqnarray*}
a &=& \sum_{j=1,2} (e^{\beta_j E}-1)^{-1} \int_{S^2} d\sigma \ |[G_j(E,\sigma)]_{12}|^2,\\
b &=& \sum_{j=1,2} \frac{e^{\beta_j E}}{e^{\beta_j E}-1}\int_{S^2} d\sigma \ |[G_j(E,\sigma)]_{12}|^2.
\end{eqnarray*}
It is easily seen that $\Omega_p^{(\beta_p=0)}\propto[1,1]^T$ is in the kernel of $\Lambda_0$, as it should be. The eigenvalues of $\Lambda_0$ are thus zero and ${\rm Tr}\Lambda_0=2\pi i E^2(a+b)\neq 0$. The kernel of $\Lambda_0^*$ is spanned by ${\mathbb C}[b/a,1]^T$, so $\zeta^*\propto [b/a,1]^T=\frac{b}{a}\varphi_1\otimes\varphi_1+\varphi_2\otimes\varphi_2$. The normalization is given by setting $\langle\Omega_p^{(\beta_p=0)},\zeta^*\rangle =1$. This yields the expressions \fer{eqn_11.26}, \fer{eqn_11.26a}.
\end{itemize}

\end{document}